\documentstyle[11pt,aaspp4,psfig]{article}

\def\kms{km~s$^{-1}$}
\def\etal{{\it et al.} }

\begin{document}
\hskip 3.truein Accepted: Astronomical Journal
\vskip 20pt
\title{Spectroscopy of Outlying H II Regions in Spiral Galaxies: \\ Abundances and Radial Gradients}
\author{Liese van Zee\footnote{Jansky Fellow}}
\affil{National Radio Astronomy Observatory,\footnote{The National Radio 
Astronomy Observatory is a facility of the National Science Foundation,
operated under a cooperative agreement by Associated Universities Inc.}
 PO Box 0, Socorro, NM 87801}
\affil{lvanzee@nrao.edu}
\authoremail{lvanzee@nrao.edu}
\author{John J. Salzer\footnote{NSF Presidential Faculty Fellow}}
\affil{Astronomy Department, Wesleyan University, Middletown, CT 06459--0123}
\affil{slaz@parcha.astro.wesleyan.edu}
\author{Martha P. Haynes}
\affil{Center for Radiophysics and Space Research and }
\affil{National Astronomy and Ionosphere Center\footnote{The National
Astronomy and Ionosphere Center is operated by Cornell
University under a cooperative agreement with the National Science Foundation.}}
\affil{Cornell University, Ithaca, NY 14853}
\affil{haynes@astrosun.tn.cornell.edu}
\author{Aileen A. O'Donoghue}
\affil{Physics Department, St. Lawrence University, Canton, NY 13617}
\affil{aodo@music.stlawu.edu}
\author{Thomas J. Balonek}
\affil{Department of Physics and Astronomy, Colgate University}
\affil{13 Oak Drive, Hamilton, NY 13346}
\affil{tbalonek@colgate.edu}
\vfill
\eject
\begin{abstract}
We present the results of low dispersion optical spectroscopy of 186 H II regions
spanning a range of radius in 13 spiral galaxies.   Abundances for 
several elements (oxygen, nitrogen, neon, sulfur, and argon)
were determined for 185 of the H II regions.  As expected, low metallicities were found
for the outlying H II regions of these spiral galaxies.  Radial abundance gradients 
were derived for the 11 primary galaxies; similar to results for other
spiral galaxies, the derived abundance gradients are typically --0.04 to --0.07 dex kpc$^{-1}$.  
 \end{abstract}


\section{Introduction}

Radial abundance gradients, in the sense that the inner H II regions have higher
abundances than the outer ones, are found in almost all large spiral galaxies.
For instance, a gradient of approximately --0.07 dex kpc$^{-1}$ is
observed in both the stellar and gaseous components of the Milky Way (Shaver \etal 
\markcite{S83}1983; Smartt \& Rolleston \markcite{SR97}1997).  The physical basis for the 
presence of an abundance gradient is still a matter of debate;  successful chemical evolution 
models typically include a combination of (1) a radial dependence on the star formation activity, 
generally taking the form of a Schmidt law (e.g., Phillipps \& Edmunds \markcite{PE91}1991); 
(2) radial gas flows (e.g., Edmunds \markcite{E90}1990; G\"otz \& K\"oppen \markcite{GK92}1992); 
and (3) a radial dependence on stellar yields due to IMF variations (e.g., G\"usten \& Mezger 
\markcite{GM82}1982).  The relative importance of these processes can be
constrained by investigating the observed gradients for a large sample of galaxies
which span a range of morphological type (e.g., Vila--Costas \& Edmunds \markcite{VE92}1992)
or environment (e.g., Skillman \etal \markcite{SKSZ96}1996).

Extensive studies of individual galaxies either via spectroscopy (e.g., NGC 5457, 
Kennicutt \& Garnett \markcite{KG96}1996; NGC 2403, Garnett \etal \markcite{GSSSD97}1997)
or spectrophotometric imaging (e.g., NGC 628, NGC 6946, Belley \& Roy \markcite{BR92}1992;
NGC 925, NGC 1073, Martin \& Roy \markcite{MR94}1994; NGC 2366, NGC 4395, Roy \etal
\markcite{RBDM96}1996; NGC 1313, Walsh \& Roy \markcite{WR97}1997) indicate that large
numbers of H II region abundances are necessary to determine accurate abundance gradients.
While spectrophotometric imaging observations have the advantage that large samples are easily 
obtained, they are not as accurate as spectroscopic studies for individual objects. 
Recently, several large spectroscopic surveys and literature compilations have appeared
(e.g., McCall \etal \markcite{MRS85}1985; Vila--Costas \& Edmunds \markcite{VE92}1992; 
Zaritsky \etal \markcite{ZKH94}1994).  These studies indicate that the 
oxygen abundance gradients may be a function of Hubble type (e.g., Vila--Costas \& Edmunds \markcite{VE92}1992).  Furthermore, the mass surface density appears to play a critical role
in supporting an abundance gradient (e.g., Vila--Costas \& Edmunds \markcite{VE92}1992; 
Garnett \etal \markcite{GSSSD97}1997).

If the abundance gradients are constant across the full radial extent of the
galaxy, the outermost H II regions in spiral galaxies should have low abundances,
similar to those found in low luminosity dwarf galaxies.  Thus, the outlying
H II regions of spiral galaxies provide a new environment in which to investigate
metallicity effects on elemental yields (particularly of nitrogen) and other
aspects of the star formation process.  

In this paper, we present the results of new spectroscopic observations of a large 
number of outlying H II regions in 11 spiral galaxies.  In the few instances 
where the target galaxies did not have previous abundance measurements reported in
the literature, spectra of inner H II regions were obtained as well.
The observations presented in this paper complement previous spectroscopic 
observations by extending the radial coverage. 
In a previous paper (van Zee \etal \markcite{vSH98}1998, hereafter Paper I),
these H II region abundances were used to investigate the origin of nitrogen in 
low metallicity environments.  Here, we present the complete data set and discuss the derived
radial abundance gradients.  This paper is organized as follows.  In Section \ref{sec:obs}, 
the galaxy sample is defined and the imaging and spectroscopic observations are described.
The observed H II region line strengths and diagnostics are presented in
Section \ref{sec:lines}.  The abundance derivations are described in 
Section \ref{sec:calcs}; the radial abundance gradients for all 11 spiral galaxies
are discussed in Section \ref{sec:grad}.  The conclusions are summarized in
Section \ref{sec:conc}.

\section{Observations}
\label{sec:obs}

\subsection{Sample Selection}

To facilitate identification of outlying H II regions and to
minimize internal extinction, a sample of 11 nearly face--on spiral galaxies 
were selected for imaging and spectroscopic observations.  In addition, two
galaxies from the NGC 2805 group (NGC 2820 and IC 2458) were also observed; 
NGC 2820 is a nearly edge--on spiral and IC 2458 is a dwarf galaxy. 
Table \ref{tab:props} lists the physical properties of the sample galaxies.   
The morphological type, inclination, position angle, and isophotal radius 
(at the 25 mag arcsec$^{-2}$ surface brightness isophote) for each system have been 
taken from the RC3 (de Vaucouleurs \etal \markcite{RC3}1991).  The inclination
was derived by assuming an intrinsic axial ratio, $q$, of 0.2.  With the exception
of NGC 1068 and NGC 2903, all of the primary targets are classified as late--type 
(Sc or Sd) spirals.  Furthermore, all the galaxies in the sample
have low inclination angles and subtend a large angular size.  

The distances listed in Table \ref{tab:props} were derived either from Cepheid variables
(NGC 2403, Freedman \& Madore \markcite{FM88}1988; NGC 925, Silbermann \etal 
\markcite{Setal96}1996; NGC 5457, Kelson \etal \markcite{Ketal96}1996),
or were estimated from the recessional velocity based on a Virgocentric infall model
and an assumed H$_0$ of 75 \kms~Mpc$^{-1}$ (NGC 628, NGC 1068, NGC 2903, NGC 3184:
Zaritsky \etal \markcite{ZKH94}1994; NGC 4395: Wevers \etal \markcite{WvA86}1986; 
NGC 1232, NGC 1637, NGC 2805, NGC 2820, IC 2458: this paper). Absolute magnitudes 
were computed from the apparent magnitudes listed in the RC3\markcite{RC3}, corrected 
for internal and Galactic extinction, and the adopted distances. All of the primary targets have 
B--band apparent magnitudes brighter than 11.2, corresponding to absolute magnitudes ranging 
between --17.7 and --21.3.

Of the 11 primary targets, three (NGC 1232, NGC 1637, and NGC 2805)
 had no previous abundance measurements reported in the 
literature.  The other 8 were known to have moderate to steep abundance gradients
(e.g., Zaritsky \etal \markcite{ZKH94}1994), but lacked abundance measurements of 
the outermost H II regions.  The total number of new H II region abundances for
each galaxy are tabulated in Table \ref{tab:props}.  Multiple H II regions were 
observed in all galaxies except for NGC 1068, where the outlying H II regions were
discovered to be too faint for the planned abundance measurements.

\subsection{Optical Imaging}

In preparation for the spectroscopic observations, wide field images
of the selected galaxies were obtained with the Burrell Schmidt
telescope at KPNO\footnote{Observations made with the Burrell Schmidt of the 
Warner and Swasey Observatory, Case Western Reserve University.} in
1995 April and October and 1996 January.  The S2KA CCD had a read noise
of 3 $e^{-}$ and a pixel scale of 2.028\arcsec~pix$^{-1}$. A gain of 2.5 DN/$e^{-}$
was selected for the 1995 April observing run; a gain of 3.7 DN/$e^{-}$ was 
selected for all subsequent observing runs.  While the typical seeing was less than
2\arcsec, the spatial resolution of these images is dominated by the large pixel
size.  R--band and H$\alpha$ narrow band images were obtained for all program galaxies. 
During the 1995 April run, three to four 450 second observations of each program 
galaxy were obtained with the R--band filter; during the subsequent observing runs,
one 600 second exposure was obtained for each galaxy.  During all the observing runs,
the H$\alpha$ imaging consisted of several sets of ON--OFF--ON observations with
individual exposure times of 600 seconds each.  A total of three to five H$\alpha$
sets were obtained for each program galaxy (corresponding to total on--line
integration times of 3600 to 6000 seconds). The H$\alpha$ filter (KP 1468) had
a central wavelength of 6571 \AA~and a width of 84 \AA; the OFF images were 
obtained through a matched off--line filter (KP 808) centered at 6411 \AA, 
with a width of 88\AA.  

Data reduction was performed within the IRAF\footnote{IRAF 
is distributed by the National Optical Astronomy Observatories.} package
and followed standard practice. After bias subtraction and flat fielding,
images with multiple exposures were aligned and the sky background was 
subtracted.  The images were then scaled to the level of the ``most photometric'' 
(either the frame with the lowest airmass or the one taken under the best weather 
conditions) and combined with a median filter.  At the same time, the off--line
H$\alpha$ images were scaled to the level of the on--line images, based on the 
observed counts for at least six to ten stars in each frame.

Because the majority of these images were 
taken under non--photometric conditions, they were used primarily as
finding charts for the outlying (faint) H II regions.  The relevant
R--band and continuum subtracted H$\alpha$ images are presented
in Figure \ref{fig:images}.  The images are oriented with north to the
top and east to the left.  The 2\arcmin~slits on the H$\alpha$ images
indicate the spatial scale of each pair of images.

Optical scale lengths (R$_{\rm d}$), tabulated in Table \ref{tab:props}, were derived from
the R--band images.  The IRAF task ELLIPSE was used to fit ellipses of
constant surface brightness.  We note that several of the galaxies in the 
present sample are asymmetric, making such fits difficult to interpret.  
In particular, satisfactory fits were obtained only for NGC 628, NGC 925,
NGC 1232, NGC 2403, and NGC 2903.  The scale length tabulated in Table
\ref{tab:props} for NGC 1068 was derived from a fit to only the outer 
isophotes, those with $r >$ 50\arcsec.  By far the worst fit was obtained 
for the severely asymmetric spiral NGC 1637; the scale length presented here
should only be taken as a rough estimate.  The derived scaled lengths for
the remaining galaxies, NGC 2805, NGC 3184, NGC 4395, and NGC 5457, are 
reasonable, although we caution that the ellipse fitting was not ideal 
due to their modest asymmetries.
  
\subsection{Optical Spectroscopy}

Low resolution optical spectra of the selected H II regions were
obtained with the Double Spectrograph on the 5m Palomar\footnote{Observations
at the Palomar Observatory were made as part of a continuing 
cooperative agreement between Cornell University and the
California Institute of Technology.} telescope during observing runs
in 1996 May and November and 1997 January.   The long slit
(2\arcmin) was set at a 2\arcsec~aperture during all observing runs; 
the seeing was generally better than 2\arcsec.   During all observing runs, a
5500 \AA~dichroic was used to split the light to the two 
sides (blue and red), providing complete spectral 
coverage from 3500--7600 \AA.  The blue spectra were acquired 
with the 300 lines/mm diffraction grating (blazed at 3990 \AA).  
The red spectra were acquired with the 316 lines/mm diffraction 
grating (blazed at 7500 \AA).  A  thinned 800$\times$800 TI CCD with a 
read noise of 8 e$^-$, a gain of 1.5, and 9.2 \AA~effective resolution 
(2.19 \AA/pix) was used on the blue side; a thinned 1024$\times$1024 TEK CCD
with a read noise of 7.5, gain of 2.0, and an effective resolution 
of 7.8\AA~(2.46 \AA/pix) was used on the red side.  The spatial scale of 
the long slit was 0.8 arcsec/pix on the blue and 0.49 arcsec/pix on the red.
The CCD on the blue side had moderate focussing problems across
the full wavelength range.  During all observing runs, the focus was optimized
for H$\delta$ and [O III] $\lambda4363$.

 The observations were conducted via blind offsets from nearby stars.
Stellar and H II region coordinates were derived from the Schmidt narrow 
band H$\alpha$ images.  Astrometric plate solutions were calculated using 
the coordinates of the bright stars in the HST Guide Star Catalog, 
yielding positions accurate to within 1\arcsec. The telescope pointing was 
verified each run by using the same offsetting technique to move between
several stars in a field.  Offsets between stars and H II regions were
typically less than 20\arcmin.  Previous observations (e.g., van Zee \etal
\markcite{vHS97}1997) indicated that the 5m telescope could point and track
accurately; nonetheless, the newly commissioned offset guider was used to 
improve the tracking during long exposures.

To reduce light losses due to atmospheric refraction, the slit position angle was 
set close to the parallactic angle during the observations.  The
slit positions are illustrated on the H$\alpha$ images in Figure \ref{fig:images}
and tabulated in Table \ref{tab:obs}.  The slit numbers listed in Table
\ref{tab:obs} are ordered in increasing Right Ascension for each galaxy
and each slit is labelled by the offset (arcseconds east--west and north--south) 
from the galaxy center (tabulated in Table \ref{tab:props}).  The offsets do not
necessarily correspond to the location of an H II region since the telescope was 
often pointed between two (or more) H II regions in order to maximize the number 
of H II regions observed.  The astrometric pointing centers are also given in Table 
\ref{tab:obs} along with the position angle (PA) of the slit.  Finally, the 
observing run and the total integration times are listed.  Most observations 
consisted of two or three 1200 second exposures; in a few instances, primarily due 
to time constraints, the integration time was only 300 seconds. 

Flux calibration was obtained by observations of at least 5 standards per night
(Stone \markcite{S77}1977; Oke \& Gunn \markcite{OG83}1983;
 Massey \etal \markcite{MSBA88}1988), interspersed with the H II region 
observations.  Wavelength calibration was obtained by observations 
of arc lamps taken before and after each H II region observation.  
A Hollow Cathode (Fe and Ar) lamp was used to calibrate the blue 
spectra; a combination of He, Ne, and Ar lamps were used to calibrate 
the red spectra.

The spectra were reduced and analyzed with the IRAF package.
The spectral reduction included bias subtraction, scattered light corrections,
and flat fielding with both twilight and dome flats.   The 2--dimensional images were
rectified based on the arc lamp observations and the trace of stars
at different positions along the slit.  The sky background was removed
from the 2-dimensional images by fitting a low order polynomial along 
each row of the spectra.  One dimensional spectra of each H II region were 
then extracted from the rectified images.  Throughout this paper, the H II region 
nomenclature is based on east--west and north--south offsets from
the galaxy center.  These offsets were derived either from the pointing center
(if only one H II region was in the slit), or were computed
for each H II region from the pointing center, the slit position angle,
and the spatial scale of the image.  These offsets, therefore, are generally only accurate
to 1--2\arcsec.
 
The 1D spectra were corrected for atmospheric extinction
and flux calibrated.  Since the  night--to--night variation in the 
calculated sensitivity function was small,  the sensitivity 
function for each run was created using the standard stars from 
all nights.  As a final step, the individual 20 minute exposures 
were combined.  In general, there was excellent agreement in the
continuum level of the final blue and red spectra for each H II region, 
indicating that the flux calibration and extraction regions were well matched.

A few representative spectra are shown in Figure \ref{fig:spec}.
The three H II regions in Figure \ref{fig:spec} are ordered in
increasing distance from the center of NGC 1232.  Figure \ref{fig:spec}(a)
illustrates a typical spectrum from the inner galaxy; absorption features
are present and the low ionization lines, such as [O II] and [N II], are
strong while the high ionization lines, such as [O III], are extremely
weak or absent.  Figure \ref{fig:spec}(b) illustrates a typical spectrum
from the outer galaxy; the emission lines dominate the spectrum, with 
both the low and high ionization lines at moderate strength.  Finally,
Figure \ref{fig:spec}(c) illustrates the low abundance nature of the
outermost H II regions.  Here, the [O III] lines are very strong while
[N II] and [S II] are weak, indicating that this is a high temperature
(low abundance) H II region.  Furthermore, [O III] $\lambda$4363 is clearly detected
in this spectrum, permitting an accurate temperature estimate (see Section \ref{sec:temp}).

\section{Line Ratios and Diagnostic Diagrams}
\label{sec:lines}

\subsection{Extinction}

The reddening along the line of sight to each H II region was computed from
the observed strengths of the Balmer emission lines.  The intrinsic line ratios of
H$\alpha$/H$\beta$  and H$\gamma$/H$\beta$ were interpolated from
the values of Hummer \& Storey \markcite{HS87}(1987)
for case B recombination, assuming that N$_e$ = 100 cm$^{-3}$ and
T$_e$ = 10000 K. For the few H II regions where the [O III] $\lambda$4363 line
was detected, the electron temperature as derived from the [O III] lines 
(see Section \ref{sec:temp}) was used instead of the nominal 10000 K.
The reddening function, $f(\lambda)$, normalized at H$\beta$ from the galactic 
reddening law of Seaton \markcite{S79}(1979) as parameterized by 
Howarth \markcite{H83}(1983), was used, assuming a value of $R = A_V/E_{B-V} = 3.1$. 
When necessary to bring the two Balmer line ratios into agreement, an underlying 
Balmer absorption with an equivalent width of 2 \AA~was assumed.   The reddening 
coefficients, c$_{H\beta}$, derived for each H II region are listed in Table 
\ref{tab:lines} and are shown as a function of radius in Figure \ref{fig:red}.  
While there is a fair amount of scatter in the reddening plots, the general
trend is for the outer H II regions to have lower extinction than the inner ones.
Similar extinction gradients have been seen in photometric studies of the 
dust content in spiral galaxies (e.g., Peletier \etal \markcite{PVMFKB95}1995).
 
\subsection{Line Intensities}

The reddening corrected line intensities relative to H$\beta$
are given in Table  \ref{tab:lines}.  In this and subsequent tables,
each H II region is identified by its east--west and north--south offsets
from the galaxy center (in arcsec, north and east are positive) and by its slit number 
(Table \ref{tab:obs}).  Furthermore, the H II regions for each galaxy are ordered by 
increasing radial distance.  The error associated with each relative line intensity
was determined by taking into account the Poisson noise in the line, the
error associated with the sensitivity function, the contributions of
the Poisson noise in the continuum, read noise, sky noise, and flat
fielding or flux  calibration errors, the error in setting the 
continuum level (assumed to be 10\% of the continuum level),
and the error in the reddening coefficient.  For a few of the high abundance
H II regions, the [O III] lines were so weak that [O III] $\lambda$4959 was
not detected.  In those instances, the tabulated [O III] line intensities
were derived by assuming that the [O III] $\lambda$5007 line intensity was
2.88 times the intensity of [O III] $\lambda$4959. 

\subsection{Diagnostic Diagrams}
\label{sec:diag}

Since the majority of the H II regions in this sample are from the outer regions 
of spiral galaxies, they should form well defined H II region sequences in  
diagnostic diagrams (e.g., Osterbrock \markcite{O89}1989). 
A few galactic nuclei (NGC 925+002--002, NGC 1637--001--000, and NGC 4395--003--003) were
also observed, however, so it is worthwhile to examine such plots to
determine if these are stellar photoionized emission regions or AGN.
Several of the line ratios commonly used as H II region diagnostics 
are tabulated in Table \ref{tab:ratios} and are graphically shown in
Figures \ref{fig:nh} and \ref{fig:sh}.   In these and subsequent Figures, the
H II regions in each galaxy are coded by common symbols.  The well defined
H II region sequence formed by the [N II]/H$\alpha$ and [O III]/H$\beta$ line ratios 
is illustrated in Figure \ref{fig:nh}.  A theoretical curve from
the H II region model of Baldwin \etal \markcite{BPT81}(1981) is superposed
on the data in Figure \ref{fig:nh}(a), illustrating that the majority are indeed H II
regions.  Panel (b) shows the same diagnostic, but with the sum of [O II]/H$\beta$ and 
[O III]/H$\beta$ (R$_{23}$) on one axis instead of [O III]/H$\beta$ (after McCall \etal \markcite{MRS85}1985).
In both of these plots, NGC 4395--003--003 (marked by an arrow) deviates from the overall
H II region sequence.  This supports the suggestion by Ho \etal \markcite{HFS95}(1995) 
that the nucleus of NGC 4395 harbors a dwarf Seyfert.  Similarly, while the diagnostic
diagrams from [S II]/H$\alpha$ are less definitive, the nucleus of NGC 4395 again
falls outside of the H II region sequence in Figure \ref{fig:sh}.  Due to the unknown
ionization source for NGC4395--003--003, it has been excluded from the following abundance
analysis.

\section{Abundance Determinations}
\label{sec:calcs}
Several assumptions are required to convert observed line ratios
into nebular abundances.  In sections \ref{sec:density} -- \ref{sec:temp},
 we describe the methodology used to obtain the electron density and electron
 temperature for each H II region.  In section \ref{sec:icf}, we describe the
 ionization correction factors which were used to  convert the derived ionic abundances to 
nebular abundances.  Finally, in sections \ref{sec:abund} and \ref{sec:alpha}
we discuss the derived oxygen abundances and the relative enrichments of
nitrogen, neon, sulfur, and argon.

\subsection{Electron Density}
\label{sec:density}
The density sensitive line ratio [S II] $\lambda6717/6731$ is tabulated
for each H II region in Table \ref{tab:ratios} and graphically
illustrated in Figure \ref{fig:density}.  The maximum value of this
line ratio in the low density limit is marked on Figure \ref{fig:density}.  
The majority of the H II regions are within the low density limit
(I($\lambda$6717)/I($\lambda$6731) $>$ 1.35).  In the following abundance
analysis, an electron density of 100 cm$^{-3}$ was assumed for
the majority of the H II regions; in the few instances where the [S II] line
ratios were below the low density limit, a version of the FIVEL program of 
De Robertis \etal \markcite{DDH87}(1987) was used to calculate
the electron density from the observed [S II] line ratios.

\subsection{Semi--Empirical Oxygen Abundances}
\label{sec:r23}
The optical oxygen lines ([O II] $\lambda3727$ and [O III] $\lambda\lambda 4959,5007$)
are sensitive to both the oxygen abundance and electron temperature.  At low metallicities,
the dominant coolant is the collisionally excited Lyman series, and thus the total oxygen 
line intensity (R$_{23} \equiv$ ([O II]+[O III])/H$\beta$) increases as the abundance increases.  
As the metallicity increases, however, the infrared fine structure
lines, such as the 52$\mu$ and 88$\mu$ oxygen lines and the C$^+$ 158$\mu$ transition,
begin to dominate the cooling.  At an oxygen abundance of approximately 1/3 solar, 
the intensity of the optical oxygen lines reaches a maximum and then declines
as the oxygen abundance increases.  The behavior of the optical oxygen line intensities
as a function of oxygen abundance has been examined via semi--empirical
methods (e.g., Searle \markcite{S71}1971; Pagel \etal \markcite{PEBCS79}1979;
Alloin  \etal \markcite{ACHV79}1979; Edmunds \& Pagel \markcite{EP84}1984; 
Dopita \&  Evans \markcite{DE86}1986; Skillman \markcite{S89}1989) and derived  
from H II region ionization models (e.g., McGaugh \markcite{M91}1991; Olofsson 
\markcite{O97}1997).   The theoretical
models of McGaugh \markcite{M91}(1991) are shown in Figure \ref{fig:r23}.  The surface 
is double valued due to the ambiguity between high and low abundances.  Furthermore,
the geometry of the H II region (represented by the average ionization parameter, \=U,
the ratio of ionizing photon density to particle density)  introduces a further
spread in the estimated abundance for a given R$_{23}$.  In addition, aging of the
H II regions can also introduce further scatter as both the ionization parameter
and the shape of the ionizing spectrum evolve (e.g., Stasi\'nska \& Leitherer
\markcite{SL96}1996; Olofsson \markcite{O97}1997); we do not include such effects 
in the present analysis.
 
The observed line strengths for the H II regions are plotted in Figure \ref{fig:r23}.
The lowest values of R$_{23}$ occur for the innermost H II regions observed, and therefore
presumably populate the high abundance side of the R$_{23}$ surface (see below).  At these high oxygen
abundances, the ionization parameter (\={U}) is approximately 0.001 for all of the H II regions.
In contrast, the H II regions with low oxygen abundances span a wide range of \={U}, indicating
that it is important to include such ionization effects when using the strong line
method to estimate oxygen abundances.

The process of determining empirical oxygen abundances from the strong line ratios
requires several steps.  First, abundances for both the high and low abundance side
of the surface were determined.
Further information is required, however, to resolve the ambiguity between
these two values.  The degeneracy  can be broken
by looking at the line strengths of other species. For instance, the [N II]/[O II] line ratio 
forms a single parameter sequence from high to low abundance (Figure \ref{fig:no}).  
In general, H II regions with log ([N II]/[O II]) $<$ --1.0 are believed to have
low oxygen abundances, while those with log ([N II]/[O II]) greater than --1.0 are on the 
high abundance side. 
For those H II regions where the [N II]/[O II] diagnostic was conclusive 
(those with log ([N II]/[O II]) $>$ --0.8 or log ([N II]/[O II]) $<$ --1.05), the abundance 
from the appropriate side of the surface was assigned.
 
A large fraction of the H II regions in this sample fall in the ambiguous region of
the R$_{23}$ relation (the turnover region) where the [N II]/[O II] diagnostic is
inconclusive.  Thus, it was necessary to identify an alternative means of determining the
oxygen abundance for these H II regions.  Like the [N II]/[O II] diagnostic, the
[N II]/H$\alpha$ ratio increases with increasing oxygen abundance (Figure \ref{fig:oh}).
The relation between the [N II]/H$\alpha$ line ratio and the oxygen abundance was
determined by a weighted least squares fit to the low abundance H II regions from van Zee \etal
\markcite{vHS97}(1997) and the high abundance H II regions (8.8 $<$ 12 + log (O/H) $<$ 9.1)
in the present sample:   
\begin{equation}
{\rm 12 + log (O/H) =  1.02~log ([N II]/H\alpha) + 9.36.}
\end{equation}
Note that the [N II]/H$\alpha$ line ratio is only valid as a metallicity estimator for 
12 + log (O/H) $<$ 9.1 and in the absence of shock excitation.  Furthermore, 
with typical errors of 0.2 dex or more, it is not a particularly accurate
abundance estimator.  However, it does provide an additional method to break 
the ambiguity of the R$_{23}$ abundances.

For those H II regions where the oxygen abundance could not be resolved based on the
[N II]/[O II] line ratio, the [N II]/H$\alpha$ relation was used to calculated an
empirical oxygen abundance.  This empirical oxygen abundance was then compared with the
two (high and low abundance) estimates from the R$_{23}$ relation.  Whichever R$_{23}$
abundance was closest to the empirical abundance was then assigned, unless
both the R$_{23}$ abundances were significantly different from the
empirical abundance.
In the instances where neither the upper nor the lower branch R$_{23}$ abundances were
within 0.2 dex of the empirical abundance, the empirical abundance
was assigned.   This primarily occurred when the empirical abundance estimate was
between 8.2 and 8.5, corresponding to the turnover region of the R$_{23}$ relation,
a region where typical errors are several tenths of a dex.    
 
\subsection{Electron Temperature}
\label{sec:temp}

The  preferred method of obtaining an estimate of the electron temperature of the 
ionized gas is to derive it from the corrected line strengths of the [O III] lines.  
For the 16 H II regions where [O III] $\lambda$4363 was detected (9\% of the sample), the [O III] 
ratio $\lambda\lambda 4959,5007/\lambda4363$ is tabulated
in Table \ref{tab:ratios}.  A version of the FIVEL program of 
De Robertis \etal \markcite{DDH87}(1987)  was used to compute T$_e$
from this reddening--corrected line ratio.   Not unexpectedly, the H II
regions with detectable [O III] $\lambda$4363 correspond to those with high
T$_e$; the electron temperatures were typically in the range of 9000 to 14000 K,
with errors of 300 to 600 K.

In most cases, however, [O III] $\lambda$4363 was  not detected, or was
 contaminated by the nearby Hg $\lambda$4358 night sky line.  In these
instances, the electron temperature was estimated from ionization models.
Because oxygen is one of the dominant coolants, the electron temperature
depends strongly on the oxygen abundance.  Thus, if the oxygen abundance
is known, the electron temperature can be estimated.  The anticorrelation
between oxygen abundance and electron temperature (Pagel \etal \markcite{PEBCS79}1979; 
Shaver \etal \markcite{S83}1983) was
used in a self--consistent manner to estimate T$_{e}$. The errors in the
derived temperatures are quite large. For the H II regions where the 
electron temperature was derived from the assumed oxygen abundance,
the error in the electron temperature was set to yield the appropriate error
(as determined by the calibration process) in the computed oxygen abundance.  
For the abundances estimated from R$_{23}$ (71\% of the sample),
this corresponds to 0.1 dex, or $\sim$500 K for a T$_e$ of 5000 to 7000 K.
For the oxygen abundances estimated from the [N II]/H$\alpha$ relation (20\% of the
sample), errors
of 0.2 dex in the oxygen abundance were adopted, corresponding to temperature uncertainties 
of $\sim$ 2000 K
for a T$_e$ of 10000 K.  As a check of self--consistency, 
the same method was used to estimate the electron temperatures for those H II regions
where [O III] $\lambda$4363 was detected; in general, the two methods yielded
similar results, with absolute differences less than the estimated errors for the
empirical method. The electron temperatures adopted for the subsequent abundance
analysis are tabulated in Table \ref{tab:abund}.  

 Numerical models of H II regions have shown that, particularly at low abundances,
there are significant differences between the temperatures in the high--
and low--ionization zones (Stasi\'nska \markcite{S80}1980).  This difference
must be taken into account to derive accurate abundances.  We therefore
adopt the approach taken by Pagel \etal \markcite{PSTE92}(1992), who use the H II region 
models of Stasi\'nska \markcite{S90}(1990) to derive an approximation for the electron 
temperature of the O$^+$ zone:
\begin{equation}
T_e(O^+) = 2 [ T_e^{-1}(O^{++}) + 0.8]^{-1},
\end{equation}
where $T_e$ is the electron temperature in units of $10^4$ K.  

\subsection{Ionization Correction Factors}
\label{sec:icf}
For all atoms other than oxygen, the derivation of atomic 
abundances requires the use of ionization correction factors (ICFs) to account 
for the fraction 
of each atomic species which is in an unobserved ionization state.
To estimate the nitrogen abundance, we have 
assumed that N/O = N$^+$/O$^+$ (Peimbert \& Costero \markcite{PC69}1969).
 To estimate the neon abundance, we have assumed that 
Ne/O =  Ne$^{++}$/O$^{++}$ (Peimbert \& Costero \markcite{PC69}1969).
To determine the sulfur and argon abundances, we are forced to adopt an 
ICF from published H II region models to correct for the unobserved S$^{+3}$ 
and Ar$^{+3}$ states.   We have adopted  analytical forms of the ICFs as given by
 Thuan \etal \markcite{TIL95}(1995).  In H II regions where S$^{++}$  $\lambda 6312$ 
was not detected, an ICF  was assumed based on the average value of the ICF in
all other H II regions.  The use of an ``average'' ICF 
assumes that the ionization characteristics of the H II regions with
non--detections of the S$^{++}$  $\lambda 6312$ line are similar to those
where  $\lambda 6312$ was detected.  This is not necessarily
the case, but it appears to work reasonably well.
 We have assumed an  error of 50\% in this ``average'' ICF of 5.8$\pm$2.9.

\subsection{Oxygen Abundances}
\label{sec:abund}
The oxygen abundances tabulated in Table \ref{tab:abund} were derived from the
observed line strengths (Table \ref{tab:lines}) and  the relevant emissivity coefficients
(based on the electron temperature (Table \ref{tab:abund}) and electron density) 
as calculated by a version of the FIVEL program of De Robertis \etal \markcite{DDH87}(1987). 
The errors in the derived abundances are dominated by the errors inherent in the electron 
temperature estimates.  

As expected, the outermost H II regions in the spiral galaxies have low oxygen
abundances.  The new observations confirm the low abundance nature of H681 (+010+885)
in NGC 5457 (Garnett \& Kennicutt \markcite{GK94}1994), which has the lowest
oxygen abundance (12 + log (O/H) = 7.92 $\pm$ 0.03) of all the spiral galaxy H II regions
in this sample (the H II regions in the dwarf galaxy IC 2458 also have similarly
low oxygen abundances).  The highest oxygen abundances are found in the asymmetric
spiral NGC 1637, with typical abundances of 1.5 to 2 times the solar value.

The majority of the oxygen abundances were ultimately derived from the strong line
method (either from the McGaugh \markcite{M91}(1991) R$_{23}$ calibration
or the modified version of this R$_{23}$ relation as described in Section \ref{sec:r23}).
Other calibrations of the R$_{23}$ relation yield slightly different results.
In particular, the R$_{23}$ calibration of Zaritsky \etal
\markcite{ZKH94}(1994) (ZKH) will yield very similar abundances for H II regions with
12 + log (O/H) $>$ 8.5 (Table \ref{tab:abund}).  Note, however, that since the ZKH 
analytical form does not explicitly take into account the turnover in the R$_{23}$ 
relation, it can result in artificially high abundances for the outlying H II regions. 
A comparison of the derived abundances from the modified McGaugh calibration (this paper)
with those derived from the Zaritsky \etal \markcite{ZKH94}(1994) analytical 
formula is shown in Figure \ref{fig:zkh}.  Also shown in Figure \ref{fig:zkh} and tabulated
in Table \ref{tab:abund} are the R$_{23}$ abundance estimates from the calibration
of Edmunds \& Pagel \markcite{EP84}(1984).  It is clear from 
this plot that the EP84 calibration will yield
steeper abundance gradients than either the ZKH or the modified McGaugh
calibrations used here.  

\subsection{Nitrogen, Neon, Sulfur, and Argon Abundances}
\label{sec:alpha}

The abundances of nitrogen, neon, sulfur, and argon were computed from the
observed line strengths (Table \ref{tab:lines}), the emissivity coefficients,
and the ionization correction factors described in Section \ref{sec:icf}.
Their abundances relative to oxygen are tabulated in Table \ref{tab:abund}.
The original purpose of the observations described here was to investigate
the N/O ratio in the outlying, low abundance, H II regions of spiral galaxies.
Detailed analysis of the derived N/O ratios is presented in
a separate paper (van Zee \etal \markcite{vSH98}1998, Paper I).  Briefly,
as seen in other high metallicity H II regions, the N/O ratio increases linearly 
with the oxygen abundance at high abundance. However,
the N/O ratio plateaus for H II regions with metallicities less than 1/3
solar, indicating that there is a primary origin for nitrogen in low metallicity
environments.

The relative abundances of the $\alpha$ elements (Ne, S, and Ar) are expected
to be constant with increasing oxygen abundance.  Their abundances relative
to oxygen are shown in Figure \ref{fig:alpha}. Also shown in Figure \ref{fig:alpha}
are the solar values (Anders \& Grevesse \markcite{AG89}1989). As expected for primary
elements, the relative abundance ratios are essentially constant:   the
mean log (Ne/O) is --0.65 $\pm$ 0.18; the mean log (S/O) is --1.55 $\pm$ 0.15; and the
mean log (Ar/O) is --2.28 $\pm$ 0.11.  These mean ratios are similar to those derived
for low metallicity H II regions in dwarf galaxies (e.g., Thuan \etal \markcite{TIL95}1995;
van Zee \etal \markcite{vHS97}1997).
 
\section{Radial Abundance Gradients}
\label{sec:grad}
The oxygen abundances presented in this paper increase the total number 
of H II regions observed in these spiral galaxies.  Furthermore, a large fraction of
the H II regions in this sample are located at radii greater than half of the isophotal radius.  
These outlying H II regions provide a significant lever--arm for the computation of 
radial abundance gradients
(see also Ferguson \etal \markcite{FGW98}1998).  
The spiral galaxy abundance gradients are illustrated in Figure \ref{fig:rad_oh}.
All deprojected radii have been computed from the H II region offsets and the
assumed inclination and position angles (Table \ref{tab:props}); the deprojected
radii are tabulated (in arcsec) in Table \ref{tab:abund}.
In Figure \ref{fig:rad_oh}, the deprojected radii have been normalized by the 
isophotal radius (R$_{25}$), as listed in the RC3 (de Vaucouleurs \etal 1991) and tabulated in
Table \ref{tab:props}.  The filled symbols in Figure \ref{fig:rad_oh}  represent
H II regions from the present study.  The open circles represent data from 
the literature:   NGC 628 (McCall \etal  \markcite{MRS}1985), NGC 925 (Zaritsky \etal 
\markcite{ZKH94}1994), NGC 1068 (Evans \& Dopita \markcite{ED87}1987; Oey \& Kennicutt 
\markcite{OK93}1993), NGC 2403 (McCall \etal \markcite{MRS85}1985; Fierro \etal 
\markcite{FTP86}1986; Garnett \etal \markcite{GSSSD97}1997), NGC 2903 (McCall \etal 
\markcite{MRS85}1985; Zaritsky \etal \markcite{ZKH94}1994), NGC 3184 (Zaritsky \etal
\markcite{ZKH94}1994), NGC 4395 (McCall \etal \markcite{MRS85}1985), and
NGC 5457 (Kennicutt \& Garnett \markcite{KG96}1996).  To place the literature
abundances on a common scale with the abundances derived in this paper, the
analytical form of the R$_{23}$ calibration presented by Zaritsky \etal \markcite{ZKH94}(1994)
was used to compute the oxygen abundances.  As illustrated in Table \ref{tab:abund} and
Figure \ref{fig:zkh}, this analytical form of the R$_{23}$ calibration 
yields similar abundance estimates for high abundance H II regions.   The only
exceptions to this recalibration of the literature data were the few instances where
abundances were derived from an electron temperature which was directly measured, 
either through the [O III] or [O II] line ratios (e.g., NGC 5457, Garnett \& Kennicutt 
\markcite{GK94}1994; NGC 2403, Garnett \etal \markcite{GSSSD97}1997). 

The oxygen abundance gradients obtained from weighted least--squares
fits to both the literature and new data points are tabulated in Table \ref{tab:grad}. 
 For those galaxies with previous
abundance gradients reported in the literature, the newly derived abundance
gradients are generally in agreement with older results (see, e.g., the compilations
by Zaritsky \etal \markcite{ZKH94}1994; Garnett \etal \markcite{GSSSD97}1997).  
The few exceptions are NGC 925, NGC 1068, and NGC 4395.  For NGC 1068, the one 
additional H II region in the present study is at such a large radius compared to 
the previous observations that it determines the gradient.  Further observations of 
the other outlying H II regions in NGC 1068 will be necessary to confirm the derived gradient. 
For NGC 925 and NGC 4395, the new observations also result in steeper abundance gradients 
than previously reported in the literature.  The majority of the H II regions in both of 
these galaxies are at, or near, the R$_{23}$ turnover, so the derived abundances have 
large errors.  The previous spectroscopic studies of NGC 925 (Zaritsky \etal \markcite{ZKH94}1994)
and NGC 4395 (McCall \etal \markcite{MRS86}1986) were limited by the small number
of H II regions observed; in these studies, the abundance gradients were swamped by the
intrinsic errors of the abundance measurements.  Spectrophotometric imaging observations
of a large number of H II regions in NGC 925 (Martin \& Roy \markcite{MR94}1994)
suggested that a steeper abundance gradient existed; the newly derived abundance
gradient from the spectroscopic observations is in good agreement with
the spectrophotometric imaging results.  Spectrophotometric imaging
observations of NGC 4395 indicated that the abundance gradient was flat (or nonexistent)
in this low luminosity galaxy (Roy \etal \markcite{RBDM96}1996); the new spectroscopic 
results suggest that a shallow gradient exists, but the errors in the derived slope are quite large.

A large scatter in the oxygen abundance for a given radius can indicate azimuthal variations of
the abundance gradient  An apparent branching of the abundance gradient of NGC 5457 at 
radii between R/R$_{25}$ $\sim$0.2--0.5 was noticed by Kennicutt \& Garnett \markcite{KG96}(1996).
The two branches are spatially separate, with the higher abundance H II regions lying
on the northwest half of the galaxy and the lower abundance H II regions
lying on the southeast side (near NGC 5461).  As discussed by Kennicutt \& Garnett 
\markcite{KG96}(1996), these abundance variations could arise from the inherent asymmetry 
of the galaxy, from tidal interactions with its companions NGC 5474 and/or NGC 5477, or
from accretion of high--velocity gas.  Other galaxies, such as NGC 925 and NGC 2403,
also show large variations in the abundance at a given radius.  The H II regions in 
both of these galaxies lie near the turnover region of the R$_{23}$ relation, however.
Thus, the large scatter in oxygen abundance is probably due to the intrinsic errors in the 
abundance measurements.   New observations with solid detections of the temperature
sensitive lines will be necessary to address the validity of these abundance variations.

Observations of the Milky Way suggest that the abundance gradient may flatten in the outer disk
(e.g., V\'\i lchez \& Esteban \markcite{VE96}1996).  The outermost points of NGC 5457 suggest
that the abundance gradient may flatten in this galaxy as well.   However, previous claims of 
flattening in the inner regions of NGC 5457 (e.g. Zaritsky \markcite{Z92}1992; Scowen \etal \markcite{SDH92}1992; Vila--Costas \& Edmunds \markcite{VE92}1992) have not been
substantiated (Henry \& Howard \markcite{HH95}1995; Kennicutt \& Garnett \markcite{KG96}1996).  
 To determine if flattening of the abundance gradient is a common phenomena of the outer disk, 
observations of H II regions at even larger radii 
will be necessary.  Such observations are challenging due to the faintness of the outermost 
H II regions, but not impossible (e.g., Ferguson \etal \markcite{FGW98}1998).

\section{Conclusions}
\label{sec:conc}
We have presented the results of high signal--to--noise spectroscopy of 186
H II regions spanning a range of radius in 13 spiral galaxies.  Abundances 
for several elements (oxygen, nitrogen, neon, sulfur, and argon) were
derived for all except the nucleus of NGC 4395.  Below, we summarize the main conclusions.

(1) The H II region diagnostic diagrams indicate that the nucleus of NGC 4395 harbors
a dwarf Seyfert.  The remaining 185 H II regions form a well defined H II region
sequence in the Osterbrock diagnostic diagrams.  Furthermore, the majority of
the H II regions are low density, with an assumed N$_e$ of 100 cm$^{-3}$. 

(2) A modified version of the R$_{23}$ calibration of McGaugh \markcite{M91}(1991)
was used to derive the majority of the oxygen abundances.  In this method, the 
[N II]/H$\alpha$ line ratios are used to break the degeneracy between the upper and 
lower branches; furthermore, this line ratio is used to estimate the oxygen abundance 
for H II regions which fall in the R$_{23}$ turnover region (8.2 $<$ 12 + log (O/H) $<$ 8.5).

(3) The enrichment of the $\alpha$ elements (neon, sulfur, and argon) relative
to oxygen is constant for all H II regions in this sample. The mean $\alpha$--to--oxygen 
ratios are consistent with those derived for low metallicity H II regions 
(e.g., Thuan \etal \markcite{TIL95}1995; van Zee \etal \markcite{vHS97}1997).  
The mean abundance ratios are:  --0.65 $\pm$ 0.18,  --1.55 $\pm$ 0.15, and 
--2.28 $\pm$ 0.11 for log (Ne/O), log (S/O), and log (Ar/O), respectively.  

(4) The outlying H II regions presented here extend the radial range of H II region abundances
measured in spiral galaxies.  For those galaxies with previous abundance measurements, 
the newly derived radial gradients are generally consistent with those previously reported.
For the three galaxies which had no abundance measurements reported in the literature 
(NGC 1232, NGC 1637, and NGC 2805), each now has abundances measured in at least 15 H II regions.

(5) As expected, the outermost H II regions in the spiral galaxies have abundances
similar to those of low luminosity dwarf galaxies.  These outlying H II regions provide
a new environment in which to investigate metallicity effects on stellar yields and
the star formation process (see, e.g., van Zee \etal \markcite{vSH98}1998).

\acknowledgements
We thank Elizabeth Barrett for processing several of the H$\alpha$ images and
 Stacy McGaugh for providing us with the ionization models.  We thank the
anonymous referee for comments which improved the presentation of this paper.
We thank Scott Lacey, Stacey Davis, Jennifer Heldmann, Lai Man Lee, Alison Schirmer, and
Matt Pickard for their assistance in the observations at the Burrell Schmidt.
They and TJB acknowledge travel support from the W.K. Keck Foundation for their
support of astronomy through the Keck Northeast Astronomy Consortium.
AOD acknowledges travel support by the Judge Francis Bergen
Career Development Award in Astrophysics, from the Dudley Observatory, Schenectady, 
New York. We acknowledge the financial support by NSF grants AST95--53020 to JJS 
and AST90--23450 and AST95-28860 to MPH.
This research has made use of the NASA/IPAC Extragalactic Database (NED)   
which is operated by the Jet Propulsion Laboratory, California Institute   
of Technology, under contract with the National Aeronautics and Space      
Administration.

\tablenum{1}
\footnotesize
\begin{deluxetable}{rcrccccrcrc}
\tablewidth{40pc}
\tablecaption{Galaxy Properties }
\tablehead{
\colhead{}& \colhead{RA} &\colhead{Dec}& \colhead{Morph.\tablenotemark{a}}& \colhead{Distance\tablenotemark{b}} & \colhead{} & \colhead{}& \colhead{} & \colhead{R$_{\rm 25}$\tablenotemark{a}} & \colhead{R$_{\rm d}$}& \colhead{No. HII}\\
\colhead{Galaxy}& \colhead{(2000)} & \colhead{(2000)} & \colhead{Type} & \colhead{[Mpc]} & \colhead{M$_{\rm B}$\tablenotemark{c}}&\colhead{$i$\tablenotemark{a}} & \colhead{p.a.\tablenotemark{a}} & \colhead{[arcsec]} & \colhead{[arcsec]} & \colhead{Regions}
}
\startdata
NGC 0628  & 01:36:41.6 &   15:47:03 & .SAS5.. &  9.7  & --20.1 & 25 &  25 & 314. &  71.8 & 19 \nl
NGC 0925  & 02:27:16.8 &   33:34:46 & .SXS7.. &  9.29 & --19.8 & 58 & 102 & 314. &  86.2 & 44 \nl
NGC 1068  & 02:42:40.6 & --00:00:46 & RSAT3.. & 14.4  & --21.3 & 32 &  70 & 212. & 111.4 &  1 \nl
NGC 1232  & 03:09:45.3 & --20:34:41 & .SXT5.. & 21.5  & --21.2    
%(10.41) 
& 30 &  90 & 222. &  60.2 & 16 \nl
NGC 1637  & 04:41:28.0 & --02:51:27 & .SXT5.. &  8.6  & --18.5 
%(11.21) 
& 36 &  15 & 120. &  20.6 & 15 \nl
NGC 2403  & 07:36:51.0 &   65:36:04 & .SXS6.. & 3.25  & --19.1 & 60 & 126 & 656. & 117.3 & 17 \nl
NGC 2805  & 09:20:20.3 &   64:06:11 & .SXT7.. & 23.5  & --20.7
%(11.18) 
& 42 & 125 & 189. &  74.8 & 17 \nl
IC 2458   & 09:21:30.0 &   64:14:20 & .I.0.P* & 23.5  & --16.9
%(14.98) 
& \nodata & \nodata &\nodata&\nodata&  3 \nl
NGC 2820  & 09:21:45.4 &   64:15:25 & .SBS5P/ & 23.5  & --20.1
%(11.73) 
& \nodata & \nodata &\nodata&\nodata&  4 \nl
NGC 2903  & 09:32:10.0 &   21:30:18 & .SXT4.. &  6.3  & --19.9 & 64 &  17 & 378. &  85.9 &  9 \nl
NGC 3184  & 10:18:16.8 &   41:25:28 & .SXT6.. &  8.7  & --19.3 & 21 & 135 & 222. &  55.8 & 17 \nl
NGC 4395  & 12:25:48.9 &   33:32:52 & .SAS9.. &  4.5  & --17.7 & 18 & 147 & 395. & 149.9 & 11 \nl
NGC 5457  & 14:03:12.5 &   54:20:55 & .SXT6.. &  7.4  & --21.2 & 18 &  37 & 865. & 152.9 & 13 \nl
\enddata 
\tablenotetext{a}{Morphological type, inclination, position angle, and isophotal radius from 
de Vaucouleurs \etal 1991 (RC3).}
\tablenotetext{b}{Distance references: NGC 628, NGC 1068, NGC 2903, NGC 3184-- 
Zaritsky \etal 1994; NGC 925-- Silbermann \etal 1996; NGC 1232, NGC 1637, NGC 2805, IC 2458, NGC 2820-- 
Virgocentric infall model; NGC 2403-- Freedman \& Madore 1988; NGC 4395-- Wevers \etal 1986; NGC 5457-- 
Kelson \etal 1996.}
\tablenotetext{c}{Absolute magnitude calculated from the apparent magnitude listed in the RC3 and
the adopted distance to the system.}
\end{deluxetable}

\psfig{figure=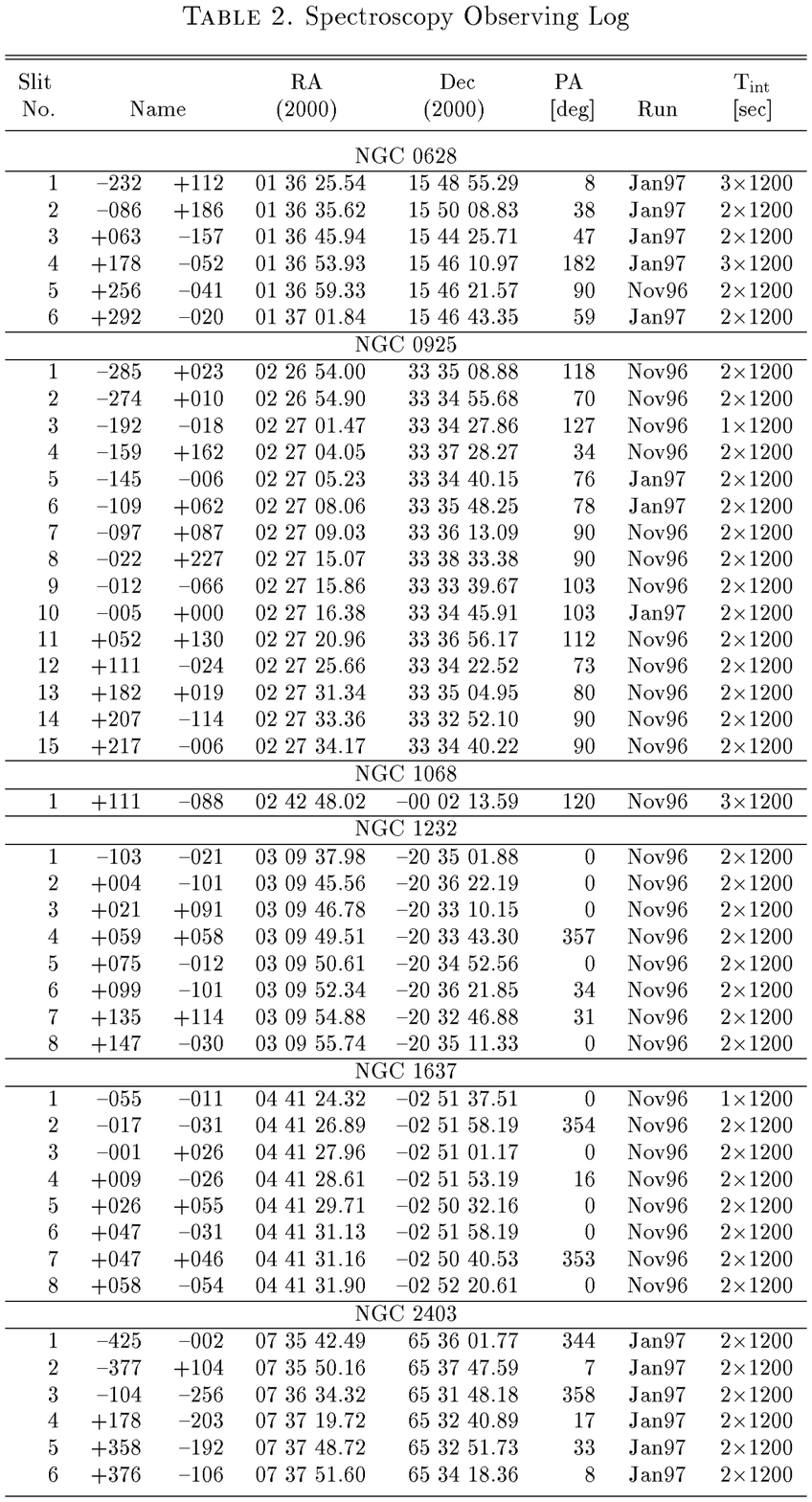,width=7.in,bbllx=50pt,bblly=50pt,bburx=600pt,bbury=750pt}

\psfig{figure=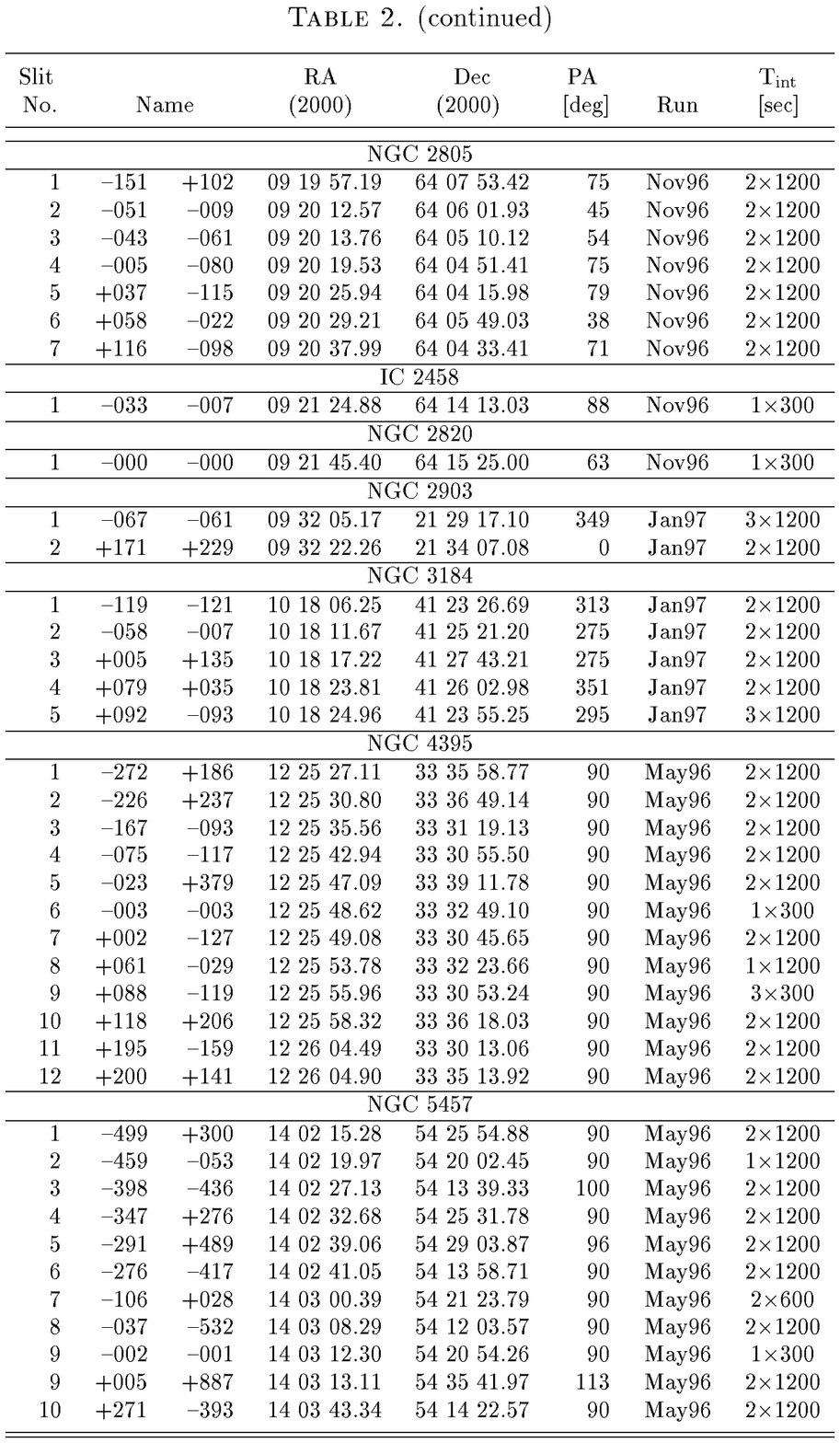,width=7.in,bbllx=50pt,bblly=50pt,bburx=600pt,bbury=750pt}

\psfig{figure=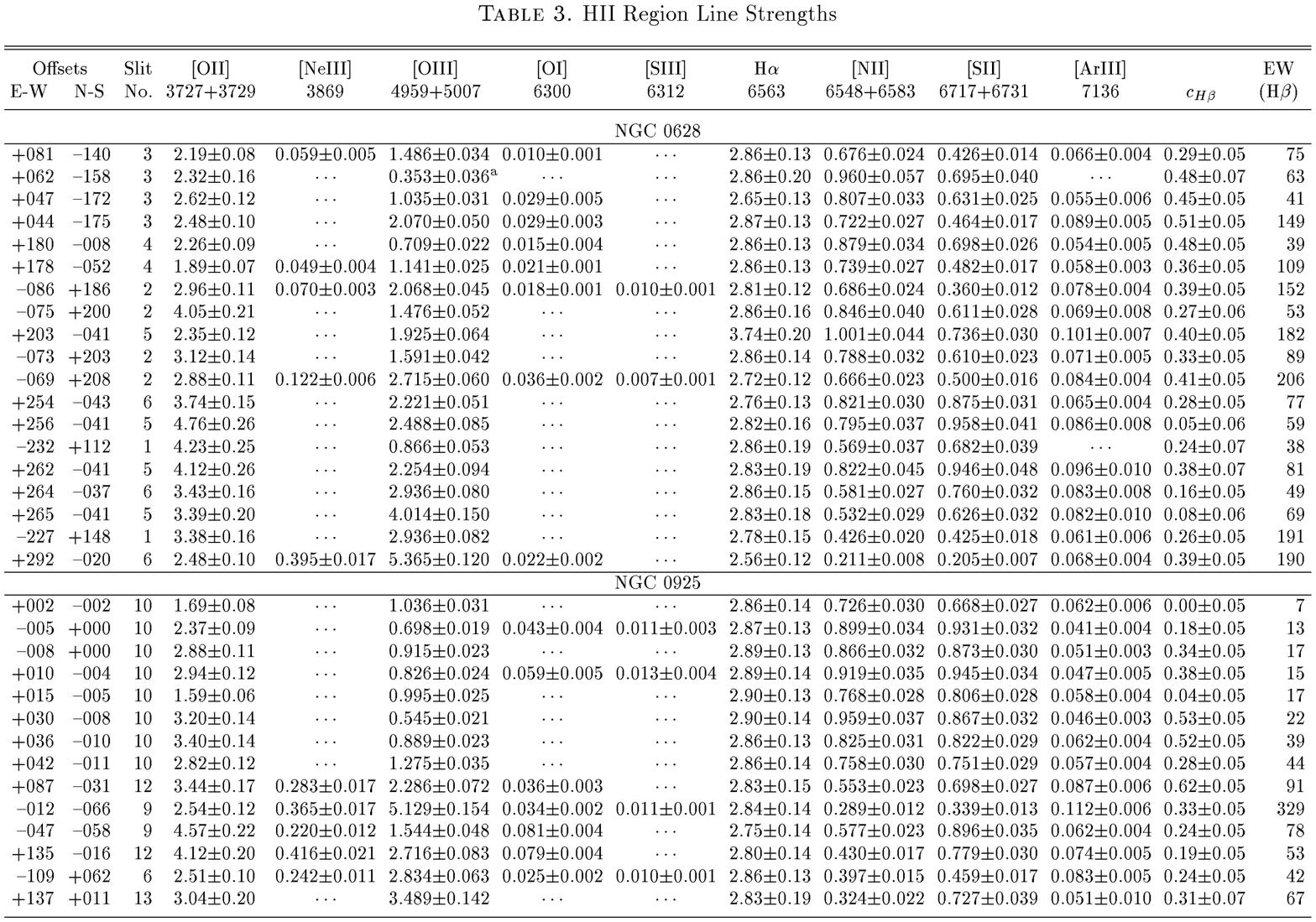,angle=90.,width=7.in,bbllx=50pt,bblly=150pt,bburx=800pt,bbury=700pt}

\psfig{figure=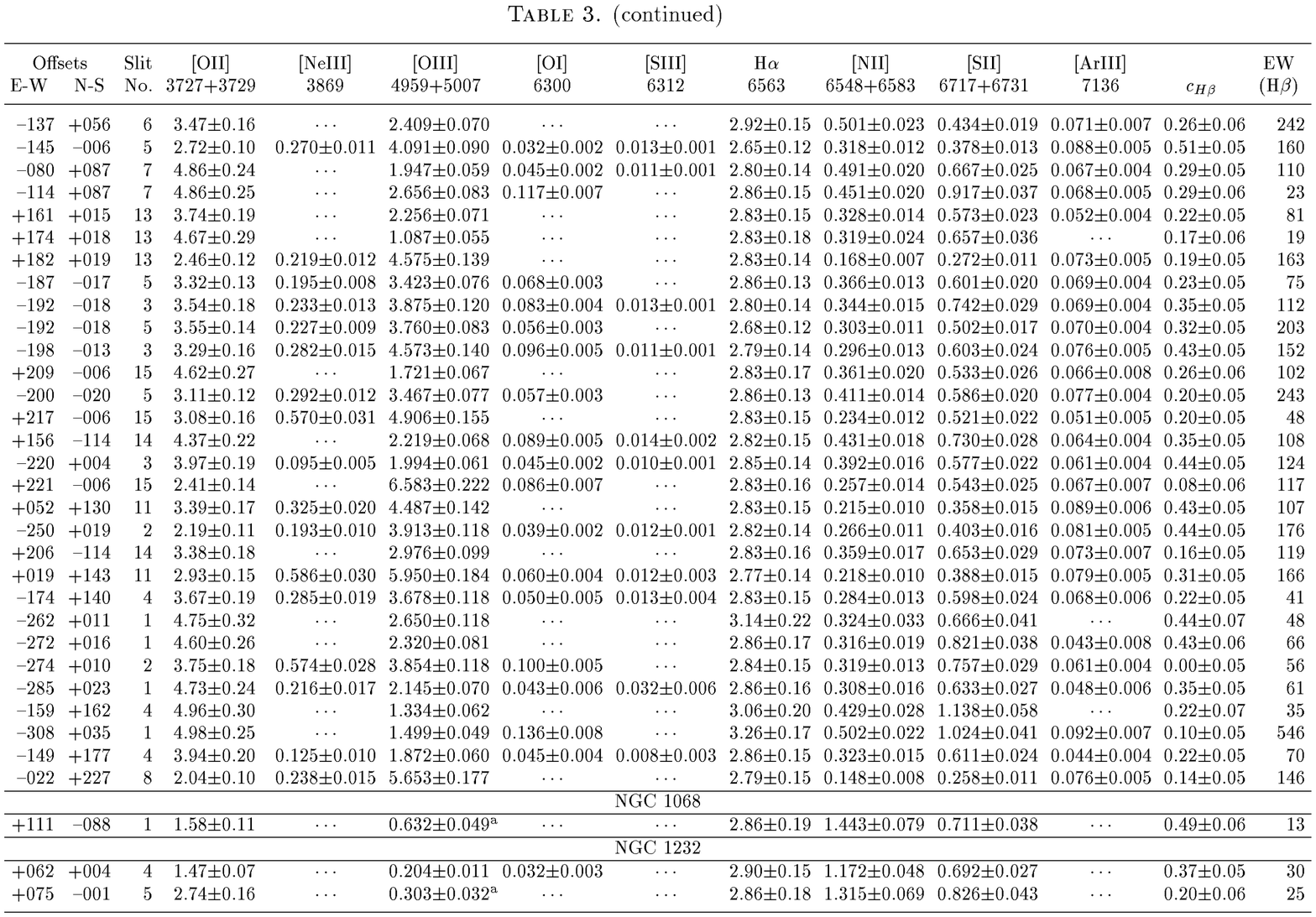,angle=90.,width=7.in,bbllx=50pt,bblly=150pt,bburx=800pt,bbury=700pt}

\psfig{figure=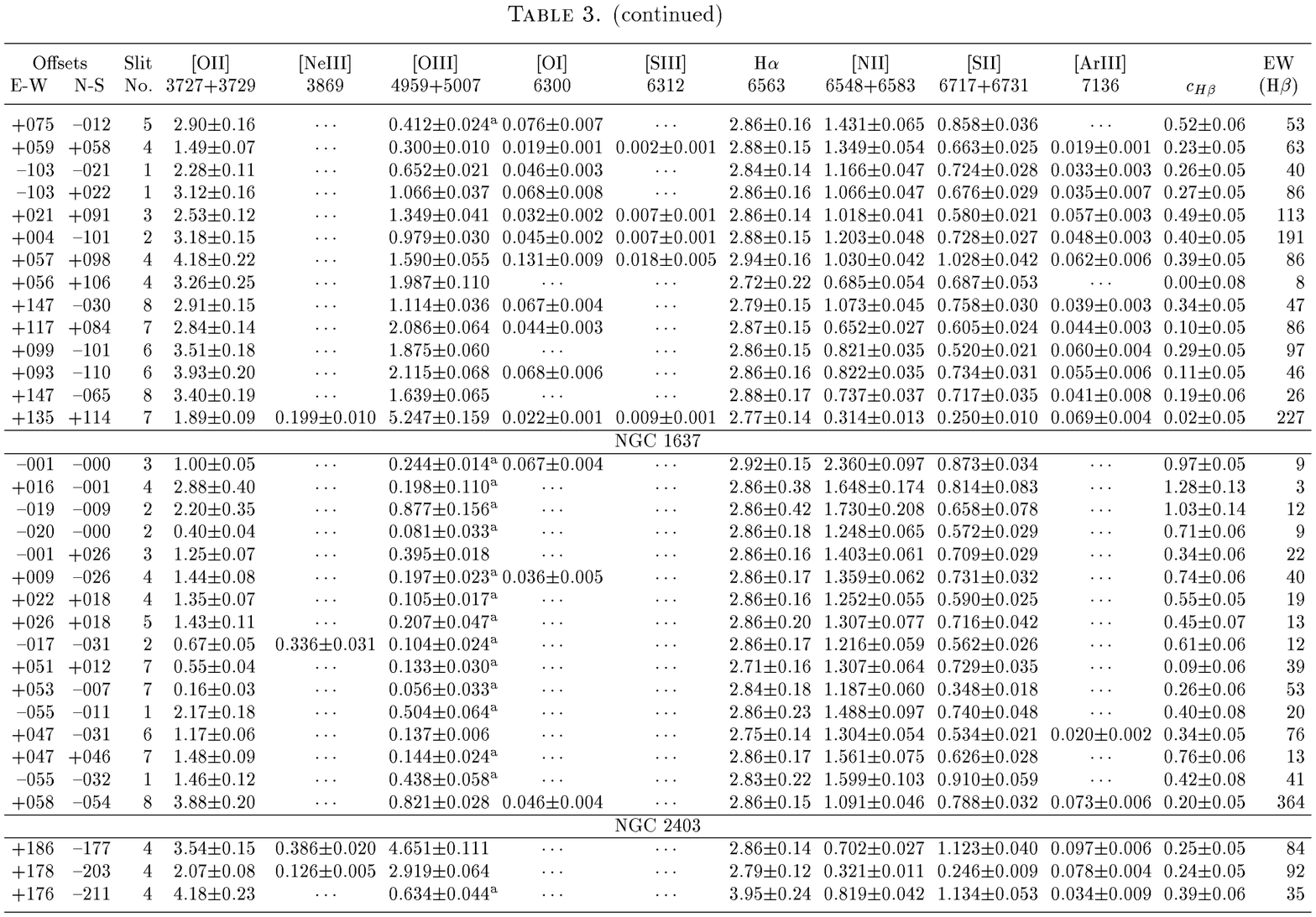,angle=90.,width=7.in,bbllx=50pt,bblly=150pt,bburx=800pt,bbury=700pt}

\psfig{figure=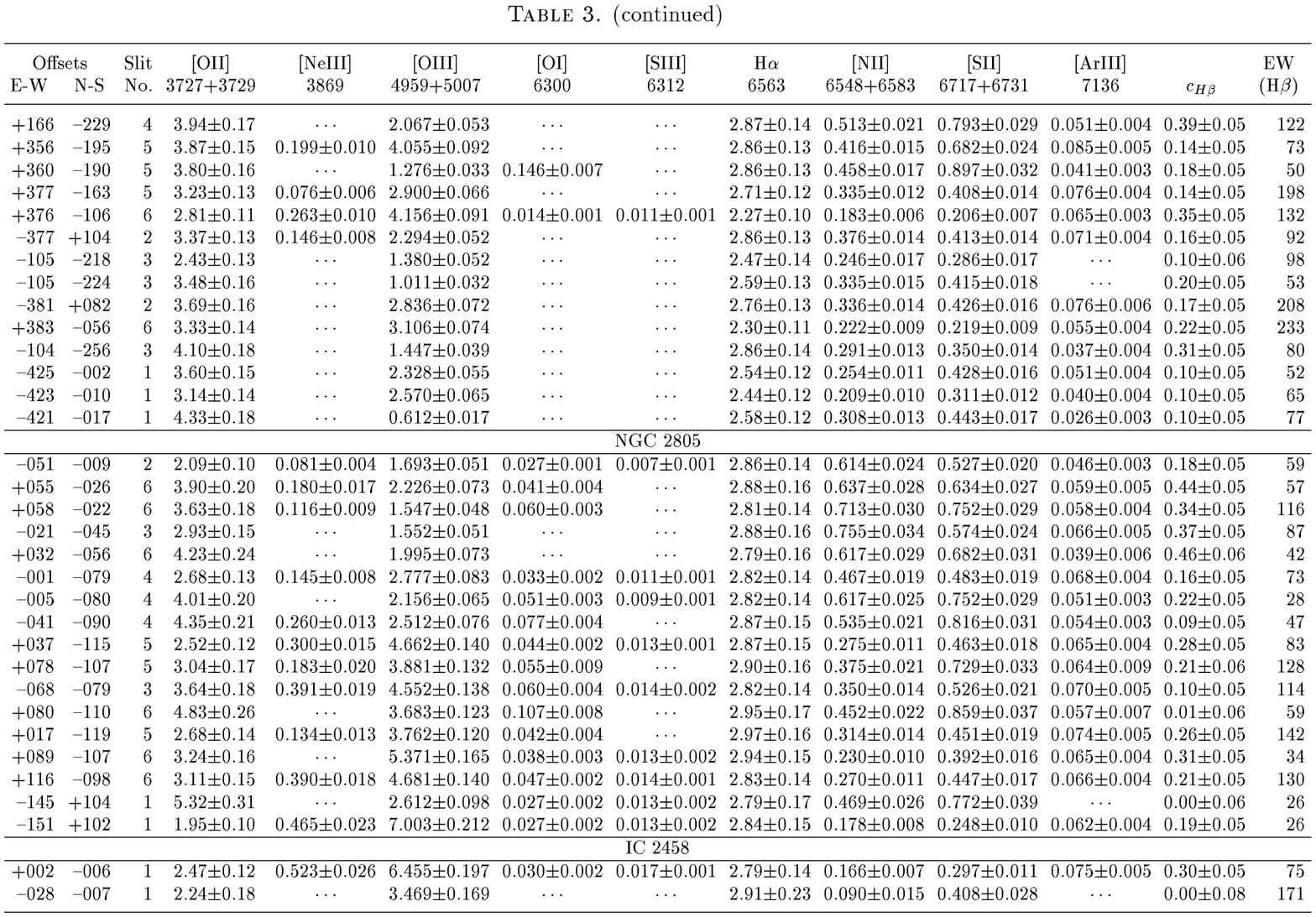,angle=90.,width=7.in,bbllx=50pt,bblly=150pt,bburx=800pt,bbury=700pt}

\psfig{figure=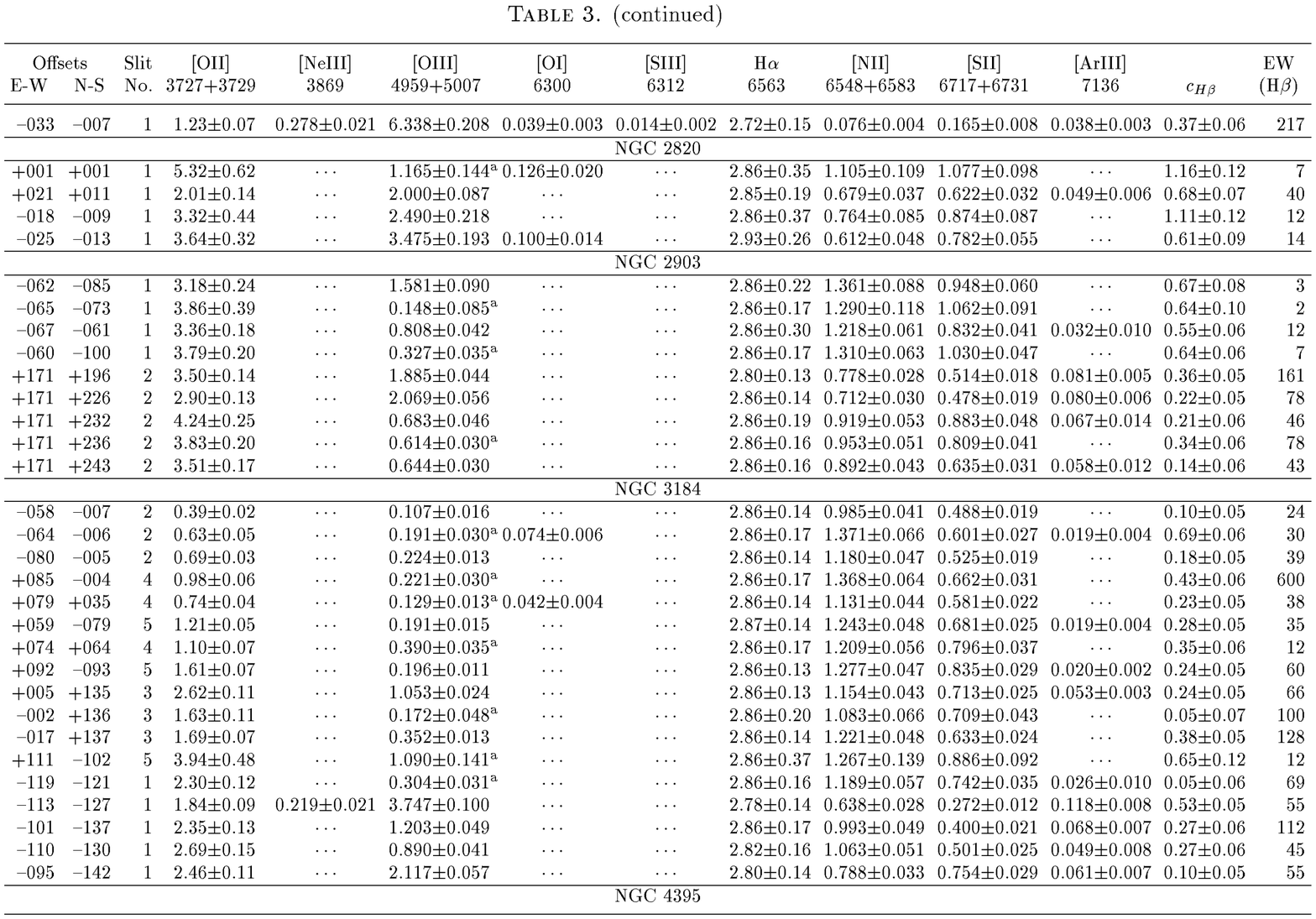,angle=90.,width=7.in,bbllx=50pt,bblly=150pt,bburx=800pt,bbury=700pt}

\psfig{figure=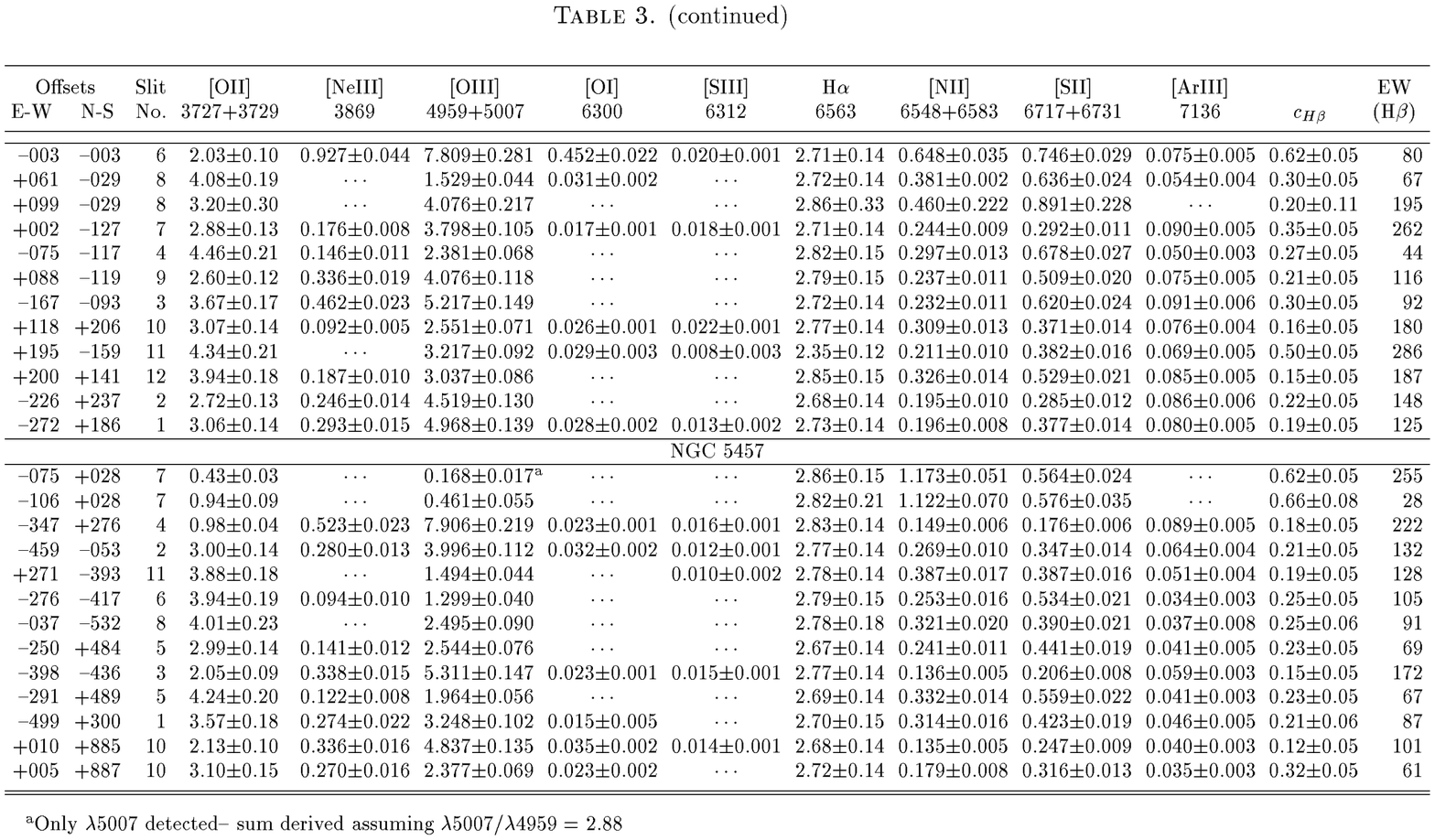,angle=90.,width=7.in,bbllx=50pt,bblly=150pt,bburx=800pt,bbury=700pt}

\psfig{figure=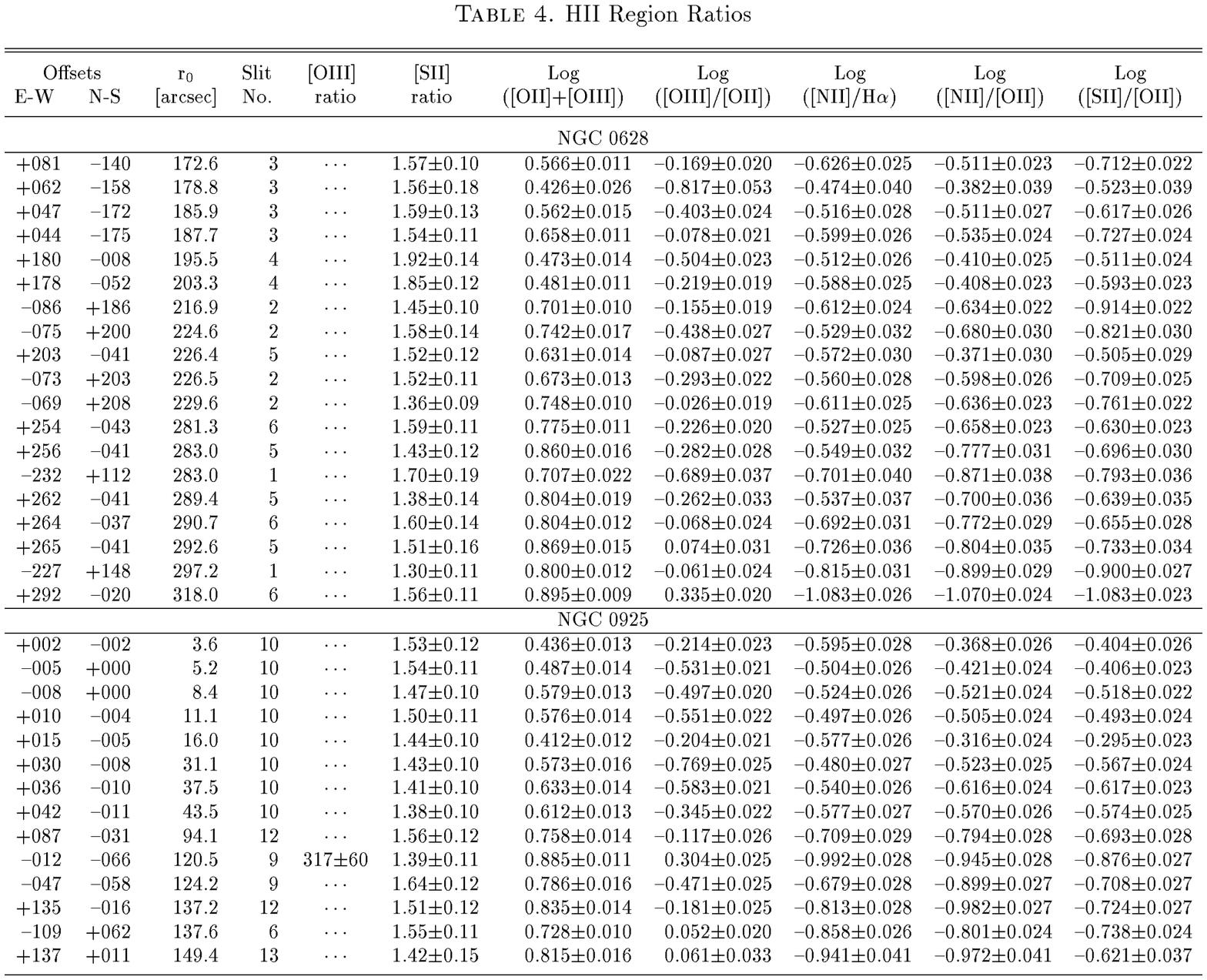,angle=90.,width=7.in,bbllx=50pt,bblly=150pt,bburx=800pt,bbury=700pt}

\psfig{figure=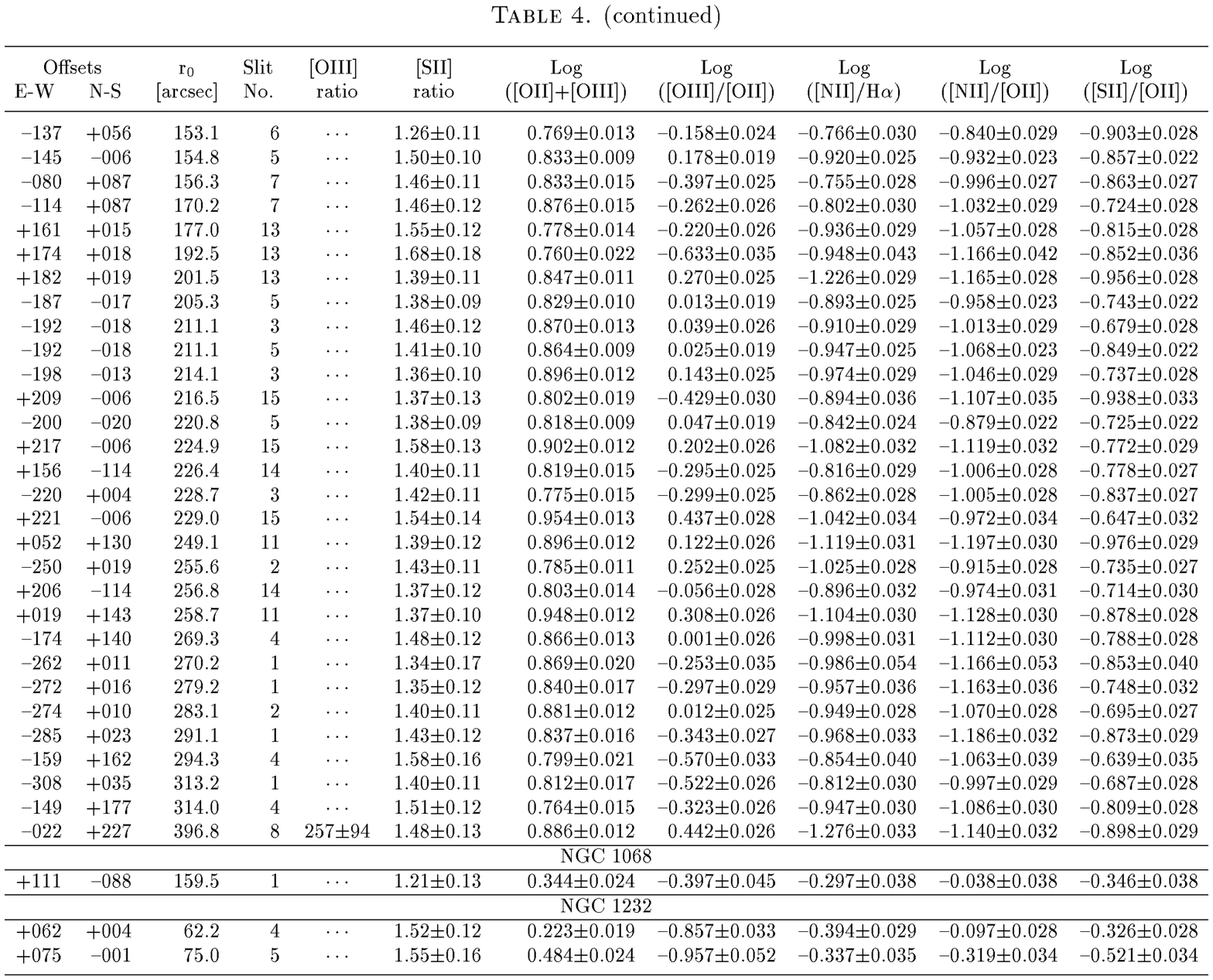,angle=90.,width=7.in,bbllx=50pt,bblly=150pt,bburx=800pt,bbury=700pt}

\psfig{figure=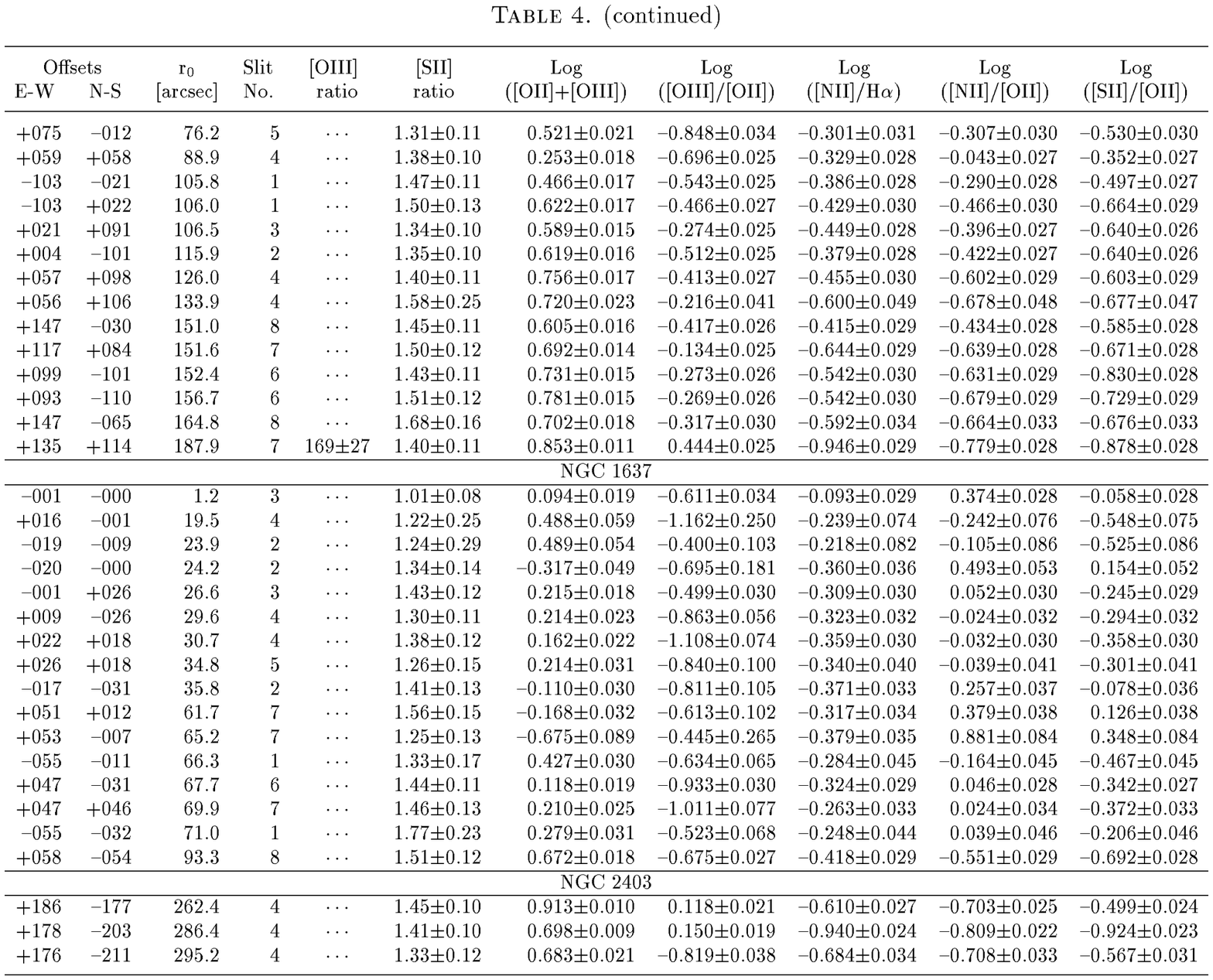,angle=90.,width=7.in,bbllx=50pt,bblly=150pt,bburx=800pt,bbury=700pt}

\psfig{figure=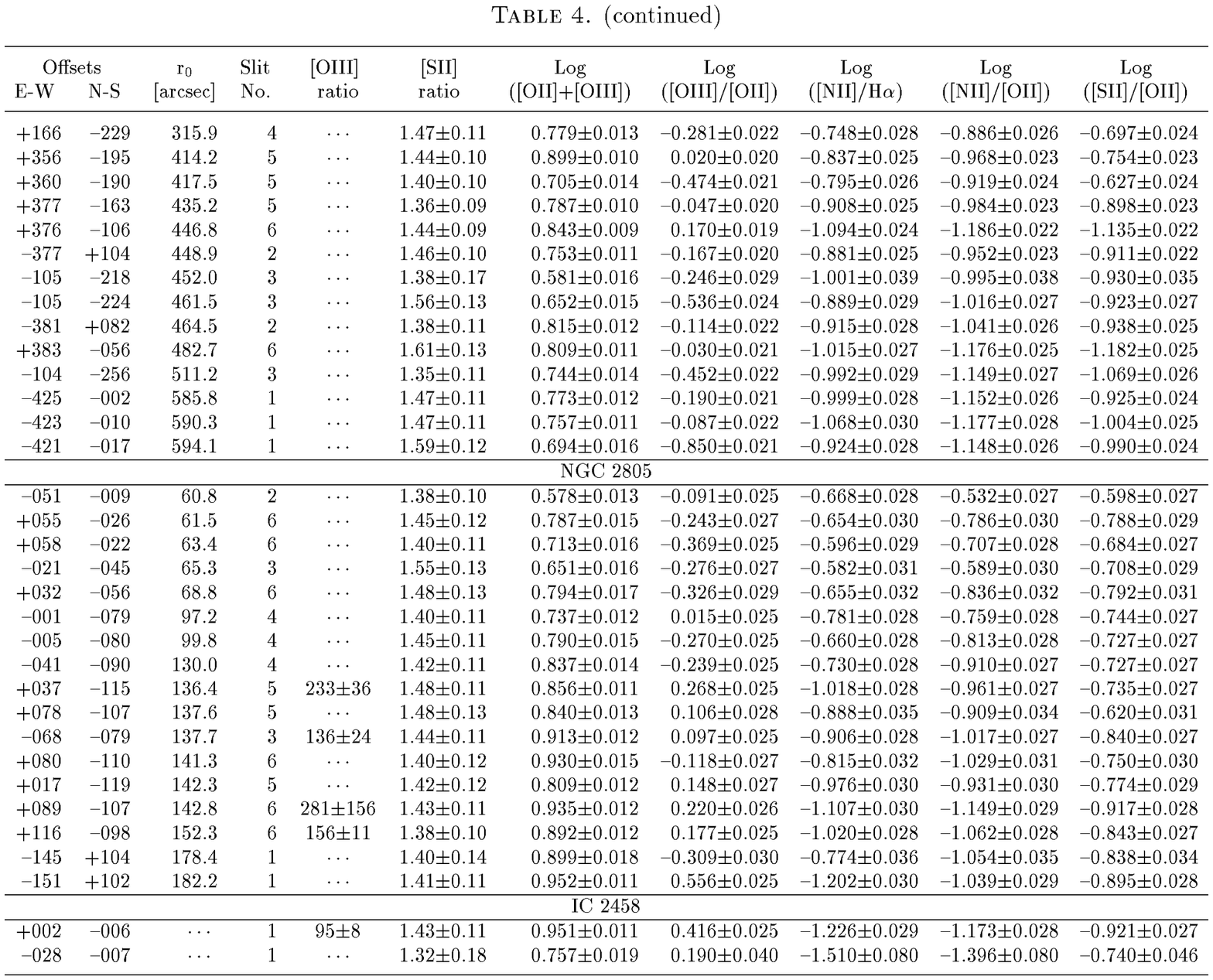,angle=90.,width=7.in,bbllx=50pt,bblly=150pt,bburx=800pt,bbury=700pt}

\psfig{figure=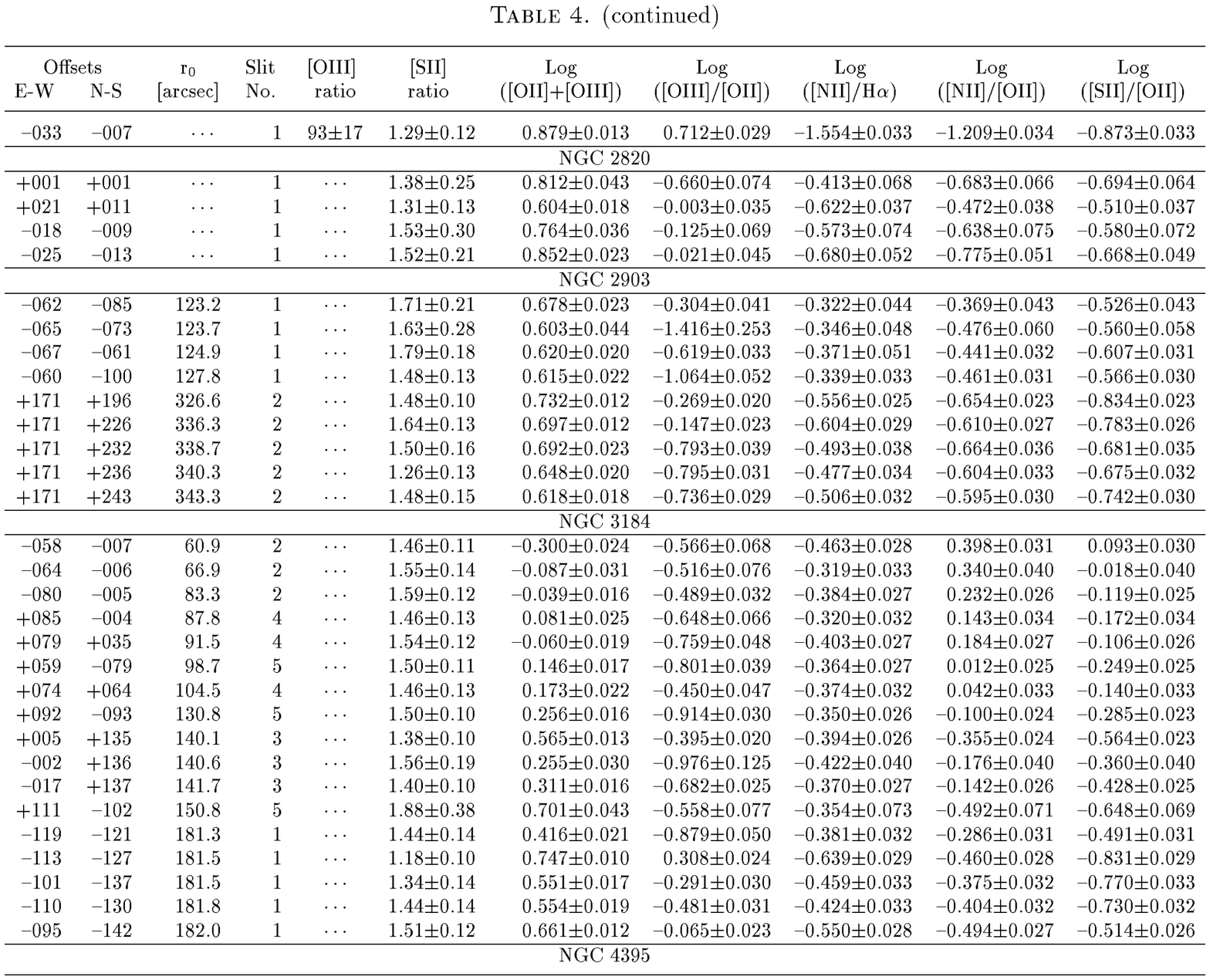,angle=90.,width=7.in,bbllx=50pt,bblly=150pt,bburx=800pt,bbury=700pt}

\psfig{figure=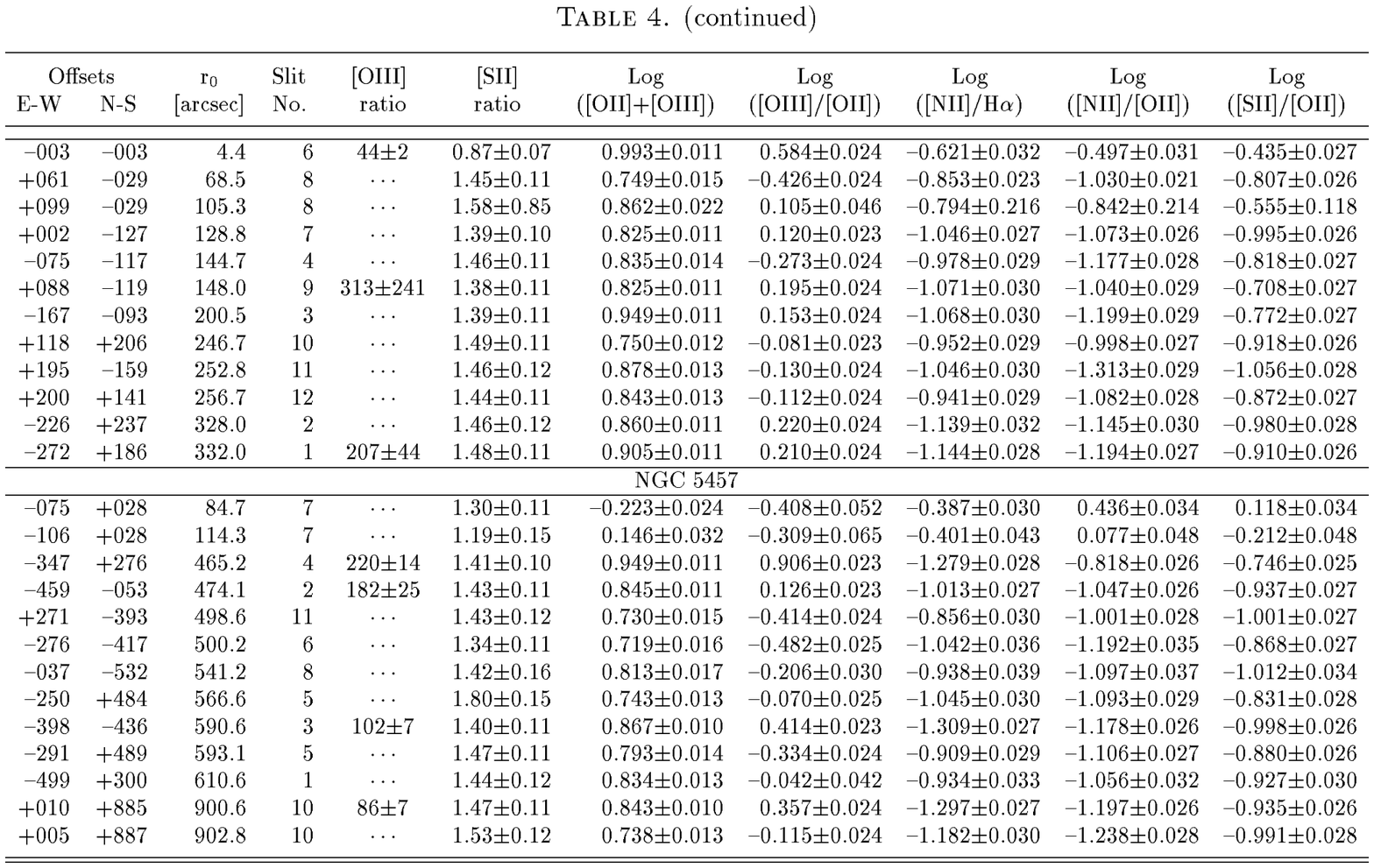,angle=90.,width=7.in,bbllx=50pt,bblly=150pt,bburx=800pt,bbury=700pt}

\tabcolsep 2pt
\tablenum{5}
\footnotesize
\begin{deluxetable}{rrrrcccccccc}
\renewcommand{\arraystretch}{0.9}
\tablewidth{40pc}
\tablecaption{HII Region Abundances}
\tablehead{
\multicolumn{2}{c}{Offsets}& \colhead{r$_0$} & \colhead{Slit} &\colhead{} & \colhead{} &  \colhead{} &  \colhead{} &\colhead{}& \colhead{} & \colhead{(O/H)} & \colhead{(O/H)}  \\
\colhead{E-W} & \colhead{N-S}& \colhead{[arcsec]} & \colhead{No.} &\colhead{T(O$^{++}$)}  &\colhead{12+Log(O/H)}& \colhead{Log(N/O)} &  \colhead{Log(Ne/O)} & \colhead{Log(S/O)}& \colhead{Log(Ar/O)} & \colhead{ZKH} & \colhead{EP84}  
}
\startdata
\multicolumn{12}{c}{NGC 0628} \nl
\hline
 +081 & --140 & 172.6 &  3 &  5700 $\pm$  500 & 8.88$\pm$0.12 & --1.15$\pm$0.17 & --0.68$\pm$0.34 & --1.83$\pm$0.26 & --2.16$\pm$0.21 & 8.94 & 8.80 \nl
 +062 & --158 & 178.8 &  3 &  4800 $\pm$  500 & 8.95$\pm$0.19 & --1.14$\pm$0.24 &     \nodata     & --1.52$\pm$0.30 &     \nodata     & 9.06 & 9.01 \nl
 +047 & --172 & 185.9 &  3 &  5500 $\pm$  500 & 8.90$\pm$0.12 & --1.17$\pm$0.18 &     \nodata     & --1.65$\pm$0.26 & --2.22$\pm$0.22 & 8.95 & 8.81 \nl
 +044 & --175 & 187.7 &  3 &  6200 $\pm$  500 & 8.82$\pm$0.10 & --1.12$\pm$0.15 &     \nodata     & --1.81$\pm$0.25 & --2.11$\pm$0.18 & 8.84 & 8.66 \nl
 +180 & --008 & 195.5 &  4 &  5100 $\pm$  500 & 8.94$\pm$0.14 & --1.15$\pm$0.21 &     \nodata     & --1.61$\pm$0.27 & --2.13$\pm$0.25 & 9.02 & 8.94 \nl
 +178 & --052 & 203.3 &  4 &  5300 $\pm$  500 & 8.95$\pm$0.14 & --1.14$\pm$0.20 & --0.60$\pm$0.40 & --1.80$\pm$0.27 & --2.16$\pm$0.24 & 9.02 & 8.93 \nl
--086 &  +186 & 216.9 &  2 &  6400 $\pm$  500 & 8.79$\pm$0.10 & --1.19$\pm$0.14 & --0.81$\pm$0.26 & --1.26$\pm$0.25 & --2.20$\pm$0.17 & 8.78 & 8.60 \nl
--075 &  +200 & 224.6 &  2 &  6700 $\pm$  500 & 8.69$\pm$0.10 & --1.23$\pm$0.13 &     \nodata     & --1.63$\pm$0.24 & --2.18$\pm$0.16 & 8.72 & 8.54 \nl
 +203 & --041 & 226.4 &  5 &  6000 $\pm$  500 & 8.85$\pm$0.11 & --0.97$\pm$0.16 &     \nodata     & --1.61$\pm$0.25 & --2.03$\pm$0.19 & 8.87 & 8.70 \nl
--073 &  +203 & 226.5 &  2 &  6200 $\pm$  500 & 8.80$\pm$0.10 & --1.18$\pm$0.15 &     \nodata     & --1.67$\pm$0.24 & --2.20$\pm$0.17 & 8.82 & 8.64 \nl
--069 &  +208 & 229.6 &  2 &  6900 $\pm$  500 & 8.73$\pm$0.10 & --1.15$\pm$0.12 & --0.73$\pm$0.23 & --1.51$\pm$0.21 & --2.20$\pm$0.15 & 8.71 & 8.53 \nl
 +254 & --043 & 281.3 &  6 &  7000 $\pm$  700 & 8.69$\pm$0.11 & --1.16$\pm$0.16 &     \nodata     & --1.50$\pm$0.25 & --2.31$\pm$0.20 & 8.67 & 8.49 \nl
 +256 & --041 & 283.0 &  5 &  8500 $\pm$  800 & 8.48$\pm$0.10 & --1.20$\pm$0.13 &     \nodata     & --1.38$\pm$0.24 & --2.20$\pm$0.16 & 8.52 & 8.36 \nl
--232 &  +112 & 283.0 &  1 &  6400 $\pm$  500 & 8.70$\pm$0.10 & --1.47$\pm$0.14 &     \nodata     & --1.56$\pm$0.25 &     \nodata     & 8.77 & 8.59 \nl
 +262 & --041 & 289.4 &  5 &  7500 $\pm$  500 & 8.60$\pm$0.10 & --1.18$\pm$0.11 &     \nodata     & --1.43$\pm$0.23 & --2.14$\pm$0.14 & 8.62 & 8.44 \nl
 +264 & --037 & 290.7 &  6 &  7400 $\pm$  500 & 8.65$\pm$0.10 & --1.26$\pm$0.11 &     \nodata     & --1.55$\pm$0.24 & --2.24$\pm$0.14 & 8.62 & 8.44 \nl
 +265 & --041 & 292.6 &  5 &  8300 $\pm$  500 & 8.55$\pm$0.10 & --1.22$\pm$0.14 &     \nodata     & --1.62$\pm$0.24 & --2.28$\pm$0.18 & 8.50 & 8.35 \nl
--227 &  +148 & 297.2 &  1 &  7400 $\pm$  500 & 8.64$\pm$0.10 & --1.39$\pm$0.11 &     \nodata     & --1.80$\pm$0.24 & --2.37$\pm$0.14 & 8.63 & 8.45 \nl
 +292 & --020 & 318.0 &  6 & 12000 $\pm$ 1500 & 8.10$\pm$0.10 & --1.39$\pm$0.13 & --0.70$\pm$0.25 & --1.87$\pm$0.24 & --2.29$\pm$0.16 & 8.45 & 7.81 \nl
\hline
\multicolumn{12}{c}{NGC 0925}  \nl
\hline
 +002 & --002 &   3.6 & 10 &  5100 $\pm$  500 & 8.98$\pm$0.15 & --1.09$\pm$0.21 &     \nodata     & --1.63$\pm$0.28 & --2.09$\pm$0.26 & 9.05 & 9.00 \nl
--005 &  +000 &   5.2 & 10 &  5100 $\pm$  500 & 8.94$\pm$0.14 & --1.14$\pm$0.21 &     \nodata     & --1.45$\pm$0.27 & --2.25$\pm$0.25 & 9.01 & 8.92 \nl
--008 &  +000 &   8.4 & 10 &  5500 $\pm$  500 & 8.89$\pm$0.12 & --1.18$\pm$0.18 &     \nodata     & --1.49$\pm$0.26 & --2.23$\pm$0.22 & 8.93 & 8.78 \nl
 +010 & --004 &  11.1 & 10 &  5500 $\pm$  500 & 8.87$\pm$0.12 & --1.16$\pm$0.18 &     \nodata     & --1.44$\pm$0.26 & --2.24$\pm$0.22 & 8.93 & 8.79 \nl
 +015 & --005 &  16.0 & 10 &  5100 $\pm$  500 & 8.96$\pm$0.15 & --1.03$\pm$0.21 &     \nodata     & --1.52$\pm$0.28 & --2.09$\pm$0.26 & 9.07 & 9.03 \nl
 +030 & --008 &  31.1 & 10 &  5400 $\pm$  500 & 8.86$\pm$0.13 & --1.19$\pm$0.19 &     \nodata     & --1.45$\pm$0.26 & --2.14$\pm$0.23 & 8.94 & 8.79 \nl
 +036 & --010 &  37.5 & 10 &  5800 $\pm$  500 & 8.82$\pm$0.11 & --1.23$\pm$0.17 &     \nodata     & --1.49$\pm$0.25 & --2.13$\pm$0.20 & 8.87 & 8.70 \nl
 +042 & --011 &  43.5 & 10 &  5800 $\pm$  500 & 8.84$\pm$0.11 & --1.18$\pm$0.17 &     \nodata     & --1.56$\pm$0.25 & --2.23$\pm$0.20 & 8.89 & 8.73 \nl
 +087 & --031 &  94.1 & 12 &  6900 $\pm$  500 & 8.71$\pm$0.10 & --1.30$\pm$0.13 & --0.28$\pm$0.23 & --1.61$\pm$0.24 & --2.18$\pm$0.15 & 8.70 & 8.51 \nl
--012 & --066 & 120.5 &  9 &  8890 $\pm$  520 & 8.48$\pm$0.05 & --1.34$\pm$0.08 & --0.61$\pm$0.13 & --1.62$\pm$0.12 & --2.11$\pm$0.10 & 8.47 & 8.32 \nl
--047 & --058 & 124.2 &  9 &  7400 $\pm$  500 & 8.60$\pm$0.10 & --1.37$\pm$0.11 & --0.25$\pm$0.20 & --1.42$\pm$0.24 & --2.27$\pm$0.13 & 8.65 & 8.47 \nl
 +135 & --016 & 137.2 & 12 &  7900 $\pm$  500 & 8.58$\pm$0.10 & --1.42$\pm$0.13 & --0.25$\pm$0.25 & --1.56$\pm$0.24 & --2.30$\pm$0.16 & 8.57 & 8.40 \nl
--109 &  +062 & 137.6 &  6 &~~7900 $\pm$ 1500 & 8.48$\pm$0.23 & --1.25$\pm$0.30 & --0.50$\pm$0.69 & --1.72$\pm$0.34 & --2.14$\pm$0.37 & 8.74 & 8.56 \nl
 +137 &  +011 & 149.4 & 13 &~~8900 $\pm$ 2000 & 8.38$\pm$0.25 & --1.37$\pm$0.33 &     \nodata     & --1.43$\pm$0.35 & --2.41$\pm$0.43 & 8.60 & 8.43 \nl
--137 &  +056 & 153.1 &  6 &  7000 $\pm$  500 & 8.68$\pm$0.10 & --1.35$\pm$0.12 &     \nodata     & --1.79$\pm$0.24 & --2.27$\pm$0.15 & 8.68 & 8.50 \nl
--145 & --006 & 154.8 &  5 & 10300 $\pm$ 2000 & 8.23$\pm$0.17 & --1.28$\pm$0.23 & --0.70$\pm$0.52 & --1.60$\pm$0.43 & --2.17$\pm$0.29 & 8.57 & 7.70 \nl
--080 &  +087 & 156.3 &  7 &  8300 $\pm$  800 & 8.48$\pm$0.10 & --1.42$\pm$0.14 & --0.43$\pm$0.26 & --1.53$\pm$0.22 & --2.26$\pm$0.17 & 8.57 & 8.40 \nl
--114 &  +087 & 170.2 &  7 &  8500 $\pm$  800 & 8.51$\pm$0.10 & --1.45$\pm$0.13 &     \nodata     & --1.44$\pm$0.24 & --2.35$\pm$0.16 & 8.49 & 8.34 \nl
 +161 &  +015 & 177.0 & 13 &~~8700 $\pm$ 2000 & 8.38$\pm$0.24 & --1.46$\pm$0.35 &     \nodata     & --1.55$\pm$0.36 & --2.36$\pm$0.43 & 8.67 & 8.48 \nl
 +174 &  +018 & 192.5 & 13 &~~8900 $\pm$ 1000 & 8.28$\pm$0.11 & --1.64$\pm$0.16 &     \nodata     & --1.39$\pm$0.25 &     \nodata     & 8.70 & 8.51 \nl
 +182 &  +019 & 201.5 & 13 & 12500 $\pm$ 1500 & 8.01$\pm$0.10 & --1.43$\pm$0.12 & --0.89$\pm$0.23 & --1.74$\pm$0.19 & --2.22$\pm$0.16 & 8.54 & 7.72 \nl
--187 & --017 & 205.3 &  5 &  7800 $\pm$  500 & 8.61$\pm$0.10 & --1.41$\pm$0.10 & --0.67$\pm$0.18 & --1.66$\pm$0.23 & --2.34$\pm$0.12 & 8.58 & 8.41 \nl
--192 & --018 & 211.1 &  3 &  8300 $\pm$  800 & 8.56$\pm$0.10 & --1.49$\pm$0.14 & --0.68$\pm$0.26 & --1.72$\pm$0.24 & --2.36$\pm$0.17 & 8.50 & 8.34 \nl
--192 & --018 & 211.1 &  5 &  8500 $\pm$  800 & 8.53$\pm$0.10 & --1.44$\pm$0.13 & --0.68$\pm$0.25 & --1.51$\pm$0.21 & --2.37$\pm$0.16 & 8.51 & 8.35 \nl
--198 & --013 & 214.1 &  3 & 10000 $\pm$ 2000 & 8.35$\pm$0.18 & --1.40$\pm$0.25 & --0.72$\pm$0.56 & --1.67$\pm$0.42 & --2.32$\pm$0.31 & 8.44 & 7.81 \nl
 +209 & --006 & 216.5 & 15 &~~8600 $\pm$ 1600 & 8.39$\pm$0.19 & --1.52$\pm$0.26 &     \nodata     & --1.55$\pm$0.30 & --2.18$\pm$0.33 & 8.63 & 8.45 \nl
--200 & --020 & 220.8 &  5 &  7500 $\pm$  500 & 8.65$\pm$0.10 & --1.34$\pm$0.11 & --0.49$\pm$0.19 & --1.68$\pm$0.23 & --2.28$\pm$0.13 & 8.60 & 8.42 \nl
 +217 & --006 & 224.9 & 15 & 11200 $\pm$ 1500 & 8.19$\pm$0.10 & --1.48$\pm$0.15 & --0.48$\pm$0.29 & --1.52$\pm$0.25 & --2.45$\pm$0.19 & 8.43 & 7.82 \nl
 +156 & --114 & 226.4 & 14 &  7700 $\pm$  600 & 8.59$\pm$0.10 & --1.46$\pm$0.12 &     \nodata     & --1.38$\pm$0.20 & --2.33$\pm$0.14 & 8.60 & 8.42 \nl
--220 &  +004 & 228.7 &  3 &  7100 $\pm$  500 & 8.67$\pm$0.10 & --1.50$\pm$0.12 & --0.71$\pm$0.22 & --1.41$\pm$0.20 & --2.33$\pm$0.14 & 8.67 & 8.49 \nl
 +221 & --006 & 229.0 & 15 & 11200 $\pm$ 1500 & 8.27$\pm$0.11 & --1.27$\pm$0.15 &     \nodata     & --1.58$\pm$0.25 & --2.35$\pm$0.19 & 8.31 & 7.91 \nl
 +052 &  +130 & 249.1 & 11 & 11000 $\pm$ 1500 & 8.20$\pm$0.10 & --1.57$\pm$0.15 & --0.68$\pm$0.30 & --1.68$\pm$0.25 & --2.21$\pm$0.19 & 8.44 & 7.81 \nl
--250 &  +019 & 255.6 &  2 &~~9300 $\pm$ 2000 & 8.31$\pm$0.23 & --1.27$\pm$0.29 & --0.77$\pm$0.70 & --1.31$\pm$0.62 & --2.15$\pm$0.38 & 8.66 & 8.48 \nl
 +206 & --114 & 256.8 & 14 &  7400 $\pm$  500 & 8.65$\pm$0.10 & --1.45$\pm$0.11 &     \nodata     & --1.62$\pm$0.24 & --2.29$\pm$0.14 & 8.62 & 8.45 \nl
 +019 &  +143 & 258.7 & 11 & 10000 $\pm$ 1000 & 8.39$\pm$0.10 & --1.47$\pm$0.12 & --0.52$\pm$0.23 & --1.71$\pm$0.21 & --2.32$\pm$0.15 & 8.32 & 7.90 \nl
--174 &  +140 & 269.3 &  4 & 11000 $\pm$ 1500 & 8.19$\pm$0.10 & --1.43$\pm$0.15 & --0.65$\pm$0.30 & --1.62$\pm$0.24 & --2.31$\pm$0.19 & 8.51 & 7.76 \nl
--262 &  +011 & 270.2 &  1 &~~9800 $\pm$ 1000 & 8.30$\pm$0.10 & --1.52$\pm$0.14 &     \nodata     & --1.45$\pm$0.24 &     \nodata     & 8.50 & 7.76 \nl
--272 &  +016 & 279.2 &  1 &~~9700 $\pm$ 1000 & 8.29$\pm$0.10 & --1.55$\pm$0.15 &     \nodata     & --1.34$\pm$0.24 & --2.44$\pm$0.18 & 8.56 & 7.71 \nl
--274 &  +010 & 283.1 &  2 & 10300 $\pm$ 2000 & 8.29$\pm$0.16 & --1.41$\pm$0.23 & --0.36$\pm$0.51 & --1.50$\pm$0.40 & --2.40$\pm$0.29 & 8.48 & 7.78 \nl
--285 &  +023 & 291.1 &  1 &~~9700 $\pm$ 1000 & 8.30$\pm$0.10 & --1.58$\pm$0.15 & --0.51$\pm$0.25 & --1.25$\pm$0.23 & --2.39$\pm$0.17 & 8.56 & 7.70 \nl
--159 &  +162 & 294.3 &  4 &  7800 $\pm$  700 & 8.49$\pm$0.10 & --1.51$\pm$0.14 &     \nodata     & --1.28$\pm$0.24 &     \nodata     & 8.63 & 8.45 \nl
--308 &  +035 & 313.2 &  1 &  8100 $\pm$  800 & 8.50$\pm$0.10 & --1.43$\pm$0.14 &     \nodata     & --1.34$\pm$0.24 & --2.10$\pm$0.17 & 8.61 & 8.43 \nl
--149 &  +177 & 314.0 &  4 &~~8600 $\pm$ 1600 & 8.38$\pm$0.18 & --1.51$\pm$0.26 & --0.63$\pm$0.40 & --1.60$\pm$0.47 & --2.40$\pm$0.32 & 8.69 & 8.50 \nl
--022 &  +227 & 396.8 &  8 &~~9440 $\pm$ 1380 & 8.39$\pm$0.09 & --1.53$\pm$0.13 & --0.87$\pm$0.24 & --1.93$\pm$0.24 & --2.23$\pm$0.16 & 8.46 & 7.79 \nl
\hline
\multicolumn{12}{c}{NGC 1068}   \nl
\hline
 +111 & --088 & 159.5 &  1 & 4700  $\pm$  300 & 9.04$\pm$0.13 & --0.82$\pm$0.15 &     \nodata     & --1.56$\pm$0.26 &     \nodata     & 9.11 & 9.13 \nl
\hline
\multicolumn{12}{c}{NGC 1232}  \nl
\hline
 +062 &  +004 &  62.2 &  4 &  4200 $\pm$  200 & 9.07$\pm$0.10 & --0.95$\pm$0.13 &     \nodata     & --1.49$\pm$0.24 &     \nodata     & 9.18 & 9.25 \nl
 +075 & --001 &  75.0 &  5 &  4900 $\pm$  300 & 8.94$\pm$0.11 & --1.07$\pm$0.14 &     \nodata     & --1.45$\pm$0.25 &     \nodata     & 9.02 & 8.92 \nl
 +075 & --012 &  76.2 &  5 &  5200 $\pm$  300 & 8.87$\pm$0.10 & --1.00$\pm$0.13 &     \nodata     & --1.42$\pm$0.24 &     \nodata     & 8.98 & 8.87 \nl
 +059 &  +058 &  88.9 &  4 &  4300 $\pm$  200 & 9.08$\pm$0.10 & --0.88$\pm$0.12 &     \nodata     & --0.93$\pm$0.30 & --2.38$\pm$0.14 & 9.16 & 9.23 \nl
--103 & --021 & 105.8 &  1 &  5100 $\pm$  400 & 8.94$\pm$0.14 & --1.00$\pm$0.26 &     \nodata     & --1.55$\pm$0.27 & --2.34$\pm$0.22 & 9.03 & 8.95 \nl
--103 &  +022 & 106.0 &  1 &  5900 $\pm$  400 & 8.83$\pm$0.10 & --1.07$\pm$0.13 &     \nodata     & --1.61$\pm$0.24 & --2.45$\pm$0.18 & 8.88 & 8.72 \nl
 +021 &  +091 & 106.5 &  3 &  5800 $\pm$  400 & 8.85$\pm$0.10 & --1.02$\pm$0.14 &     \nodata     & --1.19$\pm$0.24 & --2.24$\pm$0.16 & 8.91 & 8.77 \nl
 +004 & --101 & 115.9 &  2 &  5900 $\pm$  400 & 8.80$\pm$0.10 & --1.03$\pm$0.13 &     \nodata     & --1.20$\pm$0.23 & --2.27$\pm$0.16 & 8.88 & 8.72 \nl
 +057 &  +098 & 126.0 &  4 &  7000 $\pm$  700 & 8.67$\pm$0.10 & --1.12$\pm$0.16 &     \nodata     & --1.13$\pm$0.31 & --2.30$\pm$0.20 & 8.70 & 8.52 \nl
 +056 &  +106 & 133.9 &  4 &  6600 $\pm$  500 & 8.74$\pm$0.10 & --1.21$\pm$0.14 &     \nodata     & --1.61$\pm$0.24 &     \nodata     & 8.76 & 8.57 \nl
 +147 & --030 & 151.0 &  8 &  5800 $\pm$  500 & 8.85$\pm$0.10 & --1.05$\pm$0.17 &     \nodata     & --1.56$\pm$0.25 & --2.40$\pm$0.19 & 8.90 & 8.74 \nl
 +117 & --084 & 151.6 &  7 &  6500 $\pm$  500 & 8.77$\pm$0.10 & --1.18$\pm$0.14 &     \nodata     & --1.68$\pm$0.24 & --2.44$\pm$0.16 & 8.80 & 8.61 \nl
 +099 & --101 & 152.4 &  6 &  6700 $\pm$  500 & 8.71$\pm$0.10 & --1.16$\pm$0.13 &     \nodata     & --1.71$\pm$0.24 & --2.30$\pm$0.16 & 8.74 & 8.55 \nl
 +093 & --110 & 156.7 &  6 &  7300 $\pm$  800 & 8.64$\pm$0.11 & --1.16$\pm$0.17 &     \nodata     & --1.56$\pm$0.25 & --2.39$\pm$0.21 & 8.66 & 8.48 \nl
 +147 & --065 & 164.8 &  8 &  6400 $\pm$  500 & 8.75$\pm$0.10 & --1.22$\pm$0.14 &     \nodata     & --1.58$\pm$0.24 & --2.43$\pm$0.19 & 8.78 & 8.60 \nl
 +135 &  +114 & 187.9 &  7 & 10670 $\pm$  650 & 8.20$\pm$0.05 & --1.09$\pm$0.08 & --0.96$\pm$0.11 & --1.74$\pm$0.11 & --2.23$\pm$0.10 & 8.53 & 7.73 \nl
\hline
\multicolumn{12}{c}{NGC 1637}   \nl
\hline
--001 & --000 &   1.2 &  3 &  4000 $\pm$  200 & 9.18$\pm$0.11 & --0.55$\pm$0.14 &     \nodata     & --1.40$\pm$0.25 &     \nodata     & 9.23 & 9.34 \nl
 +016 & --001 &  19.5 &  4 &  4900 $\pm$  300 & 8.91$\pm$0.13 & --0.97$\pm$0.16 &     \nodata     & --1.42$\pm$0.26 &     \nodata     & 9.01 & 8.92 \nl
--019 & --009 &  23.9 &  2 &  5200 $\pm$  300 & 8.94$\pm$0.11 & --0.82$\pm$0.15 &     \nodata     & --1.61$\pm$0.26 &     \nodata     & 9.01 & 8.92 \nl
--020 & --000 &  24.2 &  2 &  3300 $\pm$  200 & 9.30$\pm$0.21 & --0.61$\pm$0.20 &     \nodata     & --1.46$\pm$0.31 &     \nodata     & 9.35 & 9.52 \nl
--001 &  +026 &  26.6 &  3 &  4300 $\pm$  200 & 9.10$\pm$0.10 & --0.79$\pm$0.12 &     \nodata     & --1.53$\pm$0.24 &     \nodata     & 9.18 & 9.25 \nl
 +009 & --026 &  29.6 &  4 &  4100 $\pm$  200 & 9.10$\pm$0.10 & --0.90$\pm$0.13 &     \nodata     & --1.46$\pm$0.25 &     \nodata     & 9.18 & 9.25 \nl
 +022 &  +018 &  30.7 &  4 &  3900 $\pm$  200 & 9.12$\pm$0.11 & --0.96$\pm$0.14 &     \nodata     & --1.50$\pm$0.25 &     \nodata     & 9.20 & 9.29 \nl
 +026 &  +018 &  34.8 &  5 &  4100 $\pm$  200 & 9.11$\pm$0.11 & --0.92$\pm$0.13 &     \nodata     & --1.48$\pm$0.25 &     \nodata     & 9.18 & 9.25 \nl
--017 & --031 &  35.8 &  2 &  3500 $\pm$  200 & 9.26$\pm$0.16 & --0.79$\pm$0.17 &     \nodata     & --1.52$\pm$0.28 &     \nodata     & 9.30 & 9.44 \nl
 +051 &  +012 &  61.7 &  7 &  3550 $\pm$  200 & 9.24$\pm$0.17 & --0.63$\pm$0.17 &     \nodata     & --1.42$\pm$0.28 &     \nodata     & 9.31 & 9.46 \nl
--055 & --011 &  66.3 &  1 &  4900 $\pm$  300 & 8.95$\pm$0.11 & --0.89$\pm$0.14 &     \nodata     & --1.51$\pm$0.25 &     \nodata     & 9.06 & 9.01 \nl
 +047 & --031 &  67.7 &  6 &  3900 $\pm$  200 & 9.12$\pm$0.10 & --0.88$\pm$0.14 &     \nodata     & --1.56$\pm$0.24 & --2.16$\pm$0.17 & 9.22 & 9.32 \nl
 +047 &  +046 &  69.9 &  7 &  4000 $\pm$  200 & 9.12$\pm$0.10 & --0.88$\pm$0.14 &     \nodata     & --1.52$\pm$0.25 &     \nodata     & 9.18 & 9.26 \nl
--055 & --032 &  71.0 &  1 &  4500 $\pm$  200 & 9.04$\pm$0.10 & --0.78$\pm$0.12 &     \nodata     & --1.44$\pm$0.24 &     \nodata     & 9.15 & 9.21 \nl
 +058 & --054 &  93.3 &  8 &  6200 $\pm$  400 & 8.74$\pm$0.10 & --1.14$\pm$0.12 &     \nodata     & --1.50$\pm$0.24 & --2.06$\pm$0.15 & 8.82 & 8.64 \nl
\hline
\multicolumn{12}{c}{NGC 2403}   \nl
\hline
 +186 & --177 & 262.4 &  4 &~~7600 $\pm$ 1400 & 8.74$\pm$0.24 & --1.16$\pm$0.30 & --0.50$\pm$0.69 & --1.50$\pm$0.34 & --2.26$\pm$0.38 & 8.41 & 8.28 \nl
 +178 & --203 & 286.4 &  4 &~~8200 $\pm$ 1500 & 8.40$\pm$0.21 & --1.23$\pm$0.27 & --0.81$\pm$0.62 & --1.87$\pm$0.32 & --2.12$\pm$0.35 & 8.79 & 8.60 \nl
 +176 & --211 & 295.2 &  4 &  6200 $\pm$  500 & 8.72$\pm$0.11 & --1.28$\pm$0.15 &     \nodata     & --1.31$\pm$0.25 & --2.27$\pm$0.22 & 8.81 & 8.63 \nl
 +166 & --229 & 315.9 &  4 &  7100 $\pm$  500 & 8.66$\pm$0.10 & --1.38$\pm$0.12 &     \nodata     & --1.52$\pm$0.24 & --2.40$\pm$0.14 & 8.66 & 8.48 \nl
 +356 & --195 & 414.2 &  5 &~~9100 $\pm$ 1000 & 8.44$\pm$0.10 & --1.35$\pm$0.14 & --0.80$\pm$0.28 & --1.54$\pm$0.25 & --2.27$\pm$0.17 & 8.44 & 8.30 \nl
 +360 & --190 & 417.5 &  5 &~~8000 $\pm$ 1500 & 8.44$\pm$0.18 & --1.36$\pm$0.28 & --0.20$\pm$0.67 & --1.32$\pm$0.30 & --2.39$\pm$0.34 & 8.78 & 8.59 \nl
 +377 & --163 & 435.2 &  5 &~~8500 $\pm$ 1500 & 8.42$\pm$0.18 & --1.39$\pm$0.25 & --1.04$\pm$0.56 & --1.69$\pm$0.30 & --2.22$\pm$0.31 & 8.65 & 8.47 \nl
 +376 & --106 & 446.8 &  6 & 10300 $\pm$ 2000 & 8.23$\pm$0.17 & --1.53$\pm$0.23 & --0.72$\pm$0.52 & --1.70$\pm$0.46 & --2.31$\pm$0.29 & 8.55 & 8.39 \nl
--377 &  +104 & 448.9 &  2 &~~8000 $\pm$ 1500 & 8.46$\pm$0.20 & --1.40$\pm$0.28 & --0.63$\pm$0.67 & --1.69$\pm$0.32 & --2.22$\pm$0.35 & 8.71 & 8.52 \nl
--105 & --218 & 452.0 &  3 &~~7800 $\pm$ 1500 & 8.33$\pm$0.22 & --1.48$\pm$0.31 &     \nodata     & --1.70$\pm$0.33 &     \nodata     & 8.93 & 8.78 \nl
--105 & --224 & 461.5 &  3 &~~7400 $\pm$ 1400 & 8.44$\pm$0.23 & --1.48$\pm$0.32 &     \nodata     & --1.62$\pm$0.33 &     \nodata     & 8.85 & 8.67 \nl
--381 &  +082 & 464.5 &  2 &~~8600 $\pm$ 1600 & 8.43$\pm$0.19 & --1.47$\pm$0.26 &     \nodata     & --1.69$\pm$0.30 & --2.24$\pm$0.33 & 8.60 & 8.43 \nl
 +383 & --056 & 482.7 &  6 &~~9100 $\pm$ 2000 & 8.35$\pm$0.23 & --1.56$\pm$0.31 &     \nodata     & --1.94$\pm$0.34 & --2.37$\pm$0.39 & 8.61 & 8.44 \nl
--104 & --256 & 511.2 &  3 &~~8500 $\pm$ 1500 & 8.34$\pm$0.18 & --1.58$\pm$0.25 &     \nodata     & --1.68$\pm$0.30 & --2.36$\pm$0.31 & 8.72 & 8.53 \nl
--425 & --002 & 585.8 &  1 &~~8900 $\pm$ 2000 & 8.33$\pm$0.24 & --1.58$\pm$0.32 &     \nodata     & --1.61$\pm$0.34 & --2.35$\pm$0.41 & 8.68 & 8.49 \nl
--423 & --010 & 590.3 &  1 &~~9300 $\pm$ 2000 & 8.26$\pm$0.21 & --1.56$\pm$0.29 &     \nodata     & --1.71$\pm$0.32 & --2.45$\pm$0.37 & 8.70 & 8.52 \nl
--421 & --017 & 594.1 &  1 &~~9000 $\pm$ 1000 & 8.21$\pm$0.11 & --1.55$\pm$0.15 &     \nodata     & --1.49$\pm$0.25 & --2.12$\pm$0.19 & 8.79 & 8.61 \nl
\hline
\multicolumn{12}{c}{NGC 2805}    \nl
\hline
--051 & --009 &  60.8 &  2 &  5900 $\pm$  400 & 8.84$\pm$0.10 & --1.14$\pm$0.13 & --0.62$\pm$0.25 & --1.21$\pm$0.24 & --2.33$\pm$0.16 & 8.93 & 8.78 \nl
 +055 & --026 &  61.5 &  6 &  7300 $\pm$  800 & 8.64$\pm$0.11 & --1.27$\pm$0.17 & --0.50$\pm$0.35 & --1.62$\pm$0.25 & --2.36$\pm$0.21 & 8.65 & 8.47 \nl
 +058 & --022 &  63.4 &  6 &  6500 $\pm$  500 & 8.75$\pm$0.10 & --1.25$\pm$0.14 & --0.47$\pm$0.26 & --1.56$\pm$0.24 & --2.31$\pm$0.16 & 8.77 & 8.58 \nl
--021 & --045 &  65.3 &  3 &  6200 $\pm$  500 & 8.79$\pm$0.10 & --1.18$\pm$0.15 &     \nodata     & --1.69$\pm$0.24 & --2.23$\pm$0.18 & 8.85 & 8.67 \nl
 +032 & --056 &  68.8 &  6 &  7300 $\pm$  800 & 8.62$\pm$0.11 & --1.32$\pm$0.17 &     \nodata     & --1.57$\pm$0.26 & --2.51$\pm$0.22 & 8.64 & 8.46 \nl
--001 & --079 &  97.2 &  4 &  6800 $\pm$  500 & 8.75$\pm$0.10 & --1.28$\pm$0.13 & --0.65$\pm$0.24 & --1.31$\pm$0.22 & --2.28$\pm$0.15 & 8.73 & 8.54 \nl
--005 & --080 &  99.8 &  4 &  7300 $\pm$  800 & 8.65$\pm$0.11 & --1.29$\pm$0.17 &     \nodata     & --1.47$\pm$0.29 & --2.43$\pm$0.20 & 8.65 & 8.46 \nl
--041 & --090 & 130.0 &  4 &  8000 $\pm$  800 & 8.56$\pm$0.10 & --1.35$\pm$0.15 & --0.43$\pm$0.28 & --1.49$\pm$0.24 & --2.43$\pm$0.17 & 8.56 & 8.40 \nl
 +037 & --115 & 136.4 &  5 &  9710 $\pm$  490 & 8.33$\pm$0.04 & --1.34$\pm$0.08 & --0.70$\pm$0.11 & --1.57$\pm$0.10 & --2.32$\pm$0.09 & 8.53 & 8.37 \nl
 +078 & --107 & 137.6 &  5 &  7900 $\pm$  500 & 8.61$\pm$0.10 & --1.37$\pm$0.13 & --0.76$\pm$0.25 & --1.61$\pm$0.25 & --2.37$\pm$0.17 & 8.56 & 8.39 \nl
--068 & --079 & 137.7 &  3 & 11480 $\pm$  840 & 8.19$\pm$0.05 & --1.31$\pm$0.08 & --0.62$\pm$0.13 & --1.67$\pm$0.11 & --2.35$\pm$0.10 & 8.41 & 7.84 \nl
 +080 & --110 & 141.3 &  6 & 10000 $\pm$ 1000 & 8.38$\pm$0.10 & --1.38$\pm$0.13 &     \nodata     & --1.42$\pm$0.24 & --2.48$\pm$0.16 & 8.37 & 8.26 \nl
 +017 & --119 & 142.3 &  5 &~~9700 $\pm$ 2000 & 8.28$\pm$0.20 & --1.30$\pm$0.27 & --0.96$\pm$0.62 & --1.60$\pm$0.31 & --2.24$\pm$0.34 & 8.61 & 8.44 \nl
 +089 & --107 & 142.8 &  6 &~~9180 $\pm$ 2670 & 8.49$\pm$0.13 & --1.53$\pm$0.19 &     \nodata     & --1.64$\pm$0.30 & --2.42$\pm$0.22 & 8.36 & 8.25 \nl
 +116 & --098 & 152.3 &  6 & 10970 $\pm$  270 & 8.22$\pm$0.03 & --1.37$\pm$0.06 & --0.62$\pm$0.07 & --1.64$\pm$0.07 & --2.35$\pm$0.08 & 8.45 & 7.80 \nl
--145 &  +104 & 178.4 &  1 &~~9300 $\pm$ 1000 & 8.40$\pm$0.10 & --1.42$\pm$0.14 &     \nodata     & --1.45$\pm$0.24 &     \nodata     & 8.44 & 8.30 \nl
--151 &  +102 & 182.2 &  1 & 12000 $\pm$ 1500 & 8.16$\pm$0.10 & --1.32$\pm$0.13 & --0.75$\pm$0.25 & --1.72$\pm$0.22 & --2.32$\pm$0.17 & 8.32 & 7.91 \nl
\hline
\multicolumn{12}{c}{IC 2458}   \nl
\hline
 +002 & --006 & \nodata &  1 & 13120 $\pm$  460 & 8.06$\pm$0.03 & --1.43$\pm$0.07 & --0.68$\pm$0.08 & --1.68$\pm$0.08 & --2.28$\pm$0.08 & 8.32 & 7.90 \nl
--028 & --007 & \nodata &  1 & 13100 $\pm$ 1500 & 7.87$\pm$0.10 & --1.68$\pm$0.14 &     \nodata     & --1.37$\pm$0.24 &     \nodata     & 8.70 & 7.56 \nl
--033 & --007 & \nodata &  1 & 13210 $\pm$ 1110 & 7.98$\pm$0.05 & --1.48$\pm$0.09 & --0.95$\pm$0.13 & --1.63$\pm$0.13 & --2.39$\pm$0.11 & 8.48 & 7.78 \nl
\hline
\multicolumn{12}{c}{NGC 2820}  \nl
\hline
 +001 &  +001 & \nodata &  1 &~~6000 $\pm$ 1000 & 8.95$\pm$0.25 & --1.30$\pm$0.34 &     \nodata     & --1.54$\pm$0.36 &     \nodata     & 8.61 & 8.43 \nl
--018 & --009 & \nodata &  1 &  6900 $\pm$  500 & 8.69$\pm$0.10 & --1.22$\pm$0.14 &     \nodata     & --1.49$\pm$0.24 &     \nodata     & 8.90 & 8.74 \nl
 +021 &  +011 & \nodata &  1 &  6000 $\pm$  400 & 8.83$\pm$0.10 & --1.08$\pm$0.13 &     \nodata     & --1.66$\pm$0.24 & --2.31$\pm$0.16 & 8.69 & 8.50 \nl
--025 & --013 & \nodata &  1 &  8000 $\pm$  800 & 8.60$\pm$0.10 & --1.25$\pm$0.15 &     \nodata     & --1.56$\pm$0.25 &     \nodata     & 8.53 & 8.37 \nl
\hline
\multicolumn{12}{c}{NGC 2903}  \nl
\hline
--062 & --085 & 123.2 &  1 &  6200 $\pm$  500 & 8.78$\pm$0.10 & --0.96$\pm$0.15 &     \nodata     & --1.47$\pm$0.25 &     \nodata     & 8.81 & 8.63 \nl
--065 & --073 & 123.7 &  1 &  5500 $\pm$  500 & 8.79$\pm$0.16 & --1.16$\pm$0.19 &     \nodata     & --1.31$\pm$0.28 &     \nodata     & 8.90 & 8.75 \nl
--067 & --061 & 124.9 &  1 &  5800 $\pm$  500 & 8.81$\pm$0.11 & --1.09$\pm$0.17 &     \nodata     & --1.50$\pm$0.25 & --2.40$\pm$0.24 & 8.88 & 8.72 \nl
--060 & --100 & 127.8 &  1 &  5600 $\pm$  500 & 8.79$\pm$0.14 & --1.12$\pm$0.18 &     \nodata     & --1.34$\pm$0.27 &     \nodata     & 8.89 & 8.73 \nl
 +171 &  +196 & 326.6 &  2 &  6700 $\pm$  500 & 8.71$\pm$0.10 & --1.19$\pm$0.13 &     \nodata     & --1.72$\pm$0.24 & --2.17$\pm$0.16 & 8.74 & 8.55 \nl
 +171 &  +226 & 336.3 &  2 &  6400 $\pm$  500 & 8.78$\pm$0.10 & --1.19$\pm$0.14 &     \nodata     & --1.79$\pm$0.24 & --2.17$\pm$0.17 & 8.79 & 8.60 \nl
 +171 &  +232 & 338.7 &  2 &  6400 $\pm$  500 & 8.70$\pm$0.11 & --1.26$\pm$0.14 &     \nodata     & --1.42$\pm$0.25 & --2.02$\pm$0.20 & 8.80 & 8.61 \nl
 +171 &  +236 & 340.3 &  2 &  6000 $\pm$  500 & 8.74$\pm$0.12 & --1.25$\pm$0.16 &     \nodata     & --1.45$\pm$0.26 &     \nodata     & 8.85 & 8.68 \nl
 +171 &  +243 & 343.3 &  2 &  5700 $\pm$  500 & 8.79$\pm$0.12 & --1.24$\pm$0.17 &     \nodata     & --1.57$\pm$0.26 & --2.02$\pm$0.22 & 8.89 & 8.72 \nl
\hline 
\multicolumn{12}{c}{NGC 3184}   \nl
\hline
--058 & --007 &  60.9 &  2 &  3400 $\pm$  200 & 9.27$\pm$0.15 & --0.70$\pm$0.18 &     \nodata     & --1.55$\pm$0.27 &     \nodata     & 9.35 & 9.51 \nl
--064 & --006 &  66.9 &  2 &  3700 $\pm$  200 & 9.23$\pm$0.15 & --0.67$\pm$0.16 &     \nodata     & --1.55$\pm$0.27 & --2.16$\pm$0.23 & 9.29 & 9.43 \nl
--080 & --005 &  83.3 &  2 &  3800 $\pm$  200 & 9.19$\pm$0.11 & --0.74$\pm$0.15 &     \nodata     & --1.61$\pm$0.25 &     \nodata     & 9.28 & 9.41 \nl
 +085 & --004 &  87.8 &  4 &  3900 $\pm$  200 & 9.18$\pm$0.10 & --0.78$\pm$0.14 &     \nodata     & --1.52$\pm$0.25 &     \nodata     & 9.24 & 9.34 \nl
 +079 &  +035 &  91.5 &  4 &  3600 $\pm$  200 & 9.24$\pm$0.14 & --0.81$\pm$0.16 &     \nodata     & --1.52$\pm$0.25 &     \nodata     & 9.28 & 9.42 \nl
 +059 & --079 &  98.7 &  5 &  4000 $\pm$  200 & 9.12$\pm$0.10 & --0.89$\pm$0.14 &     \nodata     & --1.48$\pm$0.24 & --2.25$\pm$0.19 & 9.21 & 9.30 \nl
 +074 &  +064 & 104.5 &  4 &  4300 $\pm$  200 & 9.11$\pm$0.11 & --0.79$\pm$0.12 &     \nodata     & --1.46$\pm$0.25 &     \nodata     & 9.20 & 9.28 \nl
 +092 & --093 & 130.8 &  5 &  4100 $\pm$  200 & 9.09$\pm$0.10 & --0.99$\pm$0.13 &     \nodata     & --1.40$\pm$0.24 & --2.18$\pm$0.16 & 9.16 & 9.22 \nl
 +005 &  +135 & 140.1 &  3 &  5000 $\pm$  600 & 9.10$\pm$0.19 & --1.08$\pm$0.28 &     \nodata     & --1.70$\pm$0.31 & --2.26$\pm$0.33 & 8.94 & 8.80 \nl
--002 &  +136 & 140.6 &  3 &  4200 $\pm$  200 & 9.05$\pm$0.10 & --1.06$\pm$0.13 &     \nodata     & --1.46$\pm$0.25 &     \nodata     & 9.16 & 9.22 \nl
--017 &  +137 & 141.7 &  3 &  4400 $\pm$  200 & 9.05$\pm$0.10 & --0.97$\pm$0.12 &     \nodata     & --1.56$\pm$0.24 &     \nodata     & 9.13 & 9.18 \nl
 +111 & --102 & 150.8 &  5 &  5400 $\pm$  300 & 9.01$\pm$0.22 & --1.21$\pm$0.28 &     \nodata     & --1.62$\pm$0.33 &     \nodata     & 8.78 & 8.60 \nl
--119 & --121 & 181.3 &  1 &  4700 $\pm$  300 & 8.97$\pm$0.11 & --1.07$\pm$0.15 &     \nodata     & --1.49$\pm$0.25 & --2.24$\pm$0.25 & 9.07 & 9.03 \nl
--113 & --127 & 181.5 &  1 &  6000 $\pm$  500 & 8.71$\pm$0.10 & --1.02$\pm$0.16 &     \nodata     & --1.62$\pm$0.25 & --2.19$\pm$0.20 & 8.72 & 8.53 \nl
--101 & --137 & 181.5 &  1 &  5500 $\pm$  500 & 8.90$\pm$0.13 & --1.05$\pm$0.18 &     \nodata     & --1.84$\pm$0.26 & --2.12$\pm$0.22 & 8.96 & 8.82 \nl
--110 & --130 & 181.8 &  1 &  5500 $\pm$  500 & 8.87$\pm$0.12 & --1.06$\pm$0.18 &     \nodata     & --1.71$\pm$0.26 & --2.23$\pm$0.23 & 8.95 & 8.82 \nl
--095 & --142 & 182.0 &  1 &  6200 $\pm$  500 & 8.82$\pm$0.10 & --1.09$\pm$0.15 &     \nodata     & --1.60$\pm$0.25 & --2.27$\pm$0.18 & 8.84 & 8.66 \nl
\hline
\multicolumn{12}{c}{NGC 4395}   \nl
\hline
 +061 & --029 &  68.5 &  8 &  7800 $\pm$  500 & 8.48$\pm$0.21 & --1.50$\pm$0.31 &     \nodata     & --1.51$\pm$0.32 & --2.28$\pm$0.37 & 8.71 & 8.53 \nl
 +099 & --029 & 105.3 &  8 &  8100 $\pm$  800 & 8.57$\pm$0.10 & --1.29$\pm$0.25 &     \nodata     & --1.48$\pm$0.35 &     \nodata     & 8.51 & 8.36 \nl
 +002 & --127 & 128.8 &  7 &~~9700 $\pm$ 2000 & 8.29$\pm$0.20 & --1.44$\pm$0.27 & --0.84$\pm$0.62 & --1.46$\pm$0.56 & --2.16$\pm$0.34 & 8.58 & 8.41 \nl
--075 & --117 & 144.7 &  4 &~~8900 $\pm$ 2000 & 8.39$\pm$0.23 & --1.60$\pm$0.32 & --0.69$\pm$0.80 & --1.47$\pm$0.34 & --2.40$\pm$0.41 & 8.57 & 8.40 \nl
 +088 & --119 & 148.0 &  9 &~~8930 $\pm ^{5790}_{1200}$ & 8.41$\pm$0.17 & --1.44$\pm$0.27 & --0.57$\pm$0.47 & --1.61$\pm$0.30 & --2.25$\pm$0.31 & 8.58 & 8.41 \nl
--167 & --093 & 200.5 &  3 & 10300 $\pm$ 2000 & 8.33$\pm$0.17 & --1.59$\pm$0.23 & --0.58$\pm$0.52 & --1.54$\pm$0.29 & --2.26$\pm$0.30 & 8.32 & 7.90 \nl
 +118 &  +206 & 246.7 & 10 &~~8500 $\pm$ 1500 & 8.39$\pm$0.17 & --1.41$\pm$0.25 & --0.91$\pm$0.56 & --1.22$\pm$0.52 & --2.19$\pm$0.31 & 8.71 & 8.52 \nl
 +195 & --159 & 252.8 & 11 & 10000 $\pm$ 2000 & 8.31$\pm$0.17 & --1.66$\pm$0.25 &     \nodata     & --1.82$\pm$0.44 & --2.32$\pm$0.31 & 8.48 & 8.33 \nl
 +200 &  +141 & 256.7 & 12 &~~9100 $\pm$ 2000 & 8.38$\pm$0.22 & --1.48$\pm$0.31 & --0.71$\pm$0.74 & --1.59$\pm$0.33 & --2.21$\pm$0.39 & 8.55 & 8.39 \nl
--226 &  +237 & 328.0 &  2 & 12000 $\pm$ 1500 & 8.06$\pm$0.10 & --1.46$\pm$0.13 & --0.83$\pm$0.25 & --1.69$\pm$0.24 & --2.17$\pm$0.16 & 8.52 & 7.74 \nl
--272 &  +186 & 332.0 &  1 & 10050 $\pm$  770 & 8.33$\pm$0.05 & --1.55$\pm$0.09 & --0.75$\pm$0.14 & --1.65$\pm$0.14 & --2.28$\pm$0.11 & 8.42 & 8.29 \nl
\hline
\multicolumn{12}{c}{NGC 5457}    \nl
\hline
--075 &  +028 &  84.7 &  7 &  3600 $\pm$  200 & 9.23$\pm$0.17 & --0.57$\pm$0.17 &     \nodata     & --1.53$\pm$0.28 &     \nodata     & 9.33 & 9.48 \nl
--106 &  +028 & 114.3 &  7 &  4250 $\pm$  250 & 9.13$\pm$0.12 & --0.80$\pm$0.15 &     \nodata     & --1.63$\pm$0.25 &     \nodata     & 9.21 & 9.30 \nl
--347 &  +276 & 465.2 &  4 &  9880 $\pm$  170 & 8.41$\pm$0.03 & --1.17$\pm$0.06 & --0.69$\pm$0.07 & --1.38$\pm$0.20 & --2.03$\pm$0.07 & 8.32 & 8.23 \nl
--459 & --053 & 474.1 &  2 & 10450 $\pm$  480 & 8.22$\pm$0.04 & --1.39$\pm$0.07 & --0.68$\pm$0.10 & --1.65$\pm$0.21 & --2.32$\pm$0.09 & 8.55 & 8.38 \nl
 +271 & --393 & 498.6 & 11 &~~7600 $\pm$ 1400 & 8.49$\pm$0.21 & --1.49$\pm$0.30 &     \nodata     & --1.43$\pm$0.63 & --2.28$\pm$0.36 & 8.74 & 8.56 \nl
--276 & --417 & 500.2 &  6 &~~8600 $\pm$ 1600 & 8.30$\pm$0.19 & --1.47$\pm$0.26 & --0.60$\pm$0.40 & --1.46$\pm$0.30 & --2.36$\pm$0.33 & 8.76 & 8.57 \nl
--037 & --532 & 541.2 &  8 &~~8900 $\pm$ 2000 & 8.38$\pm$0.24 & --1.54$\pm$0.33 &     \nodata     & --1.70$\pm$0.34 & --2.54$\pm$0.42 & 8.61 & 8.43 \nl
--250 &  +484 & 566.6 &  5 &~~9100 $\pm$ 2000 & 8.28$\pm$0.23 & --1.48$\pm$0.31 & --0.74$\pm$0.74 & --1.58$\pm$0.33 & --2.42$\pm$0.39 & 8.72 & 8.54 \nl
--398 & --436 & 590.6 &  3 & 12750 $\pm$  320 & 8.01$\pm$0.03 & --1.44$\pm$0.06 & --0.77$\pm$0.07 & --1.66$\pm$0.07 & --2.30$\pm$0.07 & 8.50 & 7.76 \nl
--291 &  +489 & 593.1 &  5 &~~8300 $\pm$ 1500 & 8.44$\pm$0.19 & --1.55$\pm$0.27 & --0.67$\pm$0.60 & --1.56$\pm$0.30 & --2.44$\pm$0.33 & 8.64 & 8.46 \nl
--499 &  +300 & 610.6 &  1 & 10300 $\pm$ 1200 & 8.09$\pm$0.10 & --1.43$\pm$0.14 & --0.23$\pm$0.27 & --1.47$\pm$0.24 & --2.24$\pm$0.17 & 8.56 & 7.70 \nl
 +010 &  +885 & 900.6 & 10 & 13630 $\pm$  440 & 7.92$\pm$0.03 & --1.45$\pm$0.06 & --0.75$\pm$0.07 & --1.69$\pm$0.39 & --2.46$\pm$0.08 & 8.55 & 7.72 \nl
 +005 &  +887 & 902.8 & 10 & 12200 $\pm$ 1500 & 7.95$\pm$0.10 & --1.51$\pm$0.13 & --0.51$\pm$0.24 & --1.53$\pm$0.24 & --2.44$\pm$0.16 & 8.73 & 7.53 \nl
\enddata 
\end{deluxetable} 

\tablenum{6}
\begin{deluxetable}{rccccccc}
\tablewidth{35pc}
\tablecaption{Radial Oxygen Abundance Gradients}
\tablehead{
\colhead{}&\colhead{No. of } & \colhead{Central}
& \multicolumn{3}{c}{\underline{\ \ \ \ \ \ \ \ \ \ \ \ \ \ \ \ \ Gradients \ \ \ \ \ \ \ \ \ \ \ \ \ \ \ \  \ \ } }\\
\colhead{Galaxy}& \colhead{HII Regions} & \colhead{12+log(O/H)} & \colhead{[dex/$r_{25}$]} & \colhead{[dex/r$_d$]} & \colhead{[dex/kpc]}}
\startdata
NGC 0628  & 26 & 9.46 $\pm$ 0.11 & --0.99 $\pm$ 0.14 & --0.23 $\pm$ 0.03 & --0.067 $\pm$ 0.009 \nl
NGC 0925  & 53 & 8.79 $\pm$ 0.06 & --0.45 $\pm$ 0.08 & --0.12 $\pm$ 0.02 & --0.032 $\pm$ 0.006 \nl
NGC 1068  & 13 & 9.26 $\pm$ 0.02 & --0.30 $\pm$ 0.07 & --0.16 $\pm$ 0.04 & --0.020 $\pm$ 0.004 \nl
NGC 1232  & 16 & 9.49 $\pm$ 0.12 & --1.31 $\pm$ 0.20 & --0.36 $\pm$ 0.05 & --0.056 $\pm$ 0.009 \nl
NGC 1637  & 15 & 9.18 $\pm$ 0.07 & --0.34 $\pm$ 0.15 & --0.06 $\pm$ 0.02 & --0.068 $\pm$ 0.028 \nl
NGC 2403  & 40 & 8.70 $\pm$ 0.07 & --0.77 $\pm$ 0.14 & --0.14 $\pm$ 0.02 & --0.074 $\pm$ 0.013 \nl
NGC 2805  & 17 & 9.11 $\pm$ 0.12 & --1.05 $\pm$ 0.17 & --0.42 $\pm$ 0.07 & --0.049 $\pm$ 0.008 \nl
NGC 2903  & 36 & 9.22 $\pm$ 0.06 & --0.56 $\pm$ 0.09 & --0.13 $\pm$ 0.02 & --0.048 $\pm$ 0.008 \nl
NGC 3184  & 32 & 9.50 $\pm$ 0.04 & --0.78 $\pm$ 0.07 & --0.20 $\pm$ 0.02 & --0.083 $\pm$ 0.007 \nl
NGC 4395  & 14 & 8.48 $\pm$ 0.13 & --0.32 $\pm$ 0.19 & --0.13 $\pm$ 0.07 & --0.037 $\pm$ 0.022 \nl
NGC 5457  & 53 & 9.29 $\pm$ 0.05 & --1.52 $\pm$ 0.09 & --0.27 $\pm$ 0.02 & --0.049 $\pm$ 0.003 \nl
\enddata
\end{deluxetable}

\begin{table}
\dummytable\label{tab:props}
%this table contains a summary of the galaxy properties
\end{table}

\begin{table}
\dummytable\label{tab:obs}
%this table contains a list of the slit positions and integration times
\end{table}

\begin{table}
\dummytable\label{tab:lines}
%this table contains a list of H II region line strengths
\end{table}

\begin{table}
\dummytable\label{tab:ratios}
%this table contains a list of H II region line ratios
\end{table}

\begin{table}
\dummytable\label{tab:abund}
%this table contains a list of H II region abundances
\end{table}

\begin{table}
\dummytable\label{tab:grad}
%this table contains a summary of abundance gradients
\end{table}

\clearpage

\psfig{figure=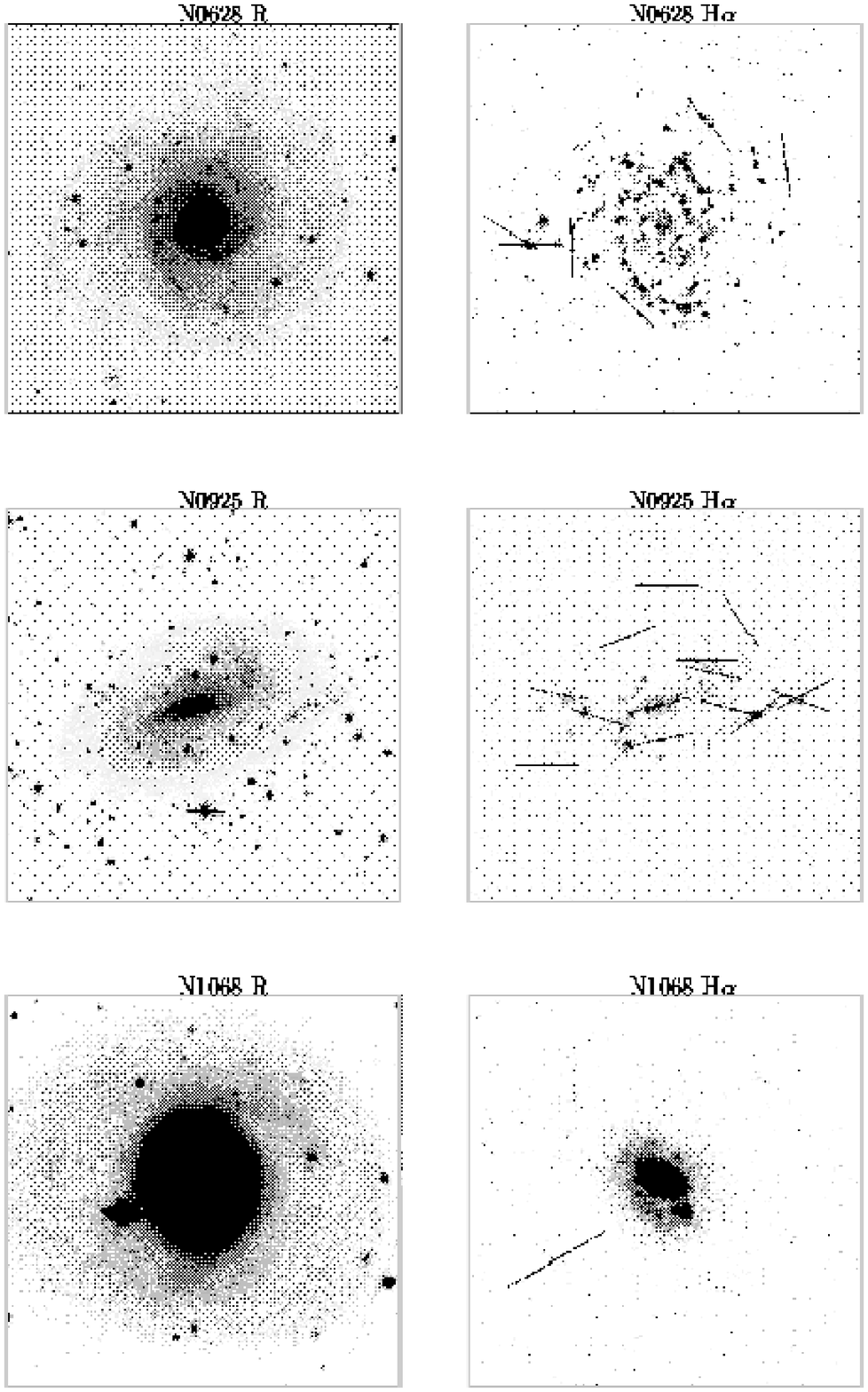,width=6.in,bbllx=-100pt,bblly=20pt,bburx=500pt,bbury=800pt}
\vskip -0.3 truein
\figcaption[] { R--band and continuum subtracted H$\alpha$ images of the galaxies in the sample.
The positions of the 2\arcmin~slits are illustrated on the H$\alpha$ images.  
The images are oriented with north to the top and east to the left.  \label{fig:images} }

\psfig{figure=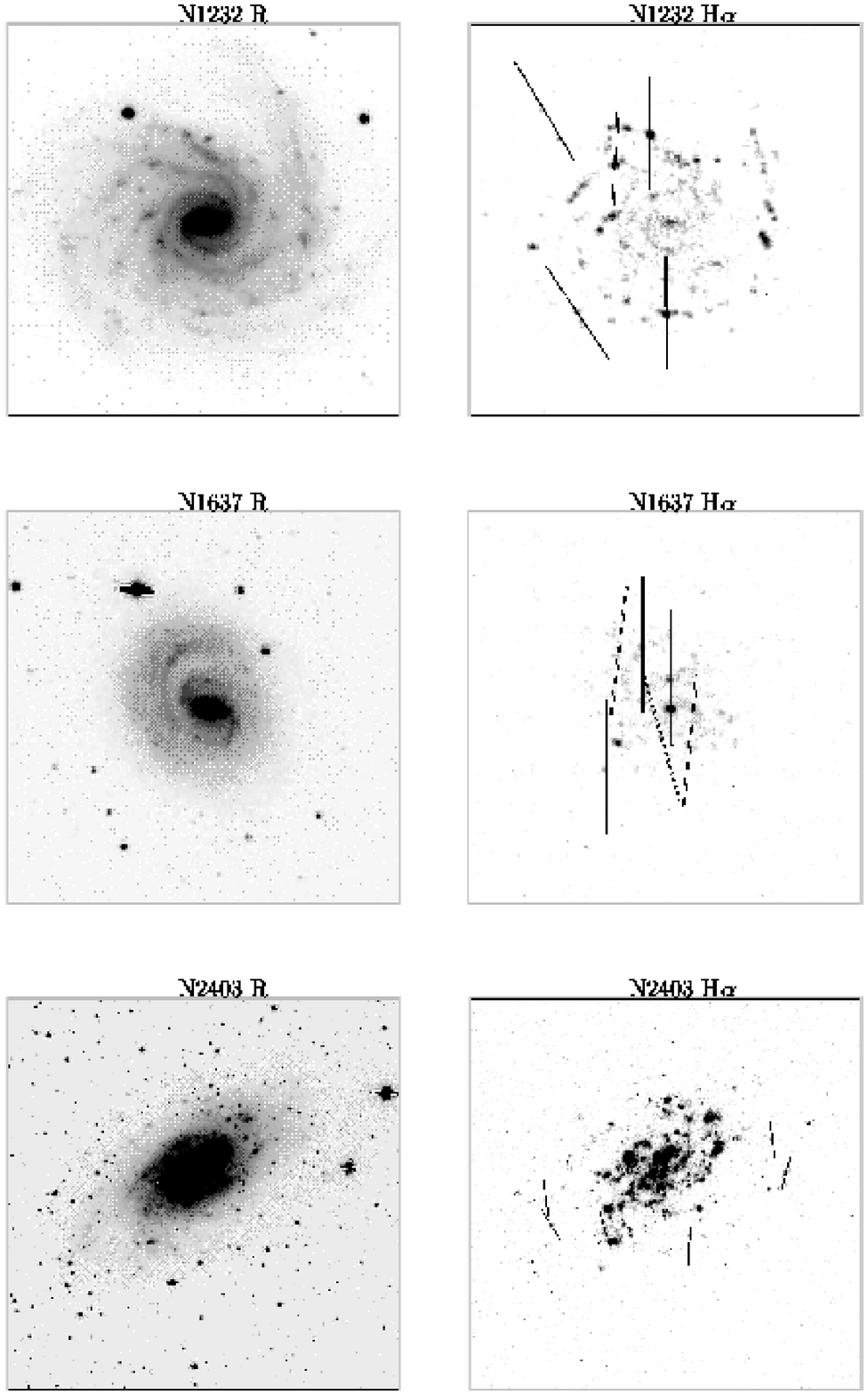,width=6.in,bbllx=0pt,bblly=20pt,bburx=600pt,bbury=800pt}
\vskip -0.3 truein
Fig. 1 continued

\psfig{figure=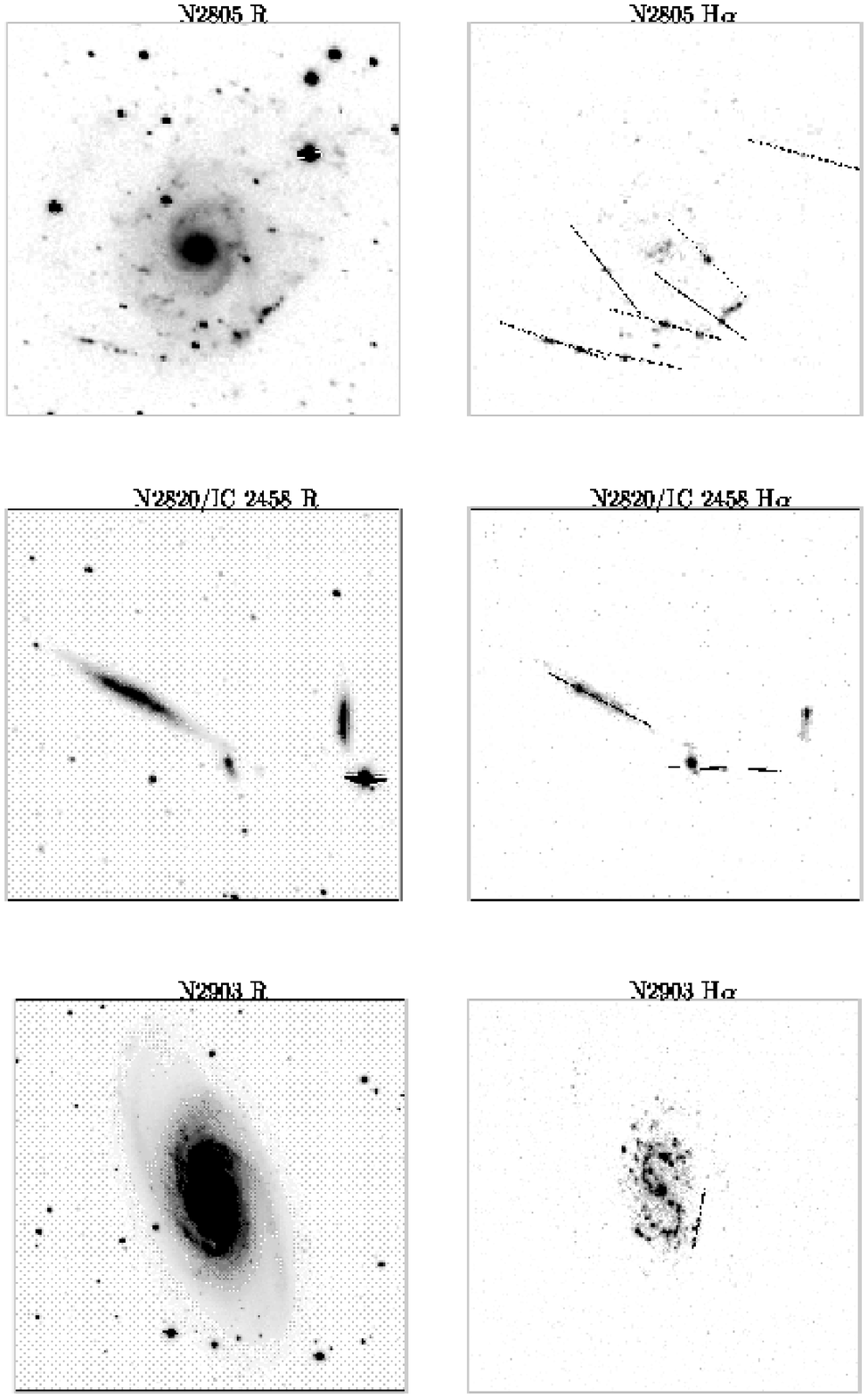,width=6.in,bbllx=0pt,bblly=20pt,bburx=600pt,bbury=800pt}
\vskip -0.3 truein
Fig. 1 continued

\psfig{figure=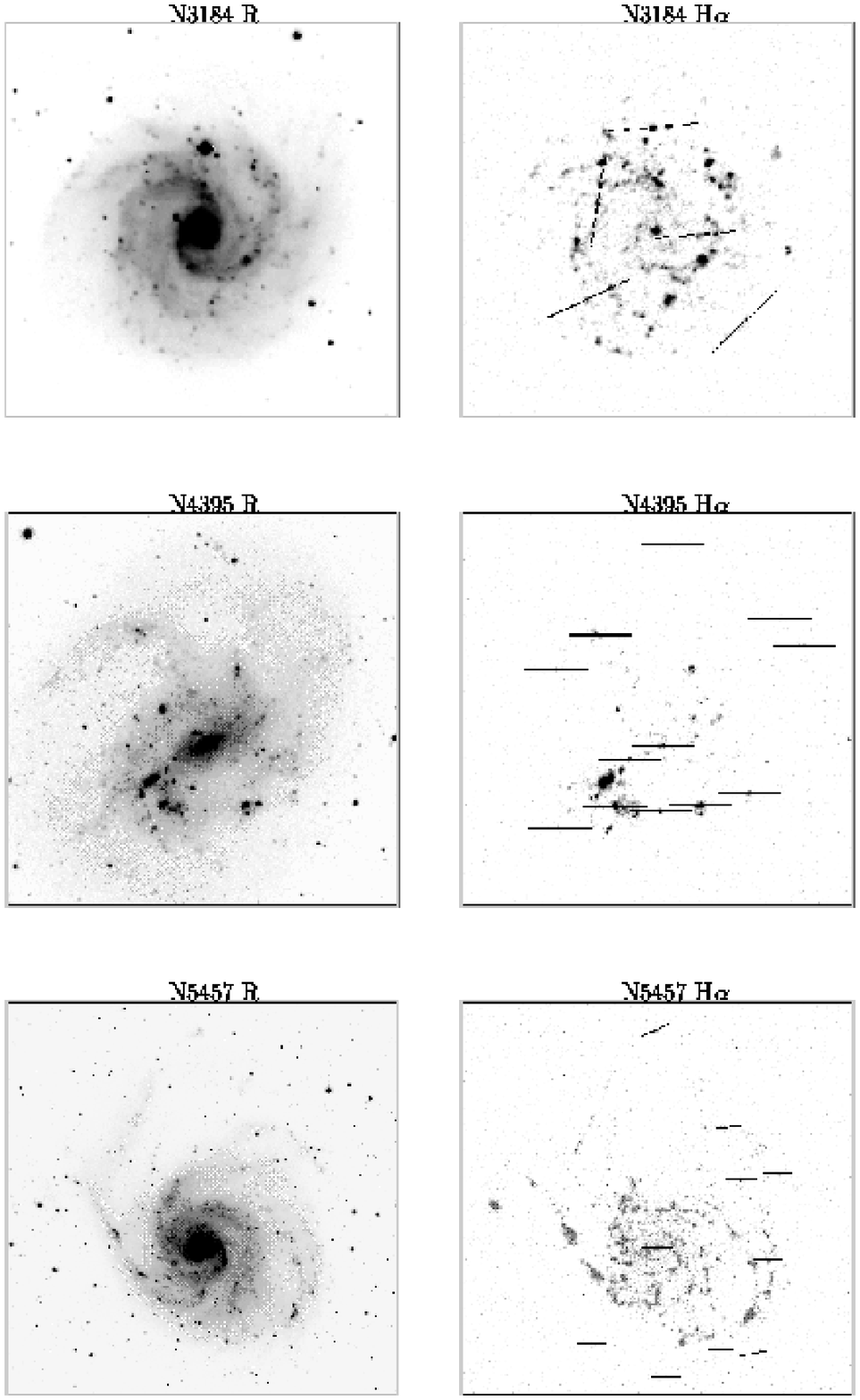,width=6.in,bbllx=0pt,bblly=20pt,bburx=600pt,bbury=800pt}
\vskip -0.3 truein
Fig. 1 continued

\psfig{figure=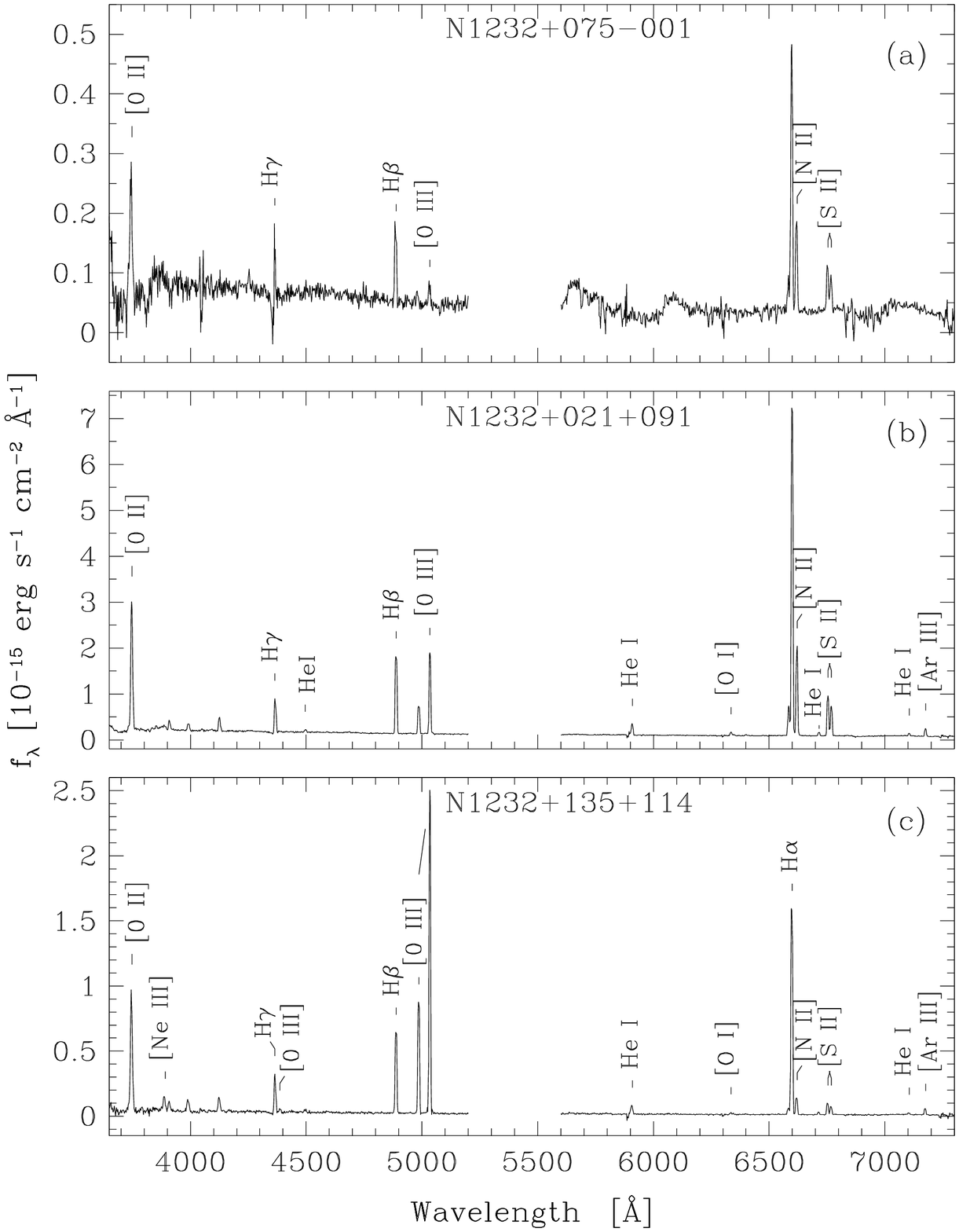,width=6.4in,bbllx=0pt,bblly=50pt,bburx=600pt,bbury=750pt}
\figcaption[] {Representative spectra of H II regions in NGC 1232.
The inner H II regions (a) and (b) are clearly
high abundance, with strong [N II] and [S II] and weak [O III] lines.  The
outermost H II region (c) is low abundance, with weak [N II] and strong [O III] lines.
\label{fig:spec} }

\psfig{figure=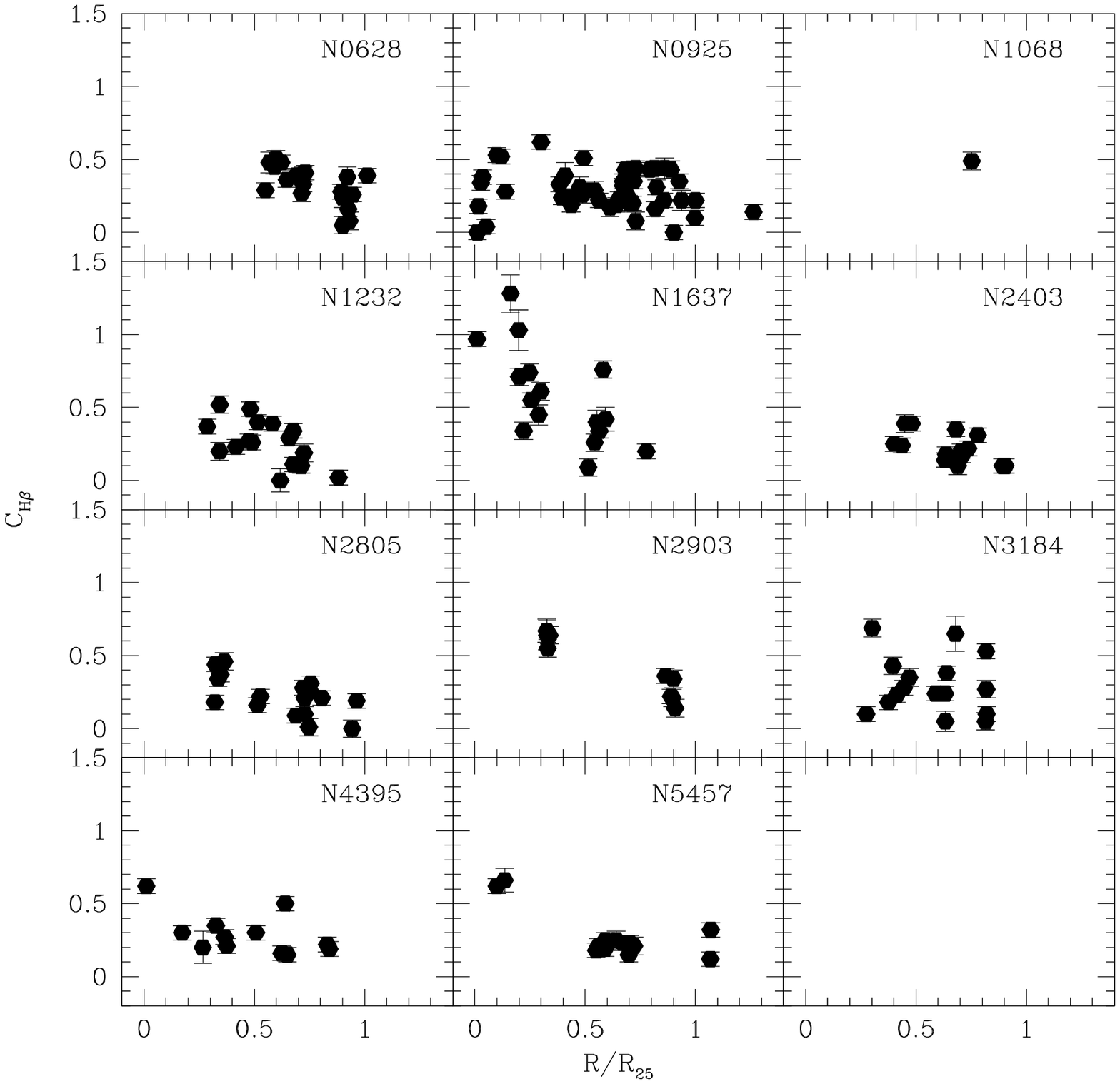,width=6.in,bbllx=0pt,bblly=100pt,bburx=600pt,bbury=700pt,clip=t}
\vskip -0.5 truein
\figcaption[]{Reddening as a function of radius for each galaxy.  In each panel
the radius has been normalized by the isophotal radius of that galaxy.  \label{fig:red}}

\psfig{figure=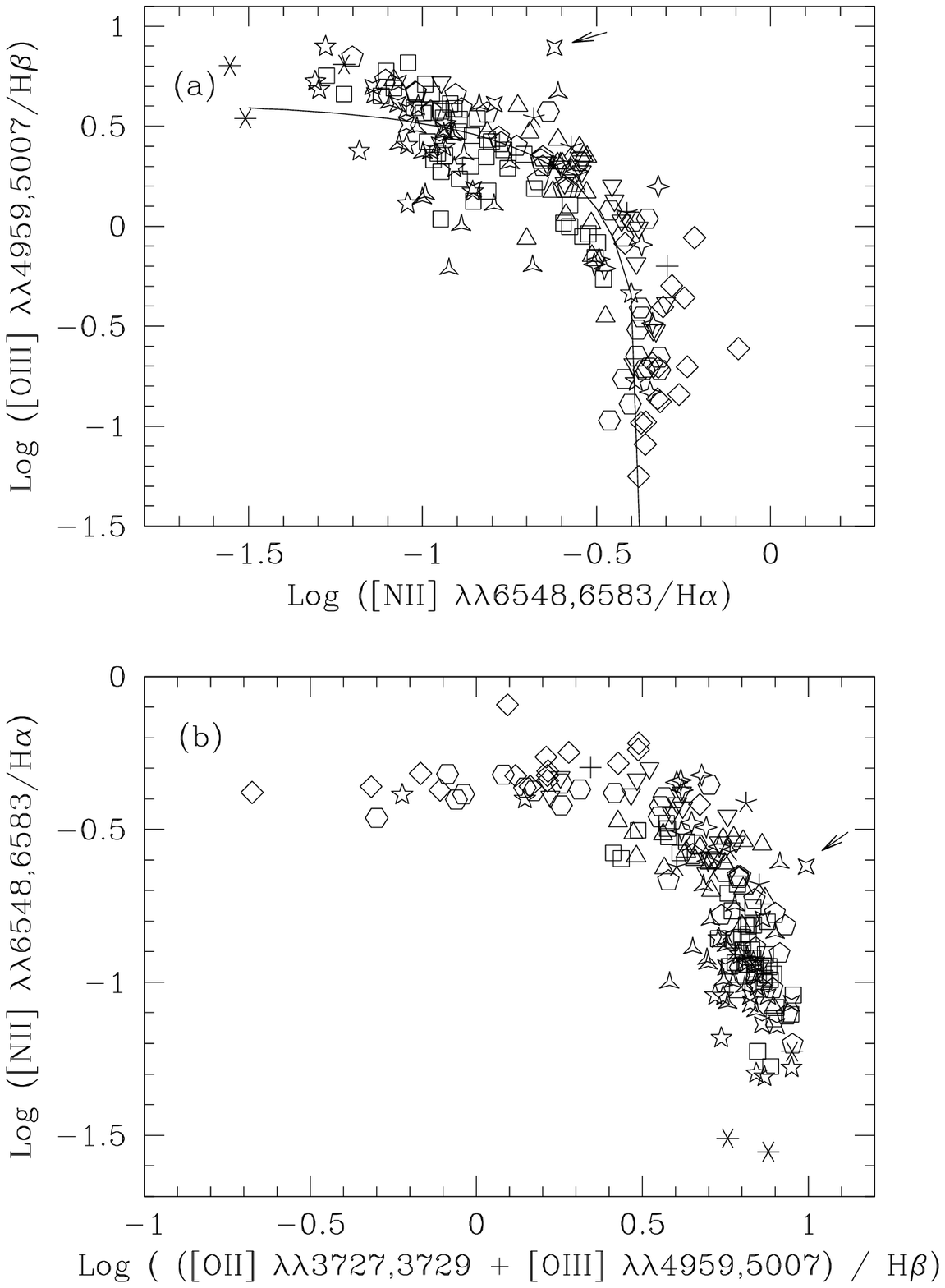,width=6.in,bbllx=0pt,bblly=100pt,bburx=600pt,bbury=700pt,clip=t}
\vskip -0.5 truein
\figcaption[]{Emission--line diagnostic diagram for [N II].  In this and subsequent
figures, H II regions from NGC 628 are denoted by upright triangles, NGC 925 by squares, NGC 1068 by
crosses, NGC 1232 by downward triangles, NGC 1637 by diamonds, NGC 2403 by upward indented
triangles, NGC 2805 by pentagons, IC 2458 by six pointed stars, NGC 2820 by five pointed stars,
NGC 2903 by indented crosses, NGC 3184 by hexagons, NGC 4395 by four pointed open stars,
and NGC 5457 by five pointed open stars. (a)  The H II region sequence formed by
[N II] and [O III] normalized by Balmer lines.  A theoretical curve from
Baldwin \etal (1981) is superposed.  (b) The H II region sequence formed by 
R$_{23}$ vs. [N II]/H$\alpha$. In both panels, NGC 4395--003--003 is marked by an arrow. 
\label{fig:nh}}

\psfig{figure=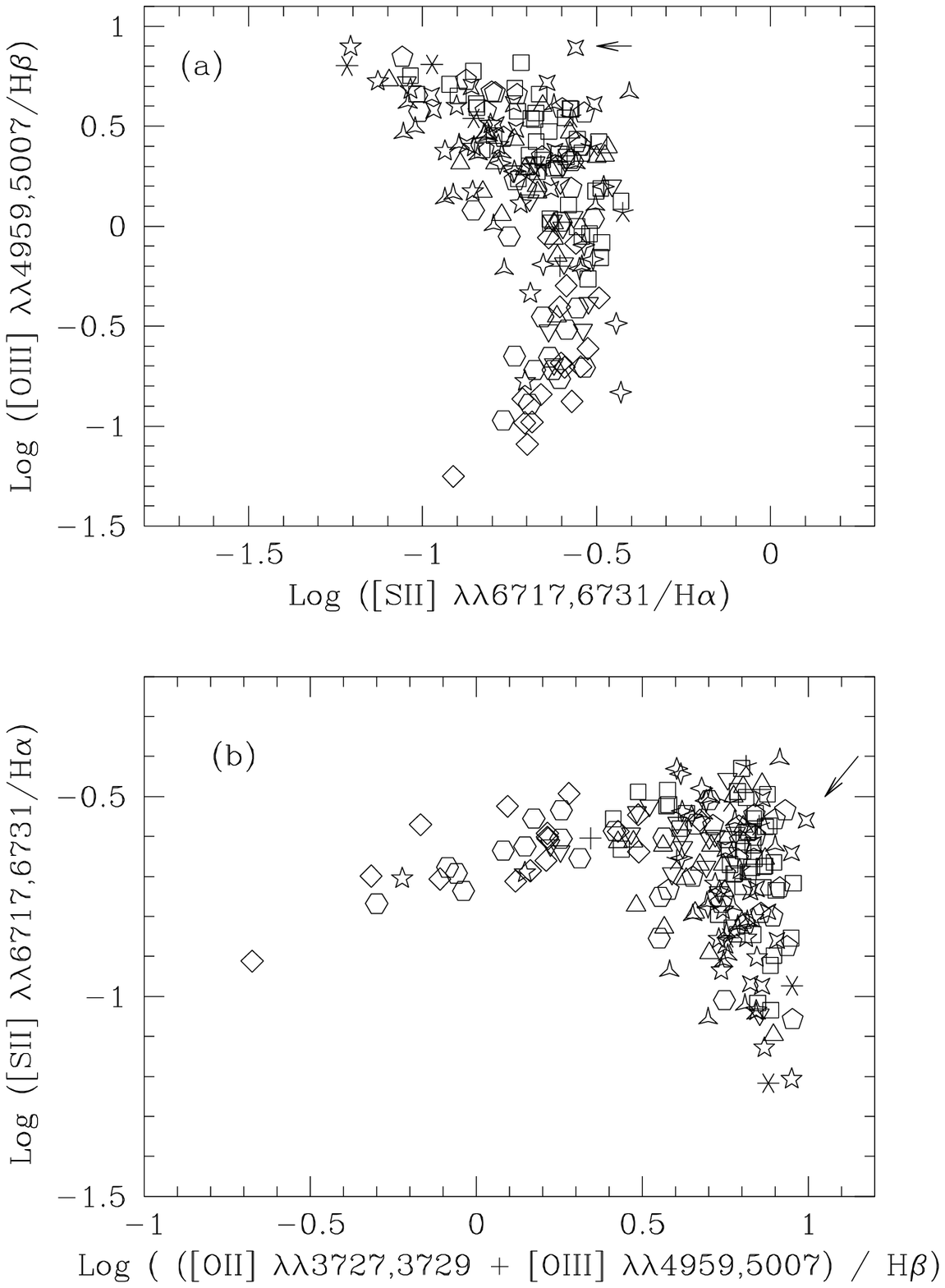,width=6.in,bbllx=0pt,bblly=100pt,bburx=600pt,bbury=700pt,clip=t}
\vskip -0.5 truein
\figcaption[]{Emission--line diagnostic diagrams for [S II]. The symbols are the
same as in Figure 4.  (a) The H II region sequence formed by [S II] and [O III] normalized
by the Balmer lines. (b) The H II region sequence formed by R$_{23}$ vs. [S II]/H$\alpha$.
In both panels, NGC 4395--003--003 is marked by an arrow. 
\label{fig:sh}}

\psfig{figure=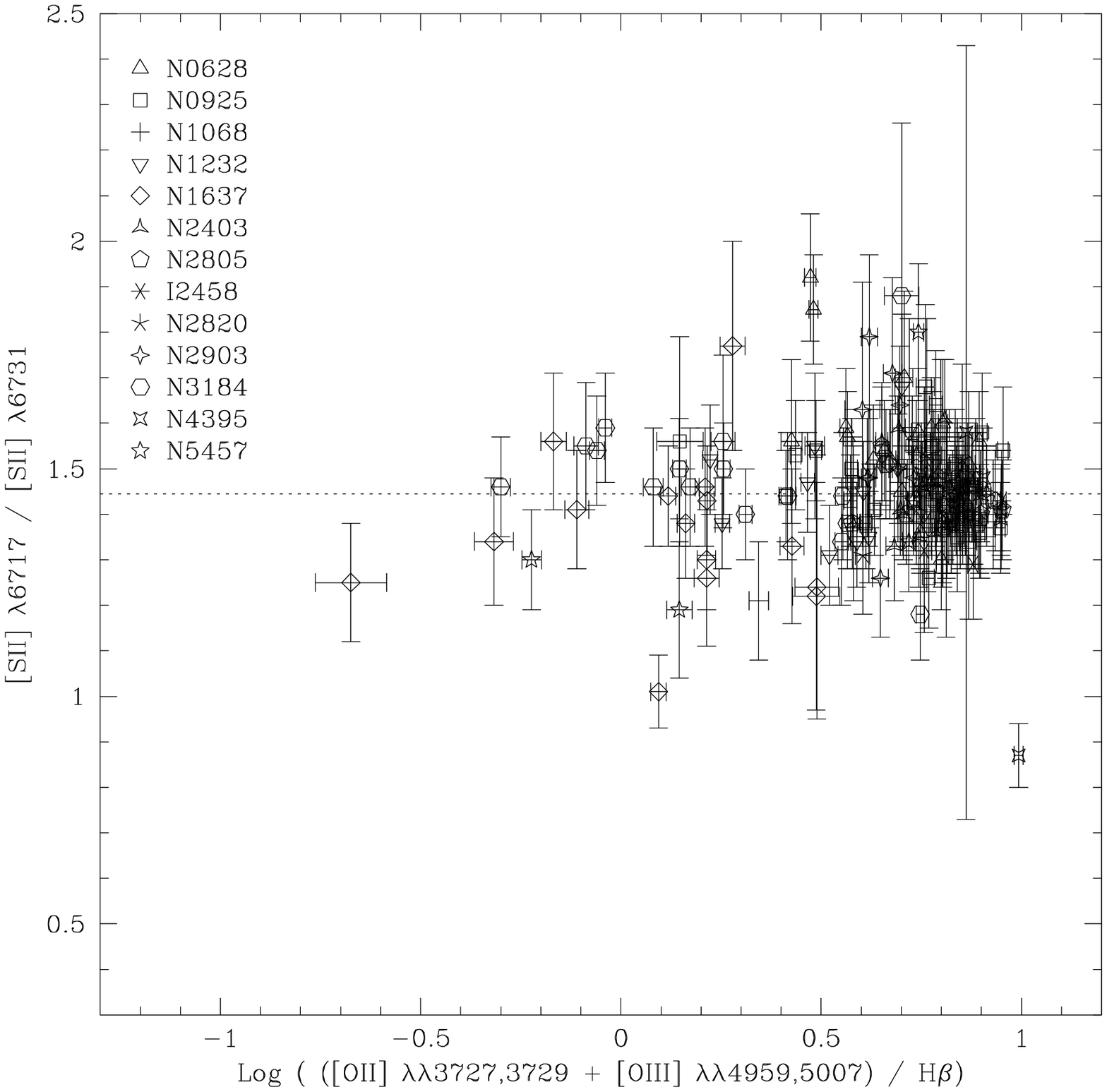,width=6.in,bbllx=0pt,bblly=100pt,bburx=600pt,bbury=700pt,clip=t}
\vskip -0.5 truein
\figcaption[]{The density sensitive line ratio [S II] $\lambda$6717/6731 as a function
of R$_{23}$. The symbols are the same as in Figure 4. The dashed line indicates the maximum
value of the [S II] ratio in the low density limit.  The majority of the H II regions
fall within the low density limit. \label{fig:density}}

\psfig{figure=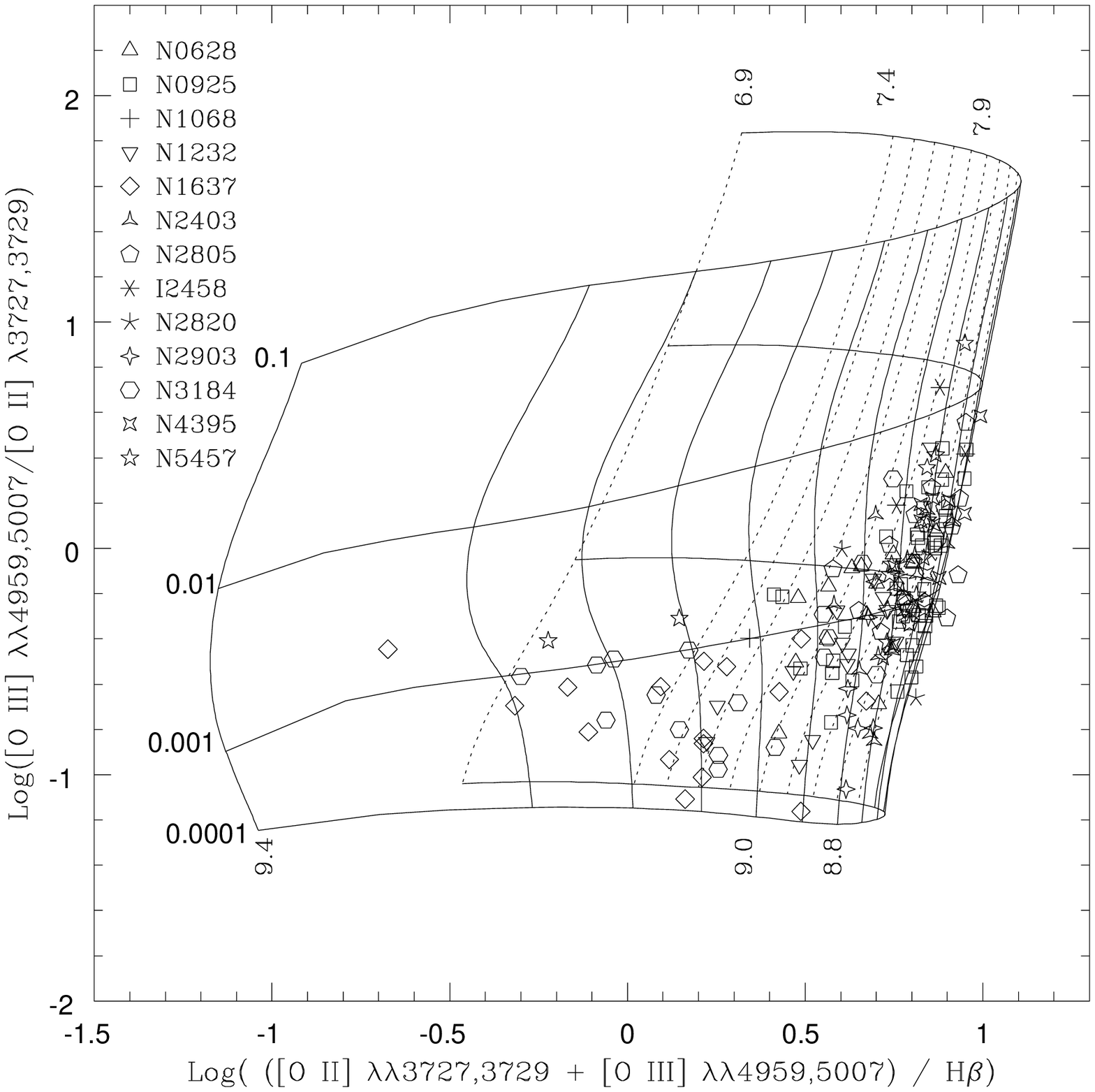,width=6.in,bbllx=0pt,bblly=100pt,bburx=600pt,bbury=700pt,clip=t}
\vskip -0.5 truein
\figcaption[]{The model grid of the R$_{23}$ relation from McGaugh (1991). The
locations of the H II regions are marked; the symbols are the same as in Figure 4. 
The ambiguity between the high and low abundance sides of the surface can be resolved 
based on the strength of the [N II] lines. \label{fig:r23}}

\psfig{figure=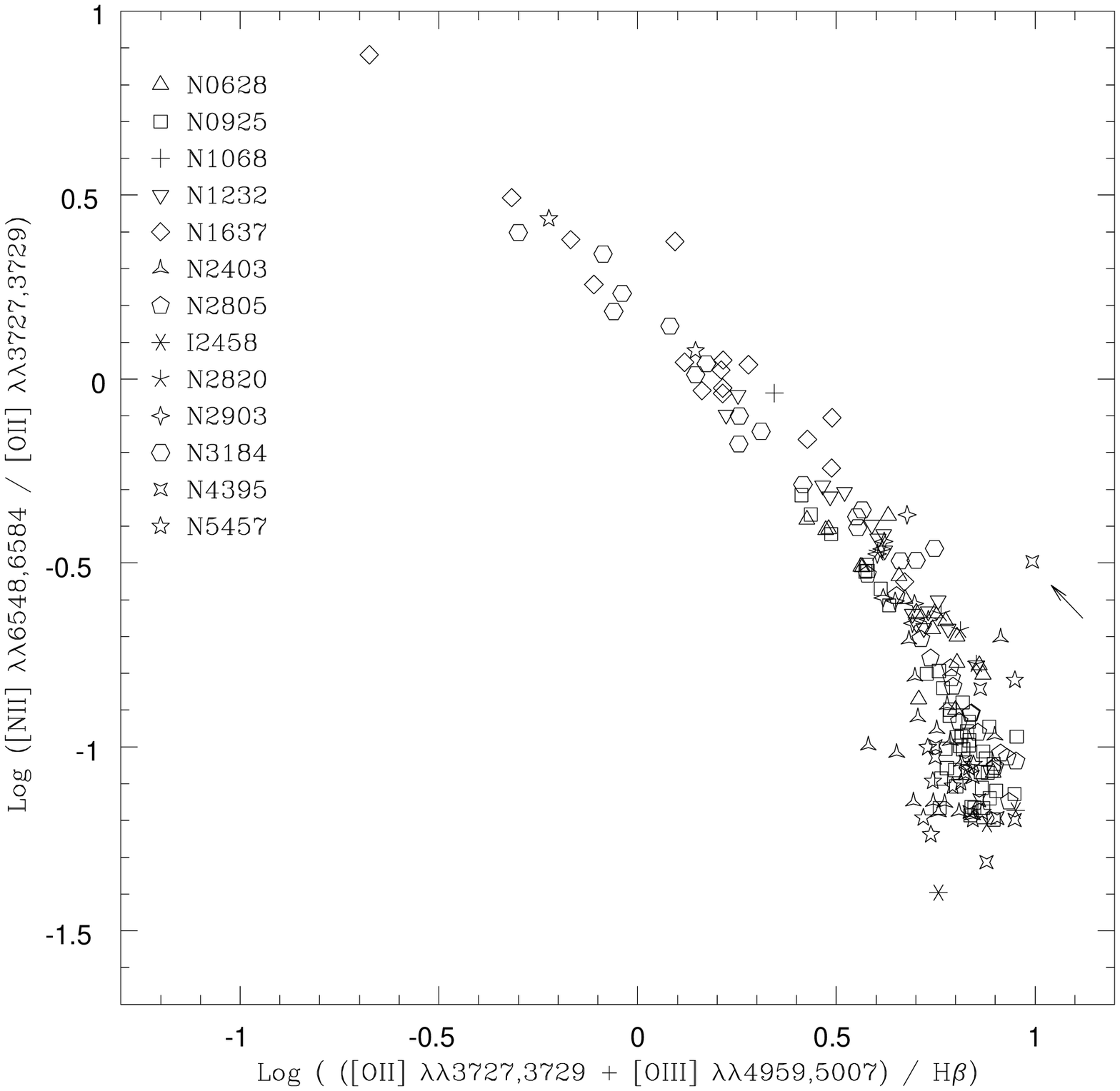,width=6.in,bbllx=0pt,bblly=100pt,bburx=600pt,bbury=700pt,clip=t}
\vskip -0.5 truein
\figcaption[]{The [N II]/[O II] diagnostic diagram.
The symbols are the same as in Figure 4. The one outlying
point is NGC 4395--003--003.  Low values of [N II]/[O II] and
high values of [O II]+[O III]/H$\beta$ are found for the outermost H II 
regions in NGC 925, NGC 2403, NGC 2805, NGC 4395, and NGC 5457, indicating that these
are low abundance H II regions. \label{fig:no} }

\psfig{figure=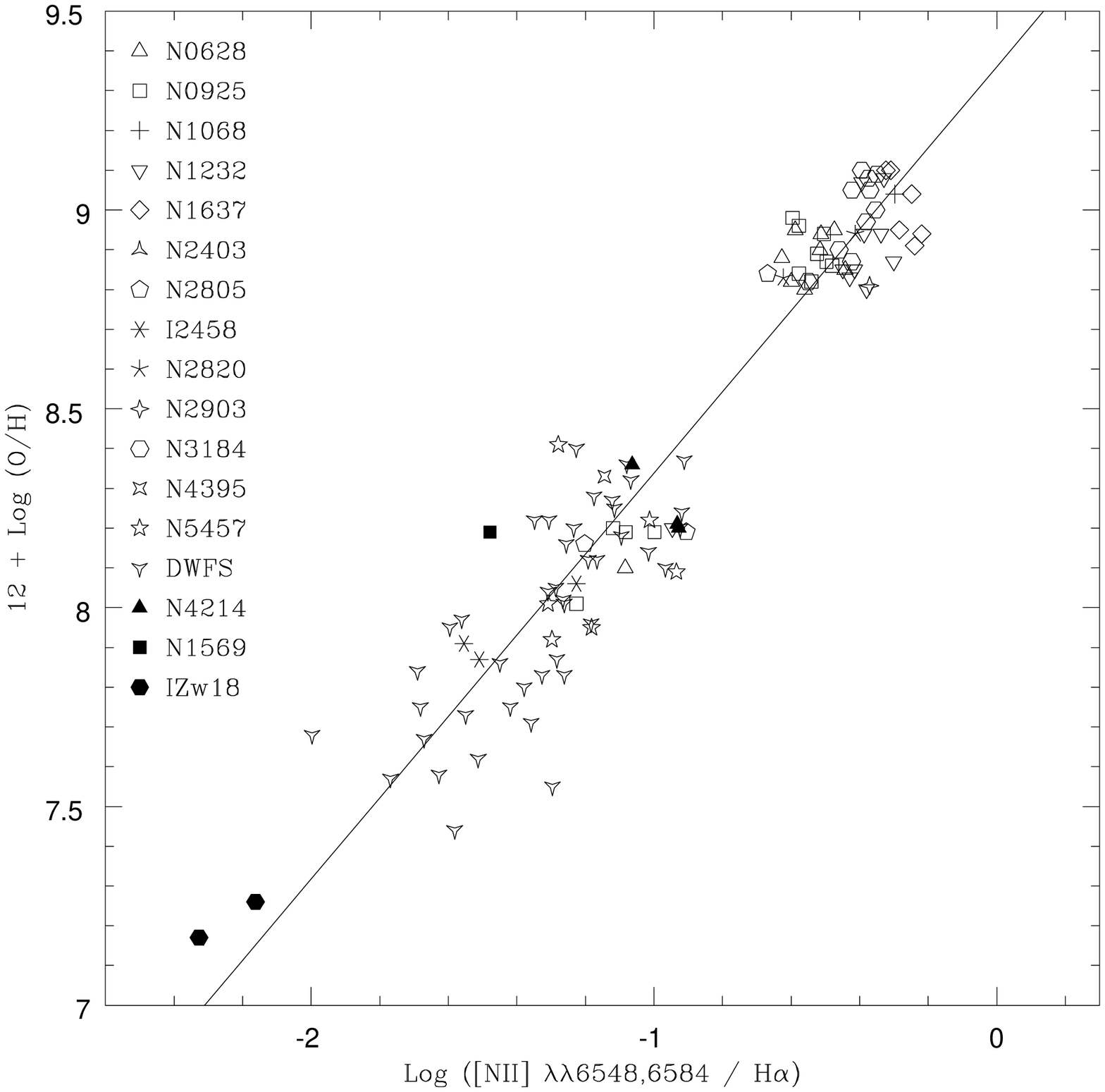,width=6.in,bbllx=0pt,bblly=100pt,bburx=600pt,bbury=700pt,clip=t}
\vskip -0.5 truein
\figcaption[]{A semi--empirical technique for determining oxygen abundances.  The
line shows the least squares fit to the low (12 + log (O/H) $<$ 8.2) and high
(8.8 $<$ 12 + log (O/H) $<$ 9.1) abundance H II regions and HII regions
from a dwarf galaxy sample (van Zee \etal 1997).  Also plotted are data points
for H II regions in NGC 4214 (Kobulnicky \& Skillman 1996), NGC 1569 (Kobulnicky \&
Skillman 1997), and I Zw 18 (Skillman \& Kennicutt 1993).   \label{fig:oh}}

\psfig{figure=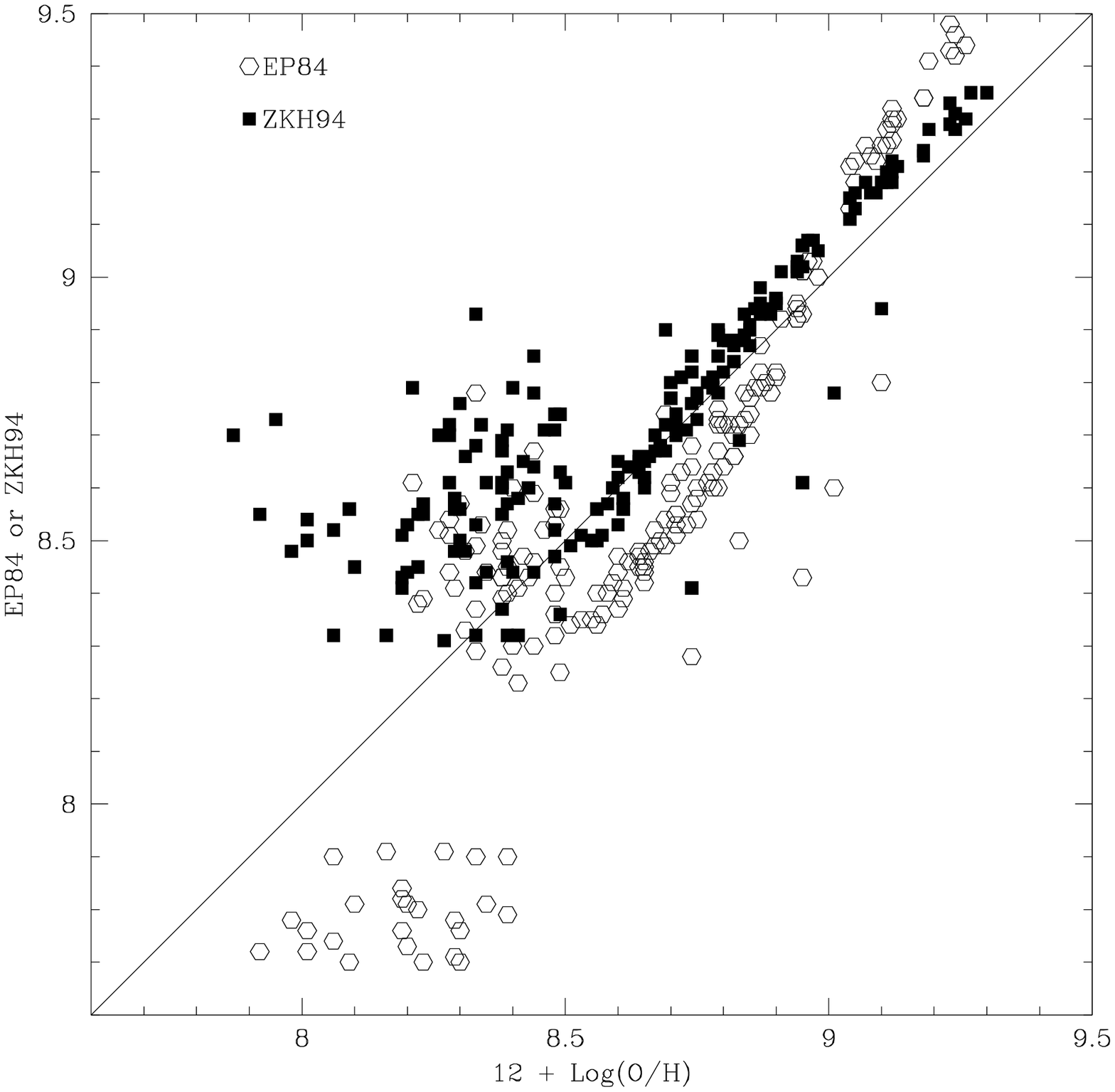,width=6.in,bbllx=0pt,bblly=100pt,bburx=600pt,bbury=700pt,clip=t}
\vskip -0.5 truein
\figcaption[]{Comparison between R$_{23}$ calibrations.  The large scatter
for 12 + log (O/H) $<$ 8.5 is due to the turnover in the R$_{23}$ relation
which has not been accounted for in the ZKH (Zaritsky \etal 1994) calibration. 
The six high abundance points which are offset from both the EP84 (Edmunds \& Pagel 1984)
and ZKH calibrations are from H II regions where the [N II]/H$\alpha$ calibration was 
used instead of the pure R$_{23}$ relation.  Two of these points are from the edge--on 
spiral NGC 2820, where the combination of uncertain extinction corrections and the 
superpositions of multiple H II regions make the empirical calibration highly uncertain.
While there are systematic differences between the calibrations, at high abundances 
the ZKH calibration is very similar to the calibration used here. 
In contrast, the calibration of EP84 would result in much steeper abundance gradients.
 \label{fig:zkh}}

\psfig{figure=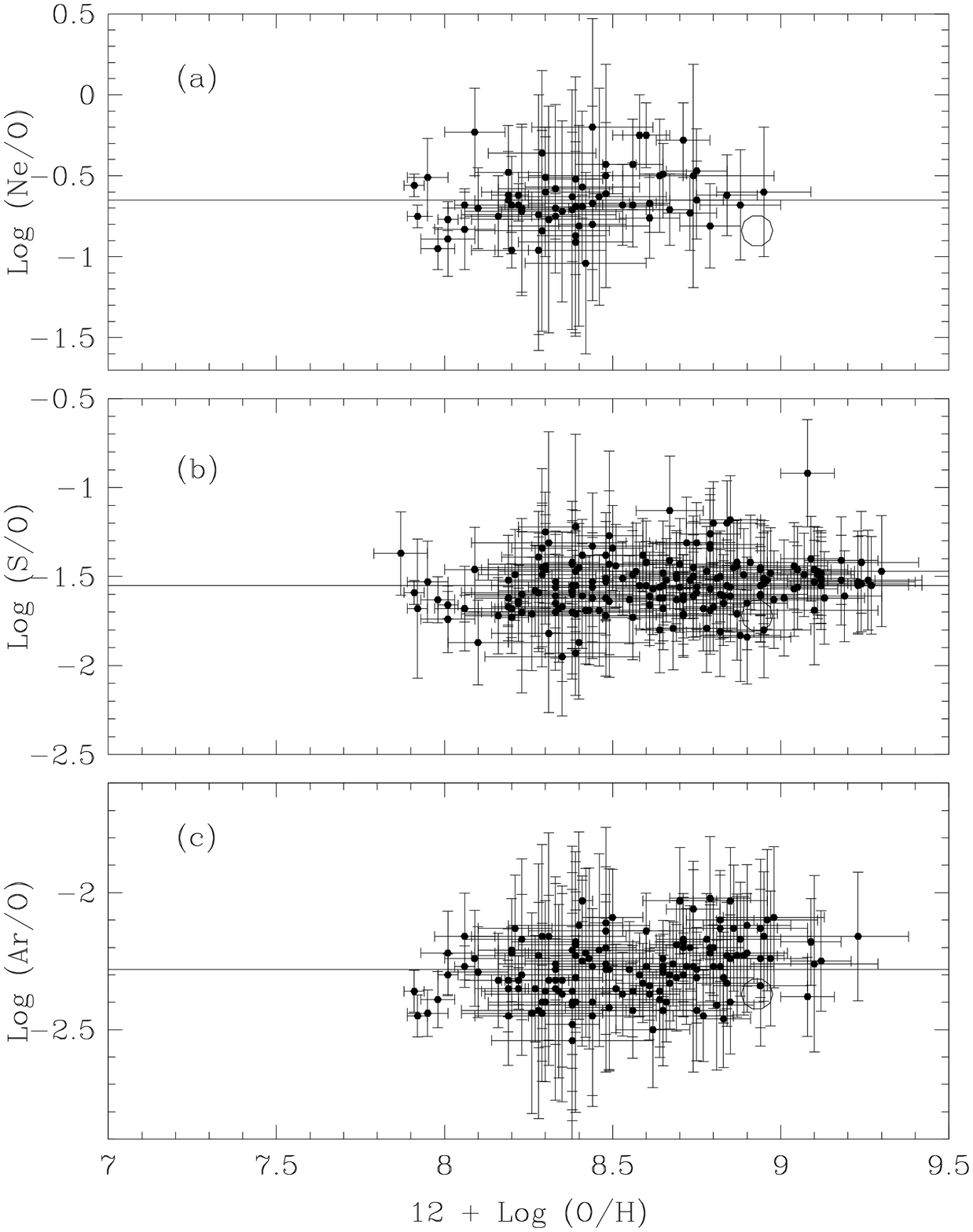,width=6.in,bbllx=0pt,bblly=0pt,bburx=600pt,bbury=750pt,clip=t}
\vskip -0.5 truein
\figcaption[]{The relative enrichment of the $\alpha$ elements. In all three panels,
the mean value is denoted by a solid line.  The solar value (Anders \& Grevesse 1989)
is represented by an open circle, where the symbol size is indicative of the errors.
(a) Log (Ne/O) as a function of oxygen abundance. (b) Log (S/O) as a function of oxygen 
abundance. (c) Log (Ar/O) as a function of oxygen abundance. \label{fig:alpha}}

\psfig{figure=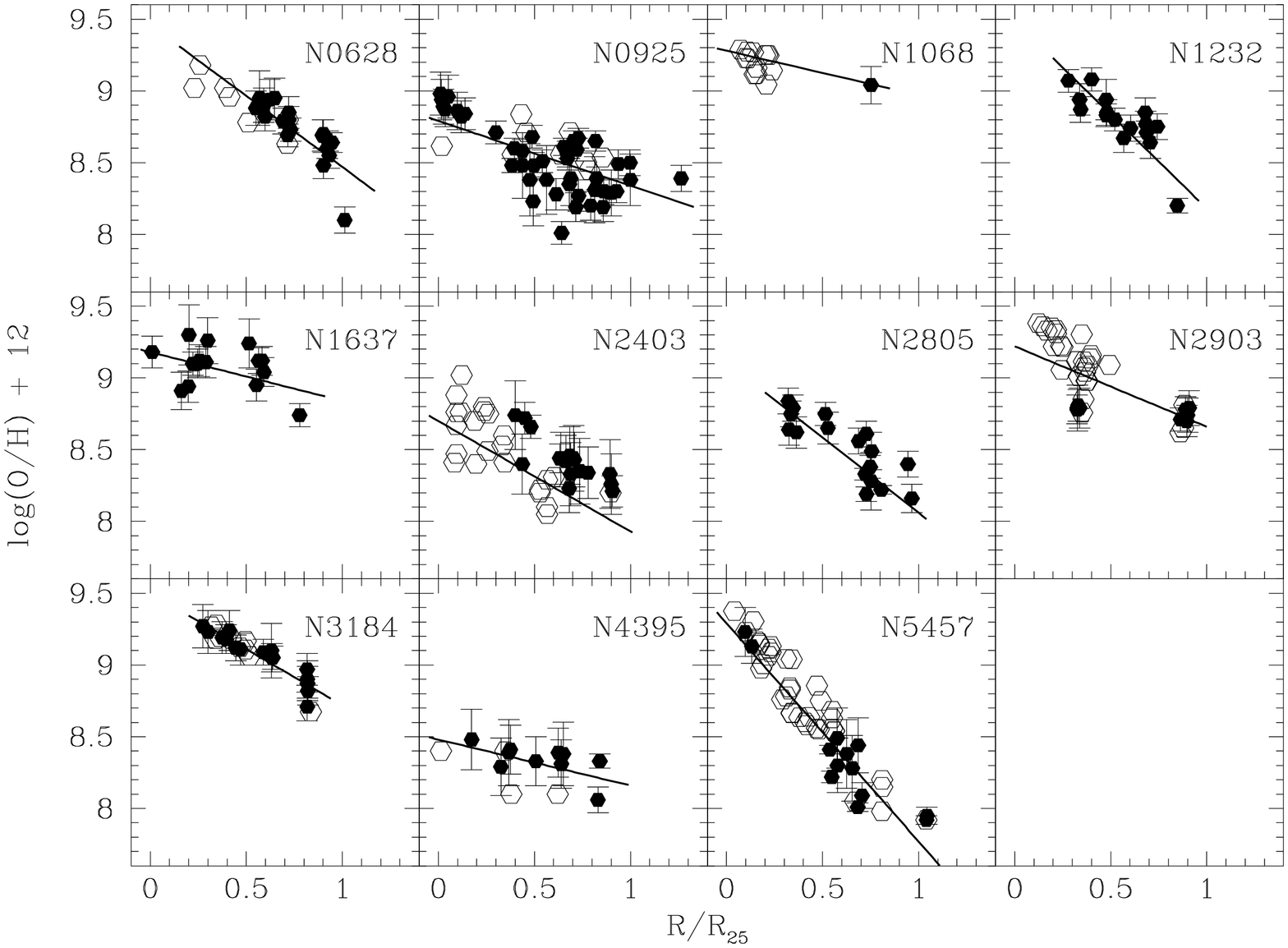,angle=-90.,width=6.in,bbllx=0pt,bblly=0pt,bburx=700pt,bbury=700pt}
\vskip -0.9 truein
\figcaption[]{The observed oxygen abundance gradients in all 11 spiral galaxies.
The filled symbols represent \ion{H}{2} regions from the present study. 
The open circles represent data from the literature:
NGC 628-- McCall \etal (1985); NGC 925-- Zaritsky \etal (1994); 
NGC 1068-- Evans \& Dopita (1987),
Oey \& Kennicutt (1993); NGC 2403-- McCall \etal (1985), Fierro \etal (1986),
Garnett \etal (1997); NGC 2903-- McCall \etal (1985), Zaritsky \etal (1994); 
NGC 3184-- Zaritsky \etal (1994); NGC 4395-- McCall \etal (1985); 
NGC 5457-- Kennicutt \& Garnett (1996). The solid lines
illustrate the derived oxygen abundance gradients. 
 \label{fig:rad_oh}}


\begin{references}
\reference{ACHV79} Alloin, D., Collin--Soufrin, S., Joly, M., \& Vigroux, L. 1979, \aap, 78, 200
\reference{AG89} Anders, E., \& Grevesse, N. 1989, Geochim. Cosmochim. Acta, 53, 197
\reference{BPT81} Baldwin, J. A., Phillips, M. M., \& Terlevich, R. 1981, \pasp, 93, 5
\reference{BR92} Belley, J., \& Roy, J.--R. 1992, \apjs, 78, 61 
\reference{DDH87} De Robertis, M. M., Dufour, R. J., \& Hunt, R. W. 1987, JRASC, 81, 195
\reference{RC3} de Vaucouleurs, G., de Vaucouleurs, A., Corwin, H. G., Buta, R., Paturel, G., 
            \& Fouqu\'e, P. 1991, Third Reference Catalogue of Bright Galaxies 
            (Springer, New York) (RC3)
\reference{DE86} Dopita, M. A., \& Evans, I. N. 1986, \apj, 307, 431
\reference{E90} Edmunds, M. G. 1990, \mnras, 246, 678
\reference{EP84} Edmunds, M. G., \& Pagel, B. E. J. 1984, \mnras, 211, 507 (EP84)
\reference{ED87} Evans, I. N., \& Dopita, M. A., 1987, \apj, 319, 662 
\reference{FGW98} Ferguson, A. M. N., Gallagher, J. S., \& Wyse, R. F. G. 1998, \aj, in press
\reference{FTP86} Fierro, J., Torres--Peimbert, S., \& Peimbert, M. 1986, \pasp, 98, 1032
\reference{FM88} Freedman, W. L., \& Madore, B. F. 1988, \apjl, 332, L63
\reference{GK94} Garnett, D. R., \& Kennicutt, R. C. 1994, 426, 123 
\reference{GSSSD97} Garnett, D. R., Shields, G. A., Skillman, E. D., Sagan, S. P., \& Dufour, R. J. 
            1997, \apj, 489, 63 
\reference{GK92} G\"otz, M., \& K\"oppen, J. 1992, \aap, 262, 455
\reference{GM82} G\"usten, R., \& Mezger, P. G. 1982, Vistas, Astro., 26, 159
\reference{HH95} Henry, R. B. C., \& Howard, J. W. 1995, \apj, 438, 170
\reference{HFS95} Ho, L. C., Filippenko, A. V., \& Sargent, W. L. W. 1995, \apjs, 98, 477
\reference{Ketal96} Kelson, D. D., Illingworth, G. D., Freedman, W. F., Graham, J. A., Hill, R.,
          Madore, B. F., Saha, A., Stetson, P. B., Kennicutt, R. C., Mould, J. R., Hughes, S. M., 
          Ferrarese, L., Phelps, R., Turner, A., Cook, K. H., Ford, H., Hoessel, J. G., \& Huchra, J.
          1996, \apj, 463, 26
\reference{KG96} Kennicutt, R. C., \& Garnett, D. R. 1996, \apj, 456, 504 
\reference{KS96} Kobulnicky, H. A., \& Skillman, E. D. 1996, 471, 211
\reference{KS97} Kobulnicky, H. A., \& Skillman, E. D. 1997, 489, 636
\reference{MR94} Martin, P., \& Roy, J.--R. 1994, \apj, 424, 599 
\reference{MRS85} McCall, M. L., Rybski, P. M., \& Shields, G. A. 1985, \apjs, 57, 1 (MRS)
\reference{M91} McGaugh, S. S. 1991, \apj, 380, 140
\reference{OK93} Oey, M. S., \& Kennicutt, R. C. 1993, \apj, 411, 137 
\reference{O97} Olofsson, K. 1997, \aap, 321, 29
\reference{O89} Osterbrock, D. E. 1989, Astrophysics of Gaseous Nebulae and Active Galactic 
        Nuclei (University Science Books, Mill Valley)
\reference{PEBCS79} Pagel, B. E. J., Edmunds, M. G., Blackwell, D. E., Chun, M. S., \& Smith, G. 
        1979, \mnras, 189, 95
\reference{PSTE92} Pagel, B. E. J., Simonson, E. A., Terlevich, R. J., \& Edmunds,
        M. G. 1992, \mnras, 255, 325
\reference{PC69} Peimbert, M., \& Costero, R. 1969, Bol. Obs. Tonantzintla y Tacubaya, 5, 3
\reference{PVMFKB95} Peletier, R. F., Valentijn, E. A., Moorwood, A. F. M., Freudling, W., 
        Knapen, J. H., \& Beckman, J. E. 1995, \aap, 300, L1
\reference{PE91} Phillipps, S., \& Edmunds, M. G. 1991, \mnras, 251, 84
\reference{RBDM96} Roy, J.--R., Belley, J., Dutil, Y., \& Martin, P. 1996, \apj, 460, 284 
\reference{SDH92} Scowen, P. A., Dufour, R. J., \& Hester, J. J. 1992, \aj, 104, 92
\reference{S71} Searle, L. 1971, \apj, 168, 327
\reference{S83} Shaver, P. A., McGee, R. X., Newton, L. M., Danks, A. C., Pottasch, S. R.
          1983, \mnras, 204, 53
\reference{Setal96} Silbermann, N. A., Harding, P., Madore, B. F., Kennicutt, R. C., Saha, A., 
          Stetson, P. B., Freedman, W. L., Mould, J. R., Graham, J. A., Hill, R. J., Turner, A., 
          Bresolin, R., Ferrarese, L., Ford, H., Hoessel, J. G., Han, M., Huchra, J., Hughes, S. M. 
          G., Illingworth, G. D., Phelps, R., \& Sakai, S. 1996, \apj, 470, 1
\reference{S89}   Skillman, E. D. 1989, \apj, 347, 883
\reference{SK93} Skillman, E. D., \& Kennicutt, R. C. 1993, \apj, 411, 655
\reference{SKS97} Skillman, E. D., Kennicutt, R. C., \& Shields, G. A., \& Zaritsky, D. 
          1996, \apj, 462, 147
\reference{SR97} Smartt, S. J. \& Rolleston, W. R. J. 1997, \apjl, 481, L47
\reference{S80} Stasi\'nska, G. 1980, \aap,  84, 320
\reference{S90}   Stasi\'nska, G. 1990, \aaps, 83, 501
\reference{SL96}  Stasi\'nska, G., \& Leitherer, C. 1996, \apjs, 107, 661
\reference{TIL95} Thuan, T. X., Izotov, Y. I., \& Lipovetsky, V. A. 1995, \apj, 445, 108
%\reference{T80} Tinsley, B. M. 1980, Fund. Cosmic Phys., 5, 287
\reference{vHS97} van Zee, L., Haynes, M. P., \& Salzer, J. J. 1997, \aj, 114, 2479
\reference{vSH98} van Zee, L., Salzer, J. J., \& Haynes, M. P. 1998, \apjl, 497, L1 (Paper I)
\reference{VE92} Vila--Costas, M. B., \& Edmunds, M. G. 1992, \mnras, 259, 121
\reference{VE96} V\'\i lchez, J. M. \& Esteban, C. 1996, \mnras, 280, 720
\reference{WR97} Walsh, J. R., \& Roy, J.--R. 1997, \mnras, 288, 715
\reference{WvA86} Wevers, B. M. H. R., van der Kruit, P. C., \& Allen, R. J. 1986, \aaps, 66, 505
\reference{Z92} Zaritsky, D. 1992, \apjl, 390, L73
\reference{ZKH94} Zaritsky, D., Kennicutt, R. C., \& Huchra, J. P. 1994, \apj, 420, 87 (ZKH)
\end{references}
\end{document}